\documentclass[aps,prx,reprint, amsmath, amssymb,superscriptaddress]{revtex4-1}

\usepackage{bm}
\usepackage[retainorgcmds]{IEEEtrantools}
\usepackage{graphicx}
\usepackage{mathrsfs}
\usepackage{amsmath}
\usepackage{amssymb}
\usepackage{color}
\usepackage{amsfonts}
\usepackage{times,txfonts}
\usepackage{nicefrac}
\usepackage[colorlinks=true,linkcolor=blue,urlcolor=blue,citecolor=blue,pdfusetitle]{hyperref}
\usepackage{physics}
\usepackage{soul}

\usepackage[dvipsnames]{xcolor}
\definecolor{mypink}{rgb}{0.858, 0.188, 0.478}
\usepackage[authormarkup=none]{changes}
\definechangesauthor[name={Gabriel}, color=red]{g}

\newcommand{\trans}{^{\text{T}}}

\begin{document}

\title{Memory kernel and divisibility of Gaussian Collisional Models}
\date{\today}
\author{Rolando Ramirez Camasca}
\email{rcamasca@if.usp.br}
\affiliation{Instituto de F\'isica da Universidade de S\~ao Paulo,  05314-970 S\~ao Paulo, Brazil.}
\author{Gabriel T. Landi}
\email{gtlandi@if.usp.br}
\affiliation{Instituto de F\'isica da Universidade de S\~ao Paulo,  05314-970 S\~ao Paulo, Brazil.}

\begin{abstract}
Memory effects in the dynamics of open systems have been the subject of significant interest in the last decades.
The methods involved in quantifying this effect, however, are often difficult to compute and may lack  analytical insight.  
With this in mind,  we consider 
Gaussian collisional models, where non-Markovianity is introduced by means of additional interactions between neighboring environmental units. 
By focusing on continuous-variable Gaussian dynamics, we are able to analytically study models of arbitrary size.
We show that the dynamics can be cast in terms of a  Markovian Embedding of the covariance matrix, which yields  closed form expressions for the  memory kernel that governs the dynamics, a quantity that can seldom be computed analytically. 
The same is also possible for
a divisibility monotone, based on the complete positivity of intermediate maps.
We analyze in detail two types of interactions, a beam-splitter implementing a partial SWAP and a two-mode squeezing, which entangles the ancillas and, at the same time, feeds excitations into the system. 
By analyzing the memory kernel and divisibility for these two representative scenarios, our results  help to shed light on the intricate mechanisms behind memory effects in the quantum domain.
\end{abstract}

\maketitle{}

%
%
\section{\label{sec:int}Introduction:}
%
%

The growing interest in quantum information processing applications has highlighted the need for furthering our knowledge on the notion of \emph{information flow}. Unlike classical systems, in the quantum realm information leaks are much more efficient, so that when a system interacts with an environment, information about the former is inevitably transferred to the latter. 
When the environment is very large and complex, this information may never return. In this case the dynamics is called Markovian. In general, however, there may be a partial backflow of information, which characterizes a non-Markovian evolution~\cite{doob1990stochastic}. From the point of view of causality, this backflow quantifies the ability of the dynamics to communicate past information to the future \cite{binder2018practical}.
Non-Markovianity therefore touches at the core of  information processing, which justifies the need for detailed studies.


Considerable attention was given in recent years on how to characterize and quantify non-Markovianity in the quantum domain (see~\cite{rivas2014quantum,breuer2016colloquium} for two recent reviews).
Due to the richness involved, however, there is no single approach capable of capturing its full essence.
The most important notion is that of map divisibility:  non-Markovianity requires that the underlying dynamical map should not be divisible~\cite{breuer2009measure,chruscinski2018divisibility}.
The notion of information flow, on the other hand, relies on information-theoretic quantifiers and is thus not uniquely defined. The most widely used measures involve the trace distance~\cite{breuer2009measure,vasile2011quantifying,laine2010measure,chruscinski2018divisibility} between different initial states or  entanglement~\cite{rivas2010entanglement} between the system and an ancilla.
Several other quantifiers have also been explored~\cite{hou2011alternative,luo2012quantifying,chruscinski2012markovianity,chruscinski2014witnessing,costa2014monogamy,strasberg2018response,souza2015gaussian,fanchini2014non,lu2010quantum}. 

A much older notion of non-Markovianity is that of a memory kernel, as present already in the seminal works of Nakajima and Zwanzig. 
The basic idea is that the open dynamics of a system's density matrix $\rho_S$ can, quite generally, be written as 
\begin{equation}\label{mem_kernel}
\frac{d \rho_S}{dt} =-i [H_S, \rho_S] +  \int\limits_0^t  \mathcal{K}_{t-t'}[ \rho(t') ] \; d t',
\end{equation}
where $\mathcal{K}_{t-t'}$, called the memory kernel (MK),  is a linear superoperator condensing all the information on how the evolution of $\rho$ at time $t$ depends on its past values. 
The MK has been studied intensively in the past~\cite{barnett2001hazards,shabani2005completely,hall2014canonical,Mazzola2010,Liu2019a}, as it provides clear insights onto the inner workings of non-Markovianity.
However, being a superoperator, it is generally difficult to compute analytically. 
We also mention in passing that, at a more operational level, MKs can be generalized to the notion of process tensor, which includes also all possible input and output operations performed in the system~\cite{pollock2018operational,pollock2018non,taranto2019structure}.

\begin{figure*}[!t]
\includegraphics[width=0.8\textwidth]{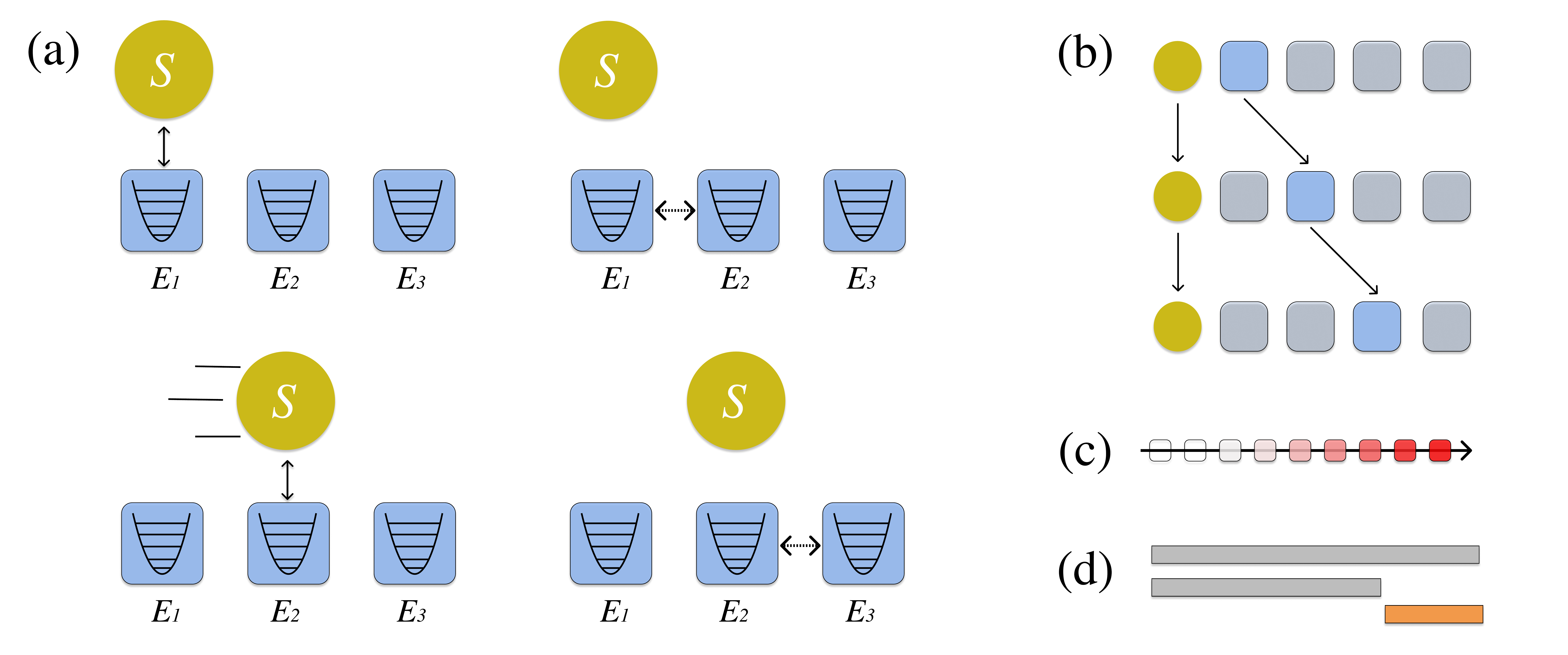}
\caption{Non-Markovian collisional models. 
(a) First few steps of the dynamics. The system-ancilla interactions $SE_n$ are interspersed by ancilla-ancilla interactions $E_nE_{n+1}$, which propagate information forward, making the dynamics non-Markovian in a fully controllable way. 
(b) Basic structure of the Markovian embedding dynamics~\eqref{gamma_embedding}, which is a map from the Hilbert space of $SE_n$ to that of $SE_{n+1}$.
(c) The memory kernel [Eq.~\eqref{mem_kernel}]  quantifies how different instants of the past affect the evolution at present times. 
(d) CP-divisibility. The maps in gray, from time 0 to $t_n$ or $t_m$ are, by construction, CPTP. But the intermediate map from $t_n$ to $t_m>t_n$ may not necessarily be. 
}
\label{fig:diagram}
\centering
\end{figure*}


Analyzing non-Markovianity for general environments is in general an extremely difficult task.
First, the calculations  quickly become impractical when the size of the bath is large. 
And second, realistic baths often have many additional features which tend to mask the effects one is interested in. 
This motivates the search for controllable models, where the degree of non-Markovianity can be finely tuned.
One way to accomplish this, which has seen an enormous surge in popularity in recent years, are through the so-called collisional models~\cite{Rau1963,Scarani2002,Ziman2002,Englert2002,Attal2006a,Pellegrini2009,karevski2009quantum,landi2014flux,Giovannetti2012,strasberg2017quantum,barra2015thermodynamic,de2018reconciliation}.
The basic idea is to replace the open dynamics of a system by  a series of sequential interactions between the system ($S$) and small environmental units $E_1, E_2, E_3\ldots$  (henceforth referred to as ancillas). 
All ancillas are prepared in the same state and each interaction only lasts for a fixed time, after which they never interact again.
This therefore leads to a stroboscopic dynamics for the system. 

The advantage of collisional models is that non-Markovianity can be introduced in a fully controllable. 
There are two main ways to do so.
The first is to consider that the ancillas already start correlated ~\cite{rybar2012simulation,bernardes2014environmental,bernardes2017coarse,mascarenhas2017quantum,man2018temperature}.
The other one is to assume information is transmitted between them during the process~\cite{ciccarello2013collision,ciccarello2013quantum,mccloskey2014non,ccakmak2017non,kretschmer2016collision,campbell2018system,lorenzo2017composite,jin2018non}.
Here we shall focus on the second case.
That is, we consider a scenario where neighboring ancillas $E_n E_{n+1}$ interact with each other in between the interactions $SE_n$ and $SE_{n+1}$ (see Fig.~\ref{fig:diagram}(a)). 
This additional interaction  \emph{signals} information from the past to the future, so that when the $SE_{n+1}$ interaction arrives, the ancilla $E_{n+1}$ will already contain some information about the system.

In this paper we overcome these difficulties by focusing on continuous-variable collisional models, undergoing Gaussian-preserving dynamics~\cite{serafini2017quantum,ferraro2005gaussian,adesso2007entanglement,adesso2014continuous,holevo2007one,caruso2006one,simon1987gaussian,simon1988gaussian,simon1994quantum}. 
The advantages that come with the Gaussian toolbox allows us to construct a complete framework for the study of non-Markovianity, which: 
(i) encompass a broad range of scenarios;
(ii) allows for the explicit construction and  computation of the memory kernel and
(iii) provides easy access to a CP-divisibility monotone, which can be directly compared with the memory kernel. 
The framework is also amenable to analytical calculations and  extremely efficient from a numerical perspective.
Thus, despite being restricted to Gaussian interactions, it offers multiple advantages over more general maps. 
Accompanying this paper, we also provide a complete numerical library for efficiently simulating Gaussian collisional models in Python 
\footnote{\href{https://github.com/gtlandi/gaussianonmark}{https://github.com/gtlandi/gaussianonmark}}
All plots in this paper were generated with this code.

The paper is divided as follows. 
The basic framework is developed in Sec.~\ref{sec:framework}, where we show that the full non-Markovian Gaussian dynamics can be converted to a set of matrix difference equations, written in terms of a Markovian embedding (Fig.~\ref{fig:diagram}(b)). 
This is the key step which makes the problem amenable to analytical calculations. 
Armed with this result, we then provide a full characterization of both the memory kernel (Sec.~\ref{sec:Memory_Kernel}) and the map divisibility (Sec.~\ref{sec:divisibility}).
Throughout the paper, our exposition will be example-oriented, with a focus on two specific types of interactions. 
The framework, however, is general and we will specify, in each part, how to properly make this generalization. 



%
%
\section{\label{sec:framework} Formal framework}
%
%

\subsection{Non-Markovian Collisional models}

We consider here the collisional model scenario presented in Fig.~\ref{fig:diagram}. 
A system $S$ is put to interact sequentially with an arbitrary number of environment ancillas $E_1, E_2, E_3, \ldots$. 
The ancillas are independent and identically prepared, each with initial density matrix $\rho_{E}$. 
The interaction between $S$ and $E_n$ is described by a unitary $U_n$.
After this, $S$ and $E_n$ never interact again.
If $U_n$ was the only interaction involved, the dynamics would be Markovian by construction. 

Here we make it non-Markovian in a controllable way, by introducing ancilla-ancilla collisions~\cite{ciccarello2013collision,ciccarello2013quantum,mccloskey2014non,ccakmak2017non,kretschmer2016collision,campbell2018system,lorenzo2017composite,jin2018non}.
That is, after collision $SE_n$,  but before $SE_{n+1}$, we put $E_nE_{n+1}$ to interact with each other by means of another unitary $V_{n,n+1}$. 
Since $E_n$ already interacted with $S$, it contains some information about it, which is then transmitted to $E_{n+1}$ via $V_{n,n+1}$. 
As a consequence, when the collision $SE_{n+1}$ starts, they will already contain some information about each other, obtained from $E_n$. 
Past information about $S$ can thus \emph{backflow} at $SE_{n+1}$, making the dynamics non-Markovian. 
This construction therefore provides a clean and controllable way of introducing non-Markovianity. In particular,  by assuming that $E_n$ only interacts with its neighbor $E_{n+1}$, we fix the memory length of the process. 
Collisional models with long-range interactions were discussed in~\cite{ccakmak2017non}.

Let $\rho^0 = \rho_S \otimes \rho_E \otimes \rho_E \otimes \ldots$ denote the initial state of the composite system $SE_1E_2\ldots$. 
We count time in integer steps, such that at time $n$ the collisions $SE_n$ and $E_nE_{n+1}$  already took place. 
That is, at time $n$ the system has already interacted with its corresponding ancilla $E_n$ \emph{and} this ancilla has already passed down its information to the next one. 
The map taking the composite system $SE_1E_2\ldots$ from $n-1$ to $n$ therefore reads 
\begin{equation}\label{basic_map}
	\rho^{n} = V_{n,n+1} U_n \; \rho^{n-1} \; U_n^\dagger V_{n,n+1}^\dagger.
\end{equation}
To avoid confusion we henceforth use superscripts to denote time so that $\rho^n$ refers to the global state of $SE_1E_2\ldots$ at time $n$. 
The map~\eqref{basic_map} involves only $SE_nE_{n+1}$. All ancillas $E_m$ with $m \geqslant n+2$ did not yet participate in the process and therefore remain in a product state with everything else. 
In addition, the ancillas with $m<n$ will never participate again and hence can be traced out (discarded). 
The process~\eqref{basic_map} can thus be equivalently written as 
\begin{equation}\label{basic_map_2}
	\rho_{S,E_n,E_{n+1}}^n =  V_{n,n+1} U_n \; \big(\rho_{SE_n}^{n-1} \otimes \rho_E\big) \; U_n^\dagger V_{n,n+1}^\dagger,
\end{equation}
where $\rho_{SE_n}^{n-1}$ is the state of $SE_n$ at time $n-1$ and $\rho_E$ refers to  the initial state of $E_{n+1}$.
After this interaction one may trace out $E_n$, leading to $\rho_{S,E_{n+1}}^n = \tr_{E_n} \rho_{S,E_n,E_{n+1}}^n$, which can then be fed again to Eq.~\eqref{basic_map_2} to evolve to the next step.

\subsection{Gaussian states and Gaussian operations}

Quantifying and understanding non-Markovianity in the collisional model~\eqref{basic_map_2} is a task that often has to be tackled numerically.
This is specially the case if one is interested in arbitrarily long times.
Here we are interested in obtaining analytical results.
To accomplish this, we therefore specialize now to the case of continuous-variable systems undergoing Gaussian-preserving dynamics. 
Our exposition, in what follows, will be example-oriented. 
However, the final results will be general [Eqs.~\eqref{gamma_def}, \eqref{gamma_embedding} and \eqref{XY_general}]. 

We assume the system is described by a bosonic annihilation operator $a$ and corresponding quadratures $Q = (a+a^\dagger)/\sqrt{2}$ and $P = i(a^\dagger-a)/\sqrt{2}$. 
Similarly, the ancillas are described by bosonic annihilation operators $b_1, b_2, \ldots$, with corresponding quadratures $q_n, p_n$. 
The generalization to a multimode system, or multimode ancillas, is straightforward.
We take the system-ancilla interaction $U_n$ in Eq.~\eqref{basic_map_2} to be a simple beam-splitter-type unitary, 
\begin{equation}\label{U}
	U_n = e^{\lambda_s (a^\dagger b_n - b_n^\dagger a)},
\end{equation}
described by a parameter $\lambda_s$.
One can view~\eqref{U} as an interaction with a Hamiltonian $ig(a^\dagger b_n - b_n^\dagger a)$ that lasts for a time $\tau$ such that $g\tau = \lambda_s$. Since we are only interested in the stroboscopic dynamics, we can omit these internal details for simplicity. 
As for the $E_n E_{n+1}$ collision unitary $V_{n,n+1}$, we shall explore two possibilities. 
The first is again a beam-splitter map
\begin{equation}\label{V_BS}
	V_{n,n+1} = e^{\lambda_e (b_n^\dagger b_{n+1} - b_{n+1}^\dagger b_n)}, 
\end{equation}
with interaction strength $\lambda_e$. 
We shall henceforth refer to this as the BS dynamics. 
In addition, we shall also look at a two-mode squeezing interaction (TMS), 
\begin{equation}\label{V_TMS}
	\tilde{V}_{n,n+1} = e^{\nu_e (b_n^\dagger b_{n+1}^\dagger-b_{n+1} b_{n})}, 
\end{equation}
with strength $\nu_e$. 
The reason behind this choice is related to the fact that two-mode squeezing interactions generate stronger forms of correlations (e.g. entanglement) between the ancillas. 
By contrasting~\eqref{V_BS} and~\eqref{V_TMS} we may therefore explore the role of quantum correlations in non-Markovianity. 

The unitaries~\eqref{U}-\eqref{V_TMS} are Gaussian preserving.
If we assume that the initial state is Gaussian, the dynamics will then be completely characterized by the first and second moments. 
We assume, for simplicity, that the first moments are initially zero, so that they will remain so throughout. 
The covariance matrix (CM) is defined as $\sigma_{ij} = \frac{1}{2} \langle \{ R_i, R_j \} \rangle$ where  $\bm{R} = (Q,P,q_1,p_1,q_2,p_2,\ldots)$.
The initial state is block-diagonal, of the form
\begin{equation}\label{Gamma_0}
\sigma^0 = \text{diag}\bigg(\theta^0, \epsilon, \epsilon, \epsilon, \ldots\bigg), 
\end{equation}
where each block is $2\times 2$: $\theta^0$ is the arbitrary initial CM of the system and $\epsilon$ is the initial CM of the ancillas (which are all the same, since we are assuming the ancillas are iid). 
In the analyses below we will usually take $\epsilon = \mathbb{I}_2/2$ (i.e., a vacuum state); but we leave it general for the moment.

The global dynamics of $SE_1E_2\ldots$ is unitary. As a consequence, the map~\eqref{basic_map} is translated into a symplectic evolution for the CM:
\begin{equation}\label{basic_map_sigma}
\sigma^n = S_{n,n+1} S_n \; \sigma^{n-1} S_n\trans S_{n,n+1}\trans, 
\end{equation}
where $S_n$ and $S_{n,n+1}$ are the symplectic matrices associated with the unitaries $U_n$ and $V_{n,n+1}$. 
The symplectic matrix associated to the beam-splitter interaction~\eqref{U} is remarkably simple because all entries become proportional to the $2\times2$ identity [this is partially because of the choice of phase in the exponent of~\eqref{U}].
For instance, the interaction $S_2$ between the $S$ and $E_2$ reads
\begin{equation}\label{S2_example}
	S_2 = \begin{pmatrix}
		x 		& 		0 	        	& 		y 		& 		0 	        	& 		\ldots  \\
		0 		&	1		&		0		&		0	        	&		\ldots  \\
		-y		&		0	        	&		x		&		0	        	&		\ldots  \\
		0 		&		0	        	&		0		&		1	&		\ldots  \\
		\vdots	&		\vdots	        &		\vdots	&		\vdots	        &		\ddots		
	\end{pmatrix},
\end{equation} 
where each entry is a $2\times2$ matrix, with $x = \cos(\lambda_s)$ and $y = \sin(\lambda_s)$. 
The extension to $S_n$ is straightforward.
The same structure also holds for the BS  unitary $V_{n,n+1}$ between $E_n E_{n+1}$ [Eq.~\eqref{V_BS}], except that now the position of the non-zero entries changes. For instance, 
\begin{equation}\label{S12_example}
	S_{1,2} = \begin{pmatrix}
		1 		& 		0 		& 		0 		& 		0 		& 		\ldots  \\
		0 		&		z		&		w		&		0		&		\ldots  \\
		0		&		-w		&		z		&		0		&		\ldots  \\
		0 		&		0		&		0		&		1		&		\ldots  \\
		\vdots	&		\vdots	&		\vdots	&		\vdots	&		\ddots		
	\end{pmatrix},
\end{equation}
where $z = \cos(\lambda_e)$ and $w = \sin(\lambda_e)$. 
The TMS interaction~\eqref{V_TMS} is slightly more complicated since some   entries are proportional to the identity, while others are proportional to the Pauli matrix $\sigma_z$; for instance, 
\begin{equation}\label{S12t_example}
	\tilde{S}_{1,2} =  \begin{pmatrix}
		1 		& 		0 				& 		0 				& 		0 		& 		\ldots  \\
		0 		&		\tilde{z}			&		\tilde{w}\sigma_z	&		0		&		\ldots  \\
		0		&		\tilde{w}\sigma_z	&		\tilde{z}			&		0		&		\ldots  \\
		0 		&		0				&		0				&		1		&		\ldots  \\
		\vdots	&		\vdots			&		\vdots			&		\vdots	&		\ddots		
	\end{pmatrix},
\end{equation}
with $\tilde{z} = \cosh(\nu_e)$ and $\tilde{w} = \sinh(\nu_e)$. 

The BS dynamics is completely characterized by the pair $(\lambda_s, \lambda_e)$, while the TMS dynamics is characterized by $(\lambda_s, \nu_e)$. 
On top of that, one also has the choice of ancilla initial state $\epsilon$, which in all analyzes below will be taken as the vacuum. 

More general Gaussian maps will continue to have a similar structure. 
The symplectic $S_n$ will have the form 
\begin{equation}\label{S2_general}
	S_2 = \begin{pmatrix}
		A 		& 		0 	        	& 		B 		& 		0 	        	& 		\ldots  \\
		0 		&	1		&		0		&		0	        	&		\ldots  \\
		C		&		0	        	&		D		&		0	        	&		\ldots  \\
		0 		&		0	        	&		0		&		1	&		\ldots  \\
		\vdots	&		\vdots	        &		\vdots	&		\vdots	        &		\ddots		
	\end{pmatrix},
\end{equation} 
for block matrices $A, B, C, D$. 
The matrices $S_n$ for other values of $n$ are obtained by simply placing $A,B,C,D$ at the correct positions. 
Note also that the condition that $S$ must be symplectic imposes constraints on $A,B,C,D$ which, however, are not particularly illuminating. 
Similarly, the $E_nE_{n+1}$ interaction reads
\begin{equation}\label{S12_general}
	\tilde{S}_{1,2} =  \begin{pmatrix}
		1 		& 		0 				& 		0 				& 		0 		& 		\ldots  \\
		0 		&		E			&		F	&		0		&		\ldots  \\
		0		&		G	&		J			&		0		&		\ldots  \\
		0 		&		0				&		0				&		1		&		\ldots  \\
		\vdots	&		\vdots			&		\vdots			&		\vdots	&		\ddots		
	\end{pmatrix},
\end{equation}
for block matrices $E, F, G, J$. Note that these two expressions also naturally contemplate the case where either the system or each ancilla are, individually, composed of multiple modes (which would simply affect the size of the matrices $A, \ldots, J$). 

\subsection{Matrix difference equations and Markovian embedding}

The biggest advantage of Gaussian collisional models, as we will now show, is that the full non-Markovian evolution can be converted into a simple system of matrix difference equations for only a handful of  entries of the full CM $\sigma^n$. 
As already discussed below Eq.~\eqref{basic_map_2}, the step from $\sigma^{n-1}$ to $\sigma^n$ involves only $S$, $E_n$ and $E_{n+1}$. 
At time $n-1$ the ancilla $E_{n+1}$ is still uncorrelated from the rest, whereas $S$ and $E_n$ are already correlated because of the previous step. 
Thus, the tripartite CM of $SE_nE_{n+1}$, at time $n-1$, will have the block structure
\begin{equation}\label{sigma_nm1}
\sigma_{SE_nE_{n+1}}^{n-1} = \begin{pmatrix}
	\theta^{n-1} 		&			\xi_n^{n-1}				&			0				\\[0.2cm]
	\xi_n^{n-1,\text{T}}	&			\epsilon_n^{n-1}			&			0		\\[0.2cm]
	0				&			0						&			\epsilon
\end{pmatrix},
\end{equation}
where $\epsilon_n^{n-1}$ is the state of ancilla $E_n$ at time $n-1$, which is no longer the original value $\epsilon$ because it already interacted with $E_{n-1}$ in the previous step. 
Moreover, $\xi_n^{n-1}$ are the correlations between $SE_n$ that were developed in the previous step. 

We now apply the map~\eqref{basic_map_sigma} to Eq.~\eqref{sigma_nm1}, using the matrices in Eqs.~\eqref{S2_example}-\eqref{S12t_example}. 
This will lead to a matrix $\sigma^n$ with many non-zero entries. 
However, as far as the dynamics of $S$ is concerned, only three entries are needed: the state of the system $\theta^n$, the state $\epsilon_{n+1}^n$ of ancilla $E_{n+1}$ and the correlations $\xi_{n+1}^n$ between $S$ and $E_{n+1}$. 

To gain intuition, let us first analyze the BS case, which is  simple since all blocks in Eq.~\eqref{S12_example} are proportional to the identity. 
Using Eqs.~\eqref{S2_example} and \eqref{S12_example} in~\eqref{basic_map_sigma}, one finds the following system of matrix difference equations:
\begin{IEEEeqnarray}{rCl}
	\theta^n &=& x^2 \theta^{n-1} + y^2 \epsilon_n^{n-1} +  xy (\xi_n^{n-1}+\xi_n^{n-1,\text{T}}),  \nonumber \\[0.2cm]
	\label{difference_equations_full_system}
	\epsilon_{n+1}^n &=& z^2 \epsilon + w^2 \Big[ x^2 \epsilon_n^{n-1} + y^2 \theta^{n-1} - x y (\xi_n^{n-1} + \xi_n^{n-1,\text{T}}) \Big],\\[0.2cm]
	\xi_{n+1}^n &=& w \Big[ x y (\theta^{n-1} - \epsilon_n^{n-1} ) + y^2 \xi_n^{n-1,\text{T}}-x^2 \xi_n^{n-1} \Big].\nonumber
\end{IEEEeqnarray}
The key point to bear in mind is that the quantities on the left and right-hand side refer to different ancillas: for instance, $\epsilon_{n+1}^n$ is the state of ancilla $E_{n+1}$ at time $n$, whereas $\epsilon_n^{n-1}$ is the state of $E_n$ at time $n-1$. 
Of course, one could also compute $\epsilon_n^n$, but this is not necessary for describing the dynamics of $S$.

The system of matrix difference equations~\eqref{difference_equations_full_system} contains the minimum amount of information required to fully account for the dynamics of $S$. 
These equations can also be recast in a more compact form using the notion of Markovian embedding~\cite{campbell2018system}.
The basic idea is to view Eq.~\eqref{difference_equations_full_system} as a quantum channel between different Hilbert spaces (Fig.~\ref{fig:diagram}(b)); 
more specifically, one which maps the CM of $SE_n$ to the CM of $SE_{n+1}$. 
We define the reduced CM of $SE_{n+1}$ at time $n$ as 
\begin{equation}\label{gamma_def}
\gamma_{n+1}^n \equiv \gamma^n = \begin{pmatrix}
\theta^{n} & \xi_{n+1}^n \\[0.2cm]
\xi_{n+1}^{n,\text{T}} & \epsilon_{n+1}^n 
\end{pmatrix},
\end{equation}
where the notation $\gamma^n$ will be used to simplify the expressions.
Eq.~\eqref{difference_equations_full_system} can then be written compactly as 
\begin{equation}\label{gamma_embedding}
\gamma^{n+1} = X \gamma^{n} X\trans + Y, 
\end{equation}
where the time index was shifted by 1. 
Here $X$ and $Y$ are $4\times4$ matrices with block form
\begin{equation}\label{XY_BS}
X = \begin{pmatrix}
	x 	& 	y \\[0.2cm]
	y w	&	-w x 
\end{pmatrix},
\qquad 
Y = \begin{pmatrix} 0 & 0 \\[0.2cm] 0 & z^2 \epsilon \end{pmatrix},
\end{equation}
where, again, each block is proportional to the identity.

Eq.~\eqref{gamma_embedding}  beautifully illustrates the notion of Markovian embedding.
It has the structure of a typical Gaussian CPTP map~\cite{serafini2017quantum}, being Markovian (time-local) by construction.
However, this Markovian dynamics takes place at the larger space of the system plus one  ancilla (which one, in specific, changes at each collision). 
Thus, we have embedded the non-Markovian dynamics into a Markovian dynamics at a larger space. 
Notice how the size of the space is directly related to the fact that we chose $E_n$ to only interact with its nearest neighbor $E_{n+1}$. 
That is, we fixed the memory length to be 1, which defines the size of the minimal space required for the embedding~\cite{campbell2018system}.  

The matrices~\eqref{XY_BS} refer to the beam-splitter unitary~\eqref{V_BS}. 
The generalization to the arbitrary Gaussian interactions~\eqref{S2_general} and~\eqref{S12_general} is similar, albeit more cumbersome. 
The result is 
\begin{equation}\label{XY_general}
X = \begin{pmatrix}
	A 	& 	B \\[0.2cm]
	GC 	&	GD
\end{pmatrix},
\qquad 
Y = \begin{pmatrix} 0 & 0 \\[0.2cm] 0 & J\epsilon J\trans
\end{pmatrix}.
\end{equation}
For instance, in the case of the TMS interaction, Eq.~\eqref{S12t_example}, one has $G = \tilde{w}\sigma_z$ and $J = \tilde{z}$, 
in addition to $A = D = x$, $B = y$ and $C = -y$ (which come from $S_n$ in~\eqref{S2_example}). 
One then finds that 
\begin{equation}\label{XY_TMS}
X = \begin{pmatrix}
	x 	& 	y \\[0.2cm]
	-y \tilde{w} \sigma_z	&	\tilde{w} x \sigma_z 
\end{pmatrix},
\qquad 
Y = \begin{pmatrix} 0 & 0 \\[0.2cm] 0 & \tilde{z}^2\epsilon \end{pmatrix}.
\end{equation}
The blocks in $X$ are therefore no-longer proportional to the identity, but some are proportional to $\sigma_z$. 

To summarize, the general non-Markovian dynamics will be described by the embedding~\eqref{gamma_embedding}, with $\gamma^n$ defined in~\eqref{gamma_def}, and with $X$ and $Y$ given by~\eqref{XY_general}.
This framework therefore provides a quite general platform, enabling one to study a broad range of situations.  

\begin{figure}[t]
\includegraphics[scale=0.6]{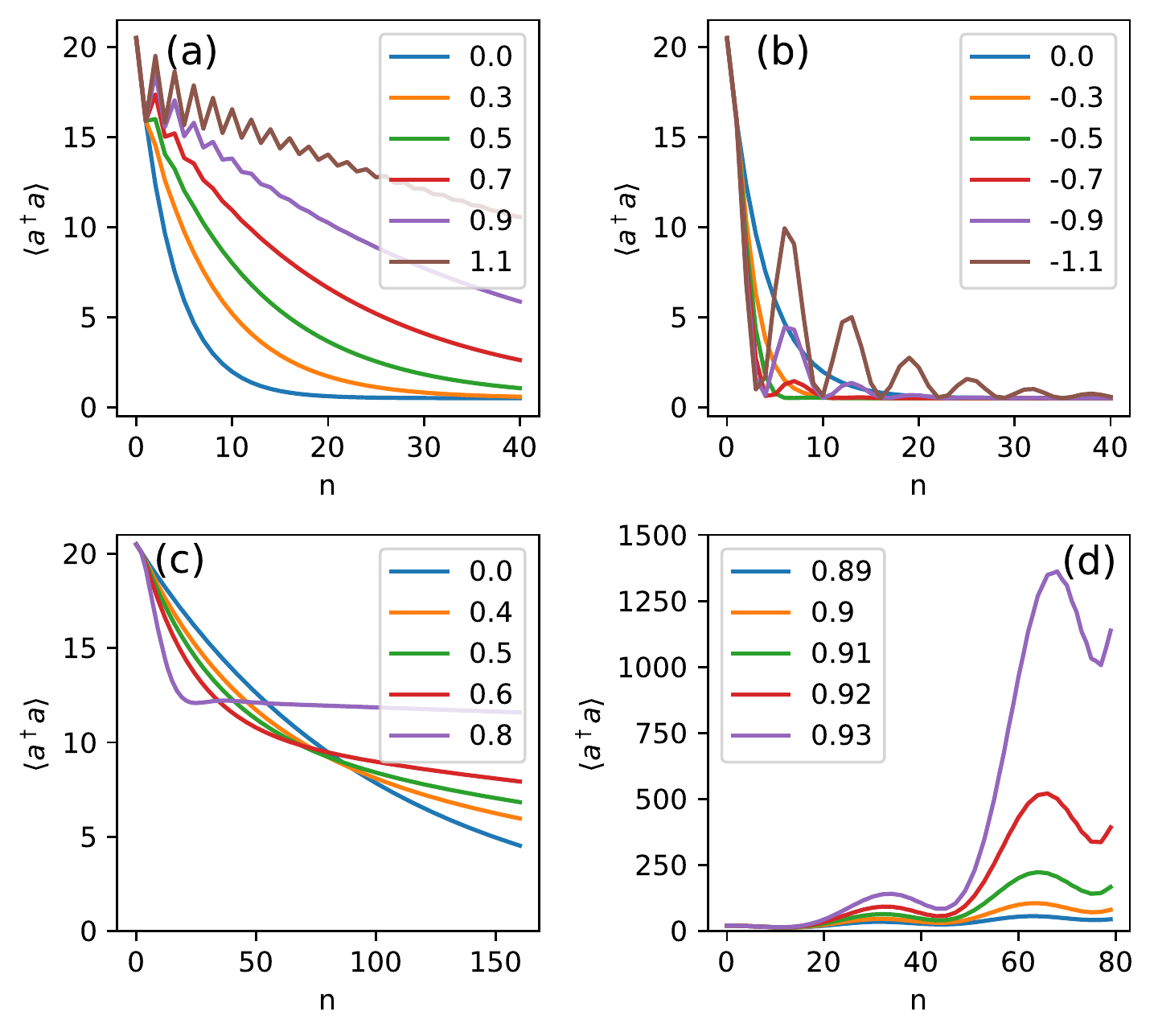}
\caption{
Number of excitations in the system as a function of time, computed from Eq.~\eqref{gamma_embedding}.
(a,b)  BS dynamics \eqref{XY_BS} with $\lambda_s = 0.5$ and different values of $\lambda_e$ (with $\lambda_e >0$ in (a) and $\lambda_e < 0$ in (b)). 
(c,d) Same, but for the TMS dynamics~\eqref{XY_TMS}, with $\lambda_s = 0.1$ and different values of $\nu_e$ (with
$\nu_e < \nu_e^\text{crit}$ in (a)  $\nu_e \geqslant \nu_e^\text{crit}$ in (b), where $\nu_e^\text{crit} = \sinh^{-1}(1)\simeq 0.8813$). 
The ancillas are assumed to start in the vacuum, and the system in a thermal state with $\langle a^\dagger a \rangle^0 = 20$. 
 }
\label{fig:ada}
\centering
\end{figure}


\subsection{Example dynamics}

Eqs~\eqref{gamma_embedding}-\eqref{XY_TMS} are the first main results of this paper. 
They provide a compact and efficient way of describing the non-Markovian dynamics of a bosonic mode in terms of a simple matrix difference equation for the augmented CM $\gamma^n$.
The reduced state of the system is always readily accessible from the first $2\times2$ block [Eq.~\eqref{gamma_def}].
Before proceeding to quantify the non-Markovianity of the process, we first illustrate the typical behavior of the BS and TMS maps,  by plotting the average system occupation $\langle a^\dagger a\rangle$ as a function of time for different values of the $E_n E_{n+1}$ interaction strength $\lambda_e$ (for the BS case) or $\nu_e$ (for the TMS case).
We choose the system to start in a thermal state with occupation number $\langle a^\dagger a \rangle^0 = 20$, while the ancillas start in the vacuum, $\epsilon = \mathbb{I}_2/2$. 
The results are summarized in Fig.~\ref{fig:ada}, for the BS (a,b) and TMS (c,d) evolutions. 

The BS dynamics is sensitive to the relative signs between $\lambda_s$ and $\lambda_e$ (and, consequently, of $y = \sin(\lambda_s)$ and $w = \sin(\lambda_e)$). 
This is an interference effect, which occurs due to the fact we are combining two beam-splitters [Eqs.~\eqref{U} and \eqref{V_BS}]. 
We emphasize this in Fig.~\ref{fig:ada}(a,b) by comparing $\lambda_e>0$ and $\lambda_e<0$, with $\lambda_s>0$. 
In both cases we see that for small $\lambda_e$ the excitations tend to decay monotonically, which is what one would expect of a Markovian BS interaction with a vacuum bath. 
For larger $\lambda_e$, on the other hand, the occupations present oscillations.
Since the interaction conserves the number of quanta, these revivals in excitations must necessarily be due to a backflow caused by the non-Markovian behavior. 
That is, some of the excitations that leave the system towards $E_n$ are transferred from $E_n$ to $E_{n+1}$ and then make it back into the system in the $SE_{n+1}$ interaction. 
The nature of these oscillations, however, is different whether $\lambda_e>0$ or $\lambda_e <0$, being fast in the former and slow in the latter. 
Irrespective of the value of $\lambda_e$, however, after an infinite time the system will always thermalize to the ancilla's state, which in this case means $\langle a^\dagger a\rangle^\infty = 0$  [the only exception is at $\lambda_e=\pm \pi/2$, which is somewhat pathological]. 

Results for the TMS interaction are shown in Fig.~\ref{fig:ada}(c,d). 
In this case the relative signs are immaterial, but the dynamics becomes more sensitive on the  magnitude of $\nu_e$, since $\tilde{z}$ and $\tilde{w}$ are hyperbolic functions. 
The TMS interaction entangles $E_n E_{n+1}$, even if both are initially in the vacuum. 
As a consequence, it also spontaneously create excitations, so that the number of quanta is not preserved.
At each $E_nE_{n+1}$ collision the net number of excitations therefore increases. 
Part of these excitations are lost when the ancillas are discarded  and part  flow to the system. 
As a consequence, depending on the rate at which excitations are created, the dynamics can be either stable or unstable. 
This occurs at the critical point $\nu_e^\text{crit} = \sinh^{-1}(1)\simeq 0.8813$, which is when $\tilde{w} = 1$, thus marking the situation where the number of excitations in the system grow unboundedly [c.f. Eq.~\eqref{XY_TMS}]. 
When $\nu_e < \nu_e^\text{crit}$ the dynamics will be stable and the system will converge to a steady-state value 
$\langle a^\dagger a \rangle = \sinh^2{\nu_e}(1-\sinh^2{\nu_e})^{-1}$ independently of $\lambda_s$ [Fig.~\ref{fig:ada}(c)].
Conversely, for $\nu_e \geqslant \nu_e^\text{crit}$, the dynamics becomes unstable and the number of excitations diverge [Fig.~\ref{fig:ada}(d)]. 
These asymptotic values can be understood from arguments of stability theory, as shown in Appendix~\ref{app:A}. 

\begin{figure}
\includegraphics[scale=0.6]{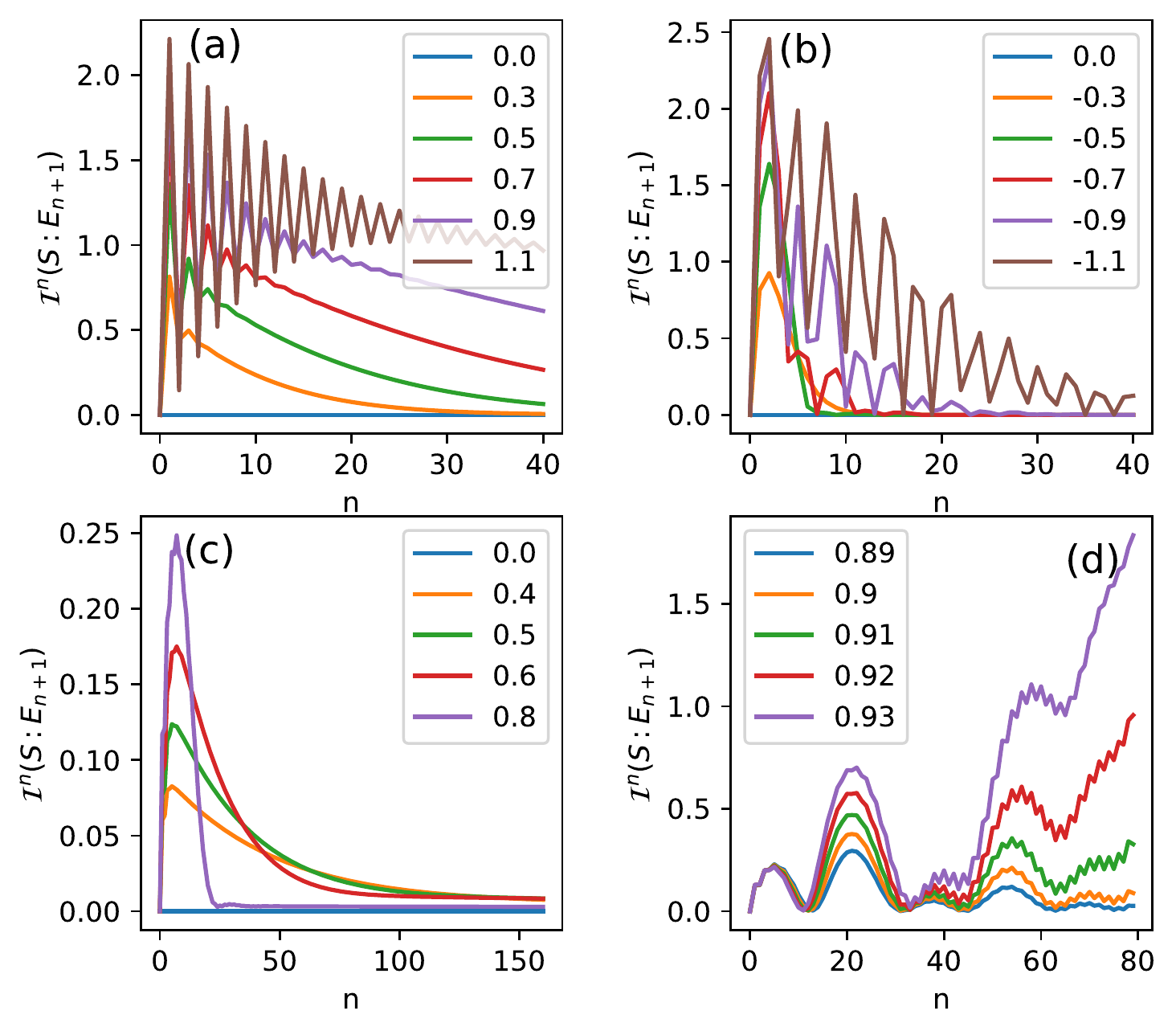}
\caption{Mutual Information~\eqref{MI} for the BS (a,b) and TMS (c,d) dynamics. Other parameters are the same as Fig.~\ref{fig:ada}.
}
\label{fig:MI}
\centering
\end{figure}

\subsection{\label{sec:MI}Mutual Information}

Before we turn to the memory kernel and divisibility, it is  useful to consider another, very simple quantifier of non-Markovianity, which is particularly suited for collisional models. 
Namely, the quantum mutual information (MI) between $S$ and the ancilla $E_{n+1}$ at the time $n$. That is, \emph{before} $S$ and $E_{n+1}$  interacted. 
The MI is defined as 
\begin{equation}\label{MI}
\mathcal{I}^n(SE_{n+1}) = S(\rho_S^n) + S(\rho_{E_{n+1}}^n) - S(\rho_{SE_{n+1}}^n),
\end{equation}
where $S(\rho) = - \tr(\rho\ln \rho)$ is the von Neumann entropy.
The ability to quantify information in this way is one of the big advantages of collisional models. 
In any non-Markovian scenario, system-bath correlations play a key role. 
But in the standard approach, where the system interacts continually with a macroscopic bath, it is not obvious which parts of these correlations actually matter. 
For instance, a correlation between the system and a part of the bath the system will never interact again is irrelevant, as far as non-Markovianity is concerned. 

The Gaussian framework used here also makes the MI readily accessible from the CM $\gamma^n$ in Eq.~\eqref{gamma_def}. 
Correlations are related to the off-diagonal blocks $\xi_{n+1}^n$ (the MI would be zero if $\gamma^n$ were block-diagonal) and can be computed in terms of the symplectic eigenvalues of  $\gamma^n$~\cite{serafini2017quantum}. 
The results are shown in Fig.~\ref{fig:MI}, for the same collection of parameters as Fig.~\ref{fig:ada}
As a sanity check, the MI is identically zero when $\lambda_e = \nu_e = 0$. 
It also tends to be larger for short times, tending to zero as $n$ grows. 
The only exception is the unstable dynamics in Fig.~\ref{fig:MI}(d), where the MI grows unboundedly.
The oscillatory patterns in $\langle a^\dagger a\rangle$ are also present in the MI. 

\begin{figure}
\includegraphics[scale=0.6]{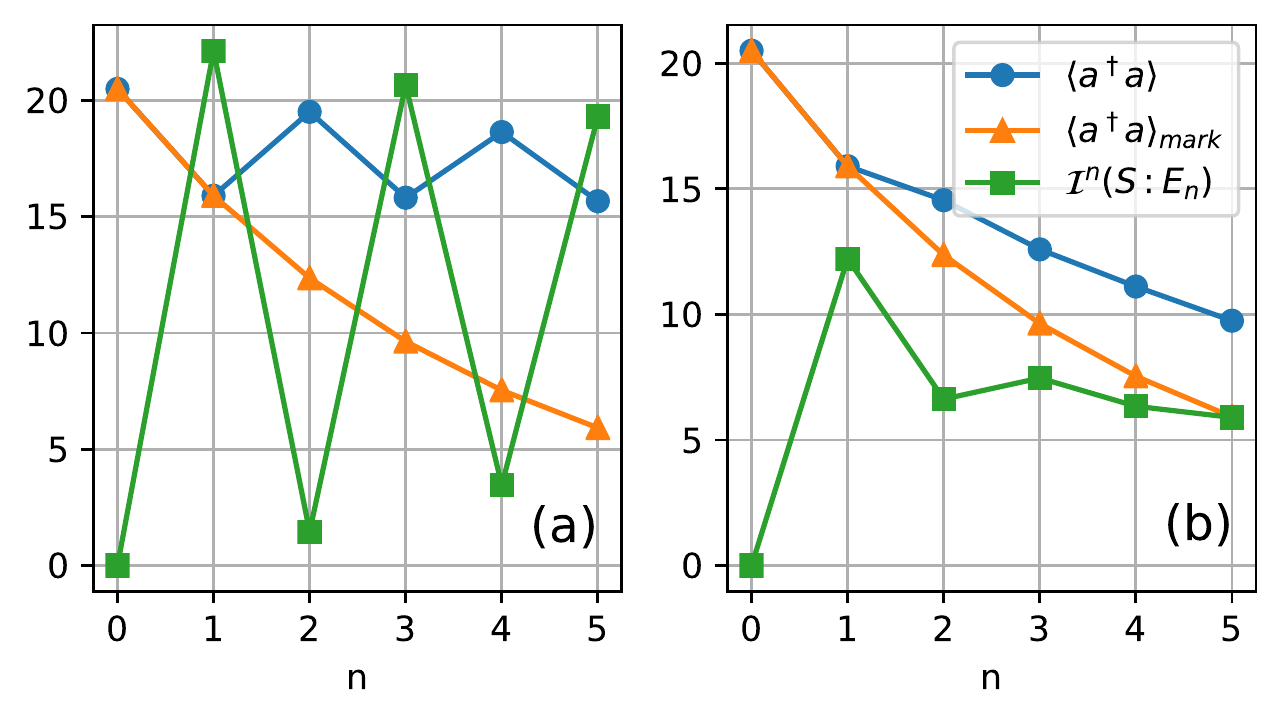}
\caption{Comparison between Markovian and non-Markovian dynamics and role of the Mutual Information. 
In blue circles we show the early dynamics of $\langle a^\dagger a\rangle$~vs.~$n$ for the BS dynamics with (a) $\lambda_e =  1.1$ and (b) $\lambda_e = 0.3$, with fixed $\lambda_s = 0.5$ [c.f.~Fig.~\ref{fig:ada}(a)].
The corresponding Markovian case ($\lambda_e = 0$) is shown in orange triangles. 
These curves are to be compared with the MI~\eqref{MI}, shown by green squares in the two cases [Fig.~\ref{fig:MI}(a)]. 
The heights of each curve were adjusted for better visibility.
}
\label{fig:ada_MI_comparison}
\centering
\end{figure}

To better understand the role of the MI in the non-Markovian dynamics we present in Fig.~\ref{fig:ada_MI_comparison} a comparison between the occupation number $\langle a^\dagger a\rangle$ of Fig.~\ref{fig:ada} and the MI of Fig.~\ref{fig:MI} for the BS dynamics. 
We focus on early times (small $n$) and also compare $\langle a^\dagger a \rangle$ with the corresponding Markovian dynamics ($\lambda_e = 0$). 
The difference between the non-Markovian (blue circles) and Markovian (orange triangles) dynamics reflects the extent to which the backflow of information affects the evolution. 
This, as can be seen in the figure, is directly correlated with the MI (green squares) \emph{of the previous step}. 
That is, a large MI in a given step implies a large difference between the blue and orange curves in the following one. 
This is particularly clear in Fig.~\ref{fig:ada_MI_comparison}(a) and serves to illustrate how the correlations built between $SE_{n+1}$, at step $n$, affect the future interaction between $S$ and $E_{n+1}$ at the next step. 

\section{\label{sec:Memory_Kernel}Memory Kernel}

The notion of a Memory Kernel (MK), discussed  in Eq.~\eqref{mem_kernel} of Sec.~\ref{sec:int}, is perhaps the most physically transparent way of analyzing non-Markovianity (see also Fig.~\ref{fig:diagram}(c)). 
Starting from any global map between system and bath, one can always write down a  differential equation for the reduced density  matrix  $\rho_S$ of the system. 
This equation, however, will in general be  time-non-local; i.e., it will be an integro-differential equation of the form~\eqref{mem_kernel}, where $\mathcal{K}_{t-t'}[\rho_S(t;)]$ describes how $d\rho_S(t)/dt$ depends on $\rho_S(t')$ in previous times $t'<t$. 
The MK therefore contains all the information about the dynamics, with non-Markovianity being related to its overall dependence on $t-t'$: 
the slower  the decay of $\mathcal{K}_{t-t'}$ with $t-t'$, the longer the memory and hence the more non-Markovian is the dynamics.
The Markovian case is recovered when $\mathcal{K}_{t-t'} \propto \delta(t-t')$.

The memory kernel $\mathcal{K}_{t-t'}$ is a superoperator acting on the full Hilbert space of the system. Computing it is thus, in general, a very difficult task. 
Within our framework, however, one may equivalently formulate a memory kernel acting only in the system's CM $\theta^n$. 
This can be accomplished starting from Eq.~\eqref{gamma_embedding} and writing down a difference equation for $\theta^n$ only. 
As we will demonstrate below, this equation  will have the form (contrast with Eq.~\eqref{mem_kernel}):
\begin{equation}\label{theta_mem_kernel}
    \theta^{n+1} = x^2 \theta^n + \sum\limits_{r=0}^{n-1} \mathcal{K}_{n-r-1}(\theta^r) + G_n, 
\end{equation}
where $G_n$ is a contribution that depends only on the initial state of the ancillas and $\mathcal{K}_n$ is the memory kernel. The way we define it, the MK is such that $\mathcal{K}_0$ measures how the step from $\theta^{n}$ to $\theta^{n+1}$ is affected by $\theta^{n-1}$ and $\mathcal{K}_{n-1}$ measures how it is affected by $\theta^0$. 
$\mathcal{K}_n$ is still a superoperator, but one which acts on the space of $2\times2$ CMs.
One can write it more explicitly in terms of a Kraus operator-sum representation~\cite{Kraus1983,Nielsen}
\begin{equation}\label{theta_mem_kernel}
    \mathcal{K}_n(\theta) = \sum\limits_{ij} \kappa_{ij}^n M_i \theta M_j\trans,
\end{equation}
where $\kappa_{ij}^n$ are coefficients that depend on time and $\{M_i\}$ are a complete set of $2\times 2$ matrices; a convenient choice is the set of Pauli matrices $\{ \mathbb{I}_2, \sigma_z, \sigma_+, \sigma_-\}$. 
A general recipe to compute the coefficients $\kappa_{ij}^n$ in Eq.~\eqref{theta_mem_kernel} is given below in Eq.~\eqref{kappa_recipe}. 
Crucially, as we show, it depends \emph{only} on the matrix $X$ of the Markovian embedding~\eqref{gamma_embedding}.

The memory itself is contained in the dependence of $\kappa_{ij}^n$ on $n$. 
The dependence on $i, j$ determines how different elements of $\theta^r$ affect $\theta^n$. 
For instance, as we will show below, in the case of the BS map [Eq.~\eqref{XY_BS}], the only non-zero coefficient will be the one proportional to $\mathbb{I}_2 \theta \mathbb{I}_2 = \theta$, which we refer to as $\kappa_{11}^n$;
that is, the memory Kernel is actually a $c$-number, $\mathcal{K}_n(\theta) = \kappa_{11}^n \theta$. 
This implies that the MK is the same for all entries of $\theta^n$ and each entry $(\theta^n)_{ij}$ is only affected by the corresponding entry $(\theta^r)_{ij}$ at past times. 
Conversely, in the TMS map there will be four non-zero coefficients, corresponding to combinations of $M_1 = \mathbb{I}_2$ and $M_2 = \sigma_z$; we refer to them as $\kappa_{11}^n$, $\kappa_{1,z}^n$, $\kappa_{z,1}^n$ and $\kappa_{z,z}^n$.
This means that the memory kernel of $(\theta^n)_{11}$ will be different from that of $(\theta^n)_{2,2}$ and so on (each entry will have its own memory kernel).
Finally, a memory kernel containing a dependence on $\sigma_\pm$ would  imply that $(\theta^n)_{11}$ would depend on the past values of other entries, such as $(\theta^r)_{12}$ and $(\theta^r)_{22}$.

\subsection{General derivation of the Memory Kernel}

We now carry out the derivation of the memory kernel for the Gaussian collisional model. 
Since we are unaware of any other papers doing this, we consider here a more general scenario, which relies only on the structure of the Markovian embedding in Eq.~\eqref{gamma_embedding}.
We also assume that the system and ancillas are each composed of an arbitrary number of modes $N_S$ and $N_E$ [Eqs.~\eqref{XY_BS} and \eqref{XY_TMS} are recovered for $N_S = N_E = 1$]. 
More specifically, we take the matrices $X$ and $Y$ to have the following block structure, 
\begin{equation}\label{MK_XY}
    X = \begin{pmatrix} X_{11} & X_{12} \\[0.2cm]
    X_{21} & X_{22} \end{pmatrix},
    \qquad
    Y = \begin{pmatrix} 0 & 0 \\[0.2cm]
    0 & Y_{22} \end{pmatrix},
\end{equation}
where e.g., $X_{11}$ and $X_{22}$ are of size $2N_S$ and $2N_E$ respectively. 
This therefore contemplates both multimode system and ancillas, as well as collisions with longer memory. 
For instance, if $E_n$ collides with $E_{n+1}$ and $E_{n+2}$, then we would have $N_S = 1$ and $N_E = 2$.


Our derivation follows the general approach of Nakajima and Zwanzig~\cite{Nakajima1958,Zwanzig1960}, but adapted to the present context. 
We begin by noting the following property: the solution of a generic difference equation of the form 
\begin{equation}\label{general_diff_equation}
\psi(n+1) = \alpha \psi(n) + g(n), 
\end{equation}
is given by 
\begin{equation}\label{general_diff_equation_solution}
    \psi(n) = \alpha^n \psi(0)_ + \sum\limits_{r=0}^{n-1} \alpha^{n-r-1} g(r). 
\end{equation}
This solution holds for arbitrary objects $\psi$, provided $\alpha$ is a linear operator. 
It therefore holds  when $\psi$ is a vector and $\alpha$ is a matrix, or when $\psi$ is a matrix and $\alpha$ is a superoperator. 
Thus, for instance, the solution of Eq.~\eqref{gamma_embedding} is 
\begin{equation}\label{evolution_complete_map}
    \gamma^n = X^n \gamma^0 (X\trans)^n + \sum\limits_{r=0}^{n-1} X^{n-r-1} Y (X\trans)^{n-r-1}.
\end{equation}
Here the notation $\gamma^n$, to denote the time index, becomes a bit ambiguous since $X^n$ is the matrix $X$ to the power $n$. 
But there is no room for confusion, since $X^n$ will be the only quantity where the superscript does not refer to the time.

We now introduce the vectorization operation~\cite{Vectorization}, which transforms a matrix $A$ into a vector $\vec{A} = \text{vec}(A)$ by stacking its columns. 
For instance, 
\begin{equation}\label{vec_def}
\text{vec} \begin{pmatrix} a& b \\ c &d \end{pmatrix} = \begin{pmatrix} a \\ c\\ b \\ d\end{pmatrix}.
\end{equation}
One may  verify that, for any three matrices $A$, $B$, $C$,
\begin{equation}\label{vec_property}
\text{vec}(A B C) = (C\trans \otimes A) \text{vec}(B).
\end{equation}
With this, the matrix difference equation~\eqref{gamma_embedding} is converted into a vector difference equation 
\begin{equation}\label{gamma_vec}
    \vec{\gamma}^{\;n+1} = (X\otimes X) \vec{\gamma}^{\; n} + \vec{Y}. 
\end{equation}

We also introduce projection matrices onto the subspaces of system and ancilla, 
\begin{equation}\label{MK_PS}
    P_S = \begin{pmatrix} 
    \mathbb{I}_{2N_S} & 0 \\[0.2cm]
    0 & 0 \end{pmatrix}, 
    \qquad 
    P_E = \begin{pmatrix} 
    0 & 0 \\[0.2cm]
    0 & \mathbb{I}_{2N_E} \end{pmatrix},
\end{equation}
which are of size $2N_S + 2N_E$.
In the larger space relevant for vectorization there are four possible projections, $P_S(\ldots)P_S$, $P_S(\ldots)P_E$ and so on. 
These operations chop the covariance matrix $\gamma^n$ in 4 blocks, as in Eq.~\eqref{gamma_def}. 
Our interest is in $P_S(\gamma^n)P_S$, as it contains the system CM $\theta^n$. 
We therefore also introduce
\begin{equation}\label{MK_P}
    P = P_S \otimes P_S, 
\end{equation}
together with its complement $ Q = 1 - P$.
Note, though, that $Q \neq P_E \otimes P_E$. 

We now multiply Eq.~\eqref{gamma_vec} by $P$ and use that $P+Q = 1$, together with the fact that $P\vec{Y} = 0$ [c.f. Eq.~\eqref{MK_XY}]. We then get
\begin{equation}\label{proj_eq_1}
P \vec{\gamma}^{\;n+1} = P (X\otimes X) P \vec{\gamma}^{\;n} + P(X\otimes X ) Q \vec{\gamma}^{\;n}.
\end{equation}
Similarly, multiplying Eq.~\eqref{gamma_vec} by $Q$ we find 
\begin{equation}\label{proj_eq_2}
    Q \vec{\gamma}^{\;n+1} = Q (X\otimes X) Q \vec{\gamma}^{\;n} + Q (X\otimes X) P \vec{\gamma}^{\;n} + \vec{Y}.
\end{equation}
Now comes the crucial idea of the  Nakajima and Zwanzig method~\cite{Nakajima1958,Zwanzig1960}.
We interpret Eqs.~\eqref{proj_eq_1} and \eqref{proj_eq_2} as two coupled equations for the variables $P\vec{\gamma}^{\;n}$ and $Q\vec{\gamma}^{\;n}$.
Since our interest is in $P\vec{\gamma}^{\;n}$, we first solve Eq.~\eqref{proj_eq_2}, assuming a given $P\vec{\gamma}^{\;n}$, and then substitute the result in Eq.~\eqref{proj_eq_1}. 
Eq.~\eqref{proj_eq_2} is of the form~\eqref{general_diff_equation}  with $\alpha = Q(X\otimes X)$ and $g(n) = Q(X\otimes X) P \vec{\gamma}^{\; n} + \vec{Y}$. 
Eq.~\eqref{general_diff_equation_solution} then gives 
\[
    Q \vec{\gamma}^{\;n} = [Q(X\otimes X)]^n Q \vec{\gamma}^{\;0} + \sum\limits_{r=0}^{n-1} [Q(X\otimes X)]^{n-r-1} \big\{Q(X\otimes X) P \vec{\gamma}^{\; n} + \vec{Y}\big\}. 
\]
Plugging this in Eq.~\eqref{proj_eq_1} we then arrive at 
\begin{equation}\label{mem_kern_partial_1}
    P \vec{\gamma}^{\;n+1} = P (X\otimes X) P \vec{\gamma}^{\;n} + \sum\limits_{r=0}^{n-1} \hat{K}_{n-r-1} P \vec{\gamma}^{\;n} + \vec{\mathcal{G}}_n, 
\end{equation}
where 
\begin{equation}
    \hat{K}_{n-r-1} = P(X\otimes X) [Q (X\otimes X)]^{n-r-1} Q(X\otimes X), 
\end{equation}
is the memory kernel in vectorized form (i.e., as a matrix of size $(2N_S+2N_E)^2$). 
The term $\vec{\mathcal{G}}_n$, on the other hand, is a function that depends only on the initial state of the ancillas and reads
\[
    \mathcal{G}_n = P(X\otimes X) [Q(X\otimes X)]^n Q \vec{\gamma}^{\; 0} + \sum\limits_{r=0}^{n-1} P(X\otimes X) [Q(X\otimes X)]^{n-r-1} \vec{Y}.
\]

What is left is to rewrite Eq.~\eqref{mem_kern_partial_1} as an equation for the evolution of the system's CM $\theta^n$ only. 
We introduce the $(2N_S)^2 \times (2N_S + 2N_E)^2$ rectangular matrix $\pi$ defined such that $\pi \vec{\gamma}^{\; n} = \vec{\theta}^{\; n}$.
For instance, in the case $N_S = N_E = 1$, the matrix $\pi$ will be $4\times 16$, of the form (for more intuition on this matrix, see Appendix~\ref{app:MK_BS})
\begin{equation}\label{MK_pi}
    \pi = 
    \begin{pmatrix} 
    1 & 0 & 0 & 0 & 0 & 0 & 0 & 0 & \ldots & 0   \\
    0 & 1 & 0 & 0 & 0 & 0 & 0 & 0 & \ldots & 0   \\
    0 & 0 & 0 & 0 & 1 & 0 & 0 & 0 & \ldots & 0   \\
    0 & 0 & 0 & 0 & 0 & 1 & 0 & 0 & \ldots & 0   
    \end{pmatrix}
\end{equation}
We also notice that $P = \pi\trans \pi$ and $\pi \pi\trans = \mathbb{I}_{(2N_S)^2}$.
Multiplying Eq.~\eqref{mem_kern_partial_1} on the left by $\pi$ we then get 
\begin{equation}\label{vec_theta}
    \vec{\theta}^{\; n+1} = (X_{11}\otimes X_{11}) \vec{\theta}^{\; n} + \sum\limits_{r=0}^{n-1}  \hat{\mathcal{K}}_{n-r-1}  \vec{\theta}^{\; n} + \vec{G}_n,
\end{equation}
where we also used the fact that $\pi (X\otimes X)\pi\trans = X_{11} \otimes X_{11}$. 
Here $\vec{G}_n = \pi \vec{\mathcal{G}}_n$ is again a term that depends only on the initial conditions of the ancillas, whereas
\[
    \hat{\mathcal{K}}_n = \pi \hat{K}_n \pi\trans = \pi (X\otimes X)\big[ Q (X\otimes X)\big]^{n+1} \pi\trans, 
\]
is the  memory kernel, now expressed as a matrix of size $(2N_S)^2 \times (2N_S)^2$ acting on $\vec{\theta}^{\; r}$. 
This can also be written more symmetrically, by exploiting the fact that $Q^2 = Q$. We can then arrange it as
\begin{equation}\label{MK_matrix}
    \hat{\mathcal{K}}_n = \pi (X\otimes X)Q\big[ Q (X\otimes X)Q\big]^{n} Q(X\otimes X)\pi\trans. 
\end{equation}
The extra $Q$'s outside the square brackets are placed simply to ensure the result also holds for $n = 0$.
This is the final form of the MK. 
Crucially, notice how it depends \emph{only} on the matrix $X$ of the Markovian embedding~\eqref{gamma_embedding}. 

\begin{figure}[!th]
    \centering
    \includegraphics[width=0.45\textwidth]{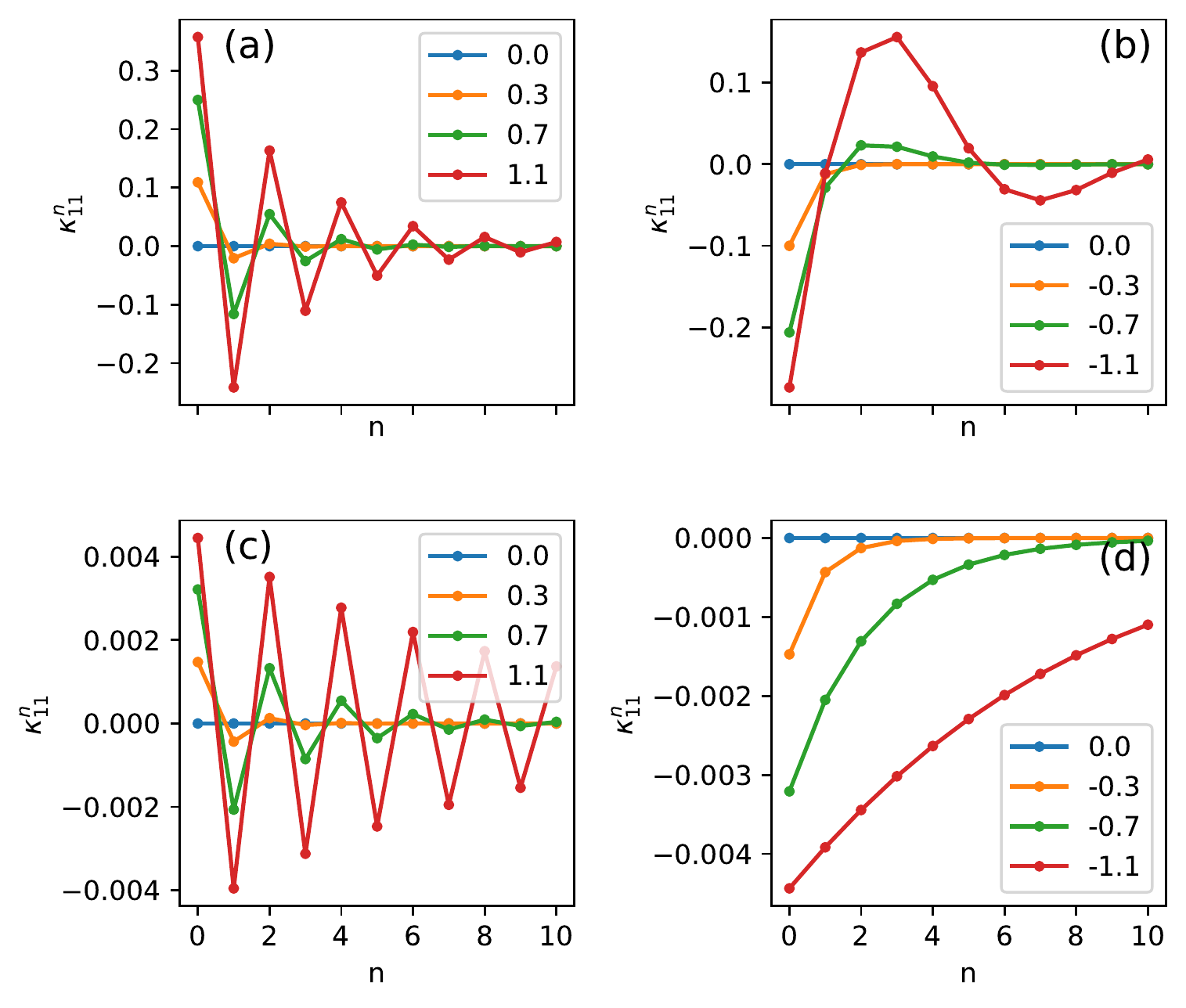}
    \caption{The memory Kernel for the BS dynamics, Eq.~\eqref{XY_BS}. In this case the only non-zero entry in Eq.~\eqref{theta_mem_kernel} is $\kappa_{11}^n$, the term proportional to the identity. The plots are for $\lambda_s = 0.5$ (upper panel) and $\lambda_s = 0.05$ (lower panel), with $\lambda_e >0$ (left) and $\lambda_e<0$ (right). 
    }
    \label{fig:MK_BS}
\end{figure}

\begin{figure*}[!t]
    \centering
    \includegraphics[width=\textwidth]{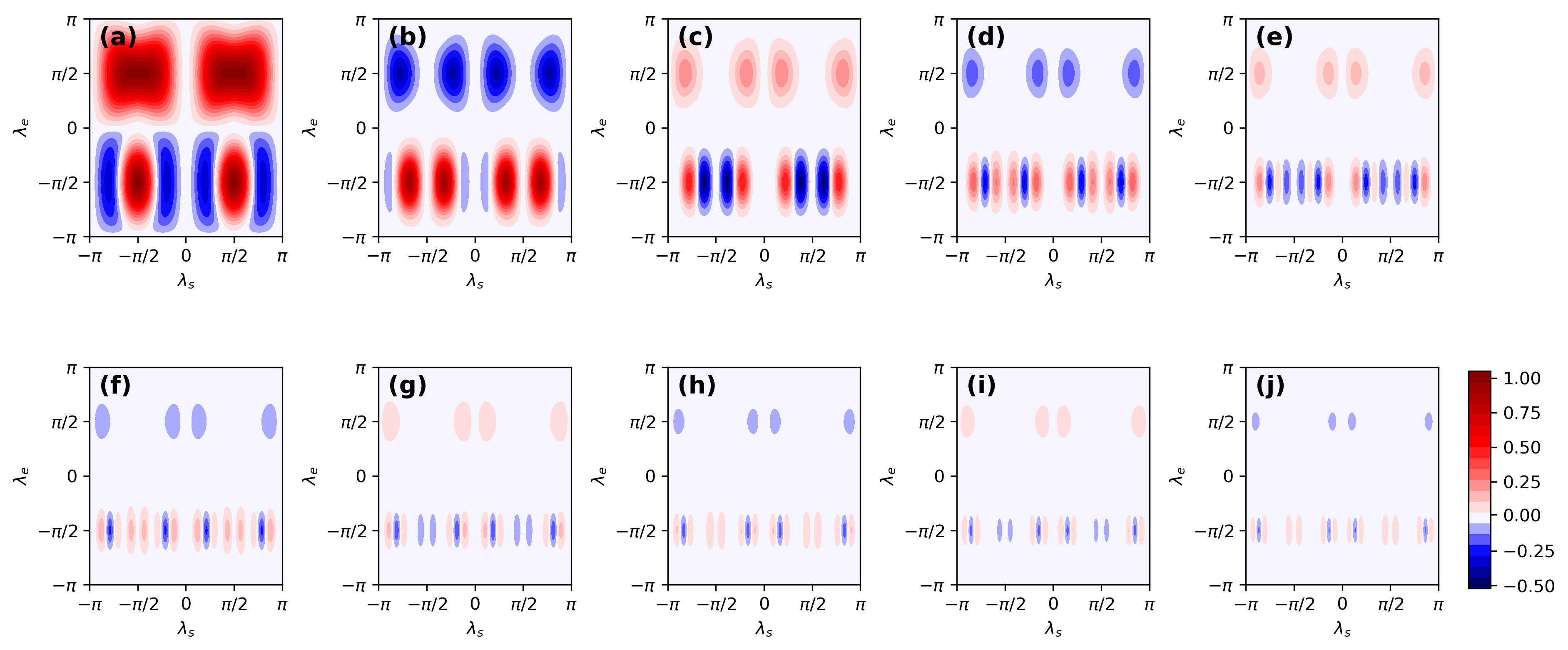}
    \caption{Diagrams for the memory kernel of the BS dynamics. Each plot shows $\kappa_{11}^n$ in the $(\lambda_s, \lambda_e)$ plane for a different value of $n$, from $n = 0$ to $n = 9$.
    \label{fig:MK_BS_diagram}
    }
\end{figure*}

To obtain a matrix difference equation for $\theta^n$ we must ``unvec'' Eq.~\eqref{vec_theta}; that is, apply the inverse map of~\eqref{vec_def}. 
Unvecking the first term is trivial since, by Eq.~\eqref{vec_property},  
\[
\text{unvec}\big[(X_{11} \otimes X_{11}) \vec{\theta}^{\; n}\big] = X_{11} \theta^n X_{11}\trans.
\]
The memory kernel~\eqref{MK_matrix}, on the other hand, cannot be unvecked as a single product of $A\theta^n B$. 
Instead, it is convenient to express it as
\begin{equation}\label{MK_decomp}
    \hat{\mathcal{K}}_n = \sum\limits_{ij} \kappa_{ij}^n M_j \otimes M_i, 
\end{equation}
where $\kappa_{ij}^n$ are real coefficients and $\{M_i\}$ are a set of operators spanning the vector space of $2N_S$-dimensional real matrices. 
Decomposed in this form, the unvecked version of the memory kernel will then be, from~\eqref{vec_property},
\begin{equation}
    \mathcal{K}_{n}(\theta) = \sum\limits_{ij} \kappa_{ij}^n M_i \theta M_j\trans. 
\end{equation}
Finally, the form of the coefficients $\kappa_{ij}^n$ can be found if we assume that the $M_i$ form an orthogonal basis with respect to the  Hilbert-Schmidt norm $(A|B) = \tr(A\trans B)$ (which is  the case of the Pauli basis, for instance). 
Multiplying Eq.~\eqref{MK_decomp} by $M_j\otimes M_i$ and tracing then yields, by orthogonality, 
\begin{equation}\label{kappa_recipe}
    \kappa_{ij}^n = \frac{\tr \big[ (M_j\trans \otimes M_i\trans) \hat{\mathcal{K}}_n\big]}{\tr(M_i\trans M_i) \tr (M_j\trans M_j)}.
\end{equation}
This, together with Eq.~\eqref{MK_matrix}, is all that is required to compute the memory kernel. 
With all these definitions, one may now finally unvec Eq.~\eqref{vec_theta}, leading to 
\begin{equation}\label{theta_general}
    \theta^{n+1} = X_{11} \theta^n X_{11}\trans + \sum\limits_{r=0}^{n-1} \mathcal{K}_{n-r-1} (\theta^r) + G_n, 
\end{equation}
where $G_n = \text{unvec}(\vec{G}_n)=\text{unvec}(\pi\vec{\mathcal{G}}_n)$ is, again, a term depending only on the initial states of the ancillas.

\subsection{Memory Kernel for the BS dynamics}

We now illustrate the memory kernel for the two maps considered in Sec.~\ref{sec:framework}, starting with the BS dynamics. 
In general, the structure of the memory kernel will be quite complicated. 
For the BS dynamics [Eq.~\eqref{XY_BS}], however, the only non-zero coefficient in Eq.~\eqref{kappa_recipe} is $\kappa_{11}^n$, the term proportional to the identity.
In this case the memory kernel is therefore rather simple, as it is just a $c$-number multiplying all entries of $\theta^r$. 
A more compact formula for the MK in this case is given in Appendix~\ref{app:MK_BS}.

Results for the BS dynamics are shown in Fig.~\ref{fig:MK_BS}. 
The upper panel corresponds to $\lambda_s = 0.5$, which is similar to Eq.~\ref{fig:ada}.
As can be seen, for $\lambda_e >0$ (Fig.~\ref{fig:MK_BS}(a)) the memory kernel's decay is oscillatory, with an exponential envelope. 
For $\lambda_e <0$, oscillations are also observed, but these are rather different in nature and more asymmetrical with respect to the horizontal axis. 
When $\lambda_s = 0.05$ the situation changes (Figs.~\ref{fig:MK_BS}(c) and (d)). 
The dynamics of $\langle a^\dagger a \rangle$ is still quite similar to that of $\lambda_s = 0.5$, shown in Fig.~\ref{fig:ada}, except that the time-scales become much longer. 
But in the MK one sees something entirely different. 
In particular, one finds that while $\kappa_{11}^n$ continues to oscillate when $\lambda_e>0$, it now becomes exclusively negative for $\lambda_e < 0$. 
In this case therefore, all past values of $\theta^r$ tend to contribute negatively to the evolution. 

Negative values in the memory kernel are rather important, as they are associated with faster convergence. 
The reason is that the CM is a positive matrix and the first term in~\eqref{theta_mem_kernel} is always positive. 
The negativities observed in Fig.~\ref{fig:MK_BS} therefore represent an accelerated draining of excitations from the system. 
This sheds light on some of the behaviors  previously observed for the  number operator (Fig.~\ref{fig:ada}) and mutual information (Fig.~\ref{fig:MI}).

It is possible to condensed a lot of information  about the memory kernel by plotting $\kappa_{11}^n$ in the $(\lambda_s, \lambda_e)$ plane, for different values of $n$. 
This is shown in Fig.~\ref{fig:MK_BS_diagram}. 
Each plot corresponds to a different value of $n$, from 0 up to 9. 
The dependence on the relative signs of $\lambda_s$ and $\lambda_e$ is clearly visible, as is the overall damping of the memory with increasing $n$. 
Particularly interesting, this map is able to very clearly pinpoint the regions have negative memory kernels, something which is found to be highly non-trivial.

\subsection{Memory Kernel for the TMS dynamics}

\begin{figure}
    \centering
    \includegraphics[width=0.45\textwidth]{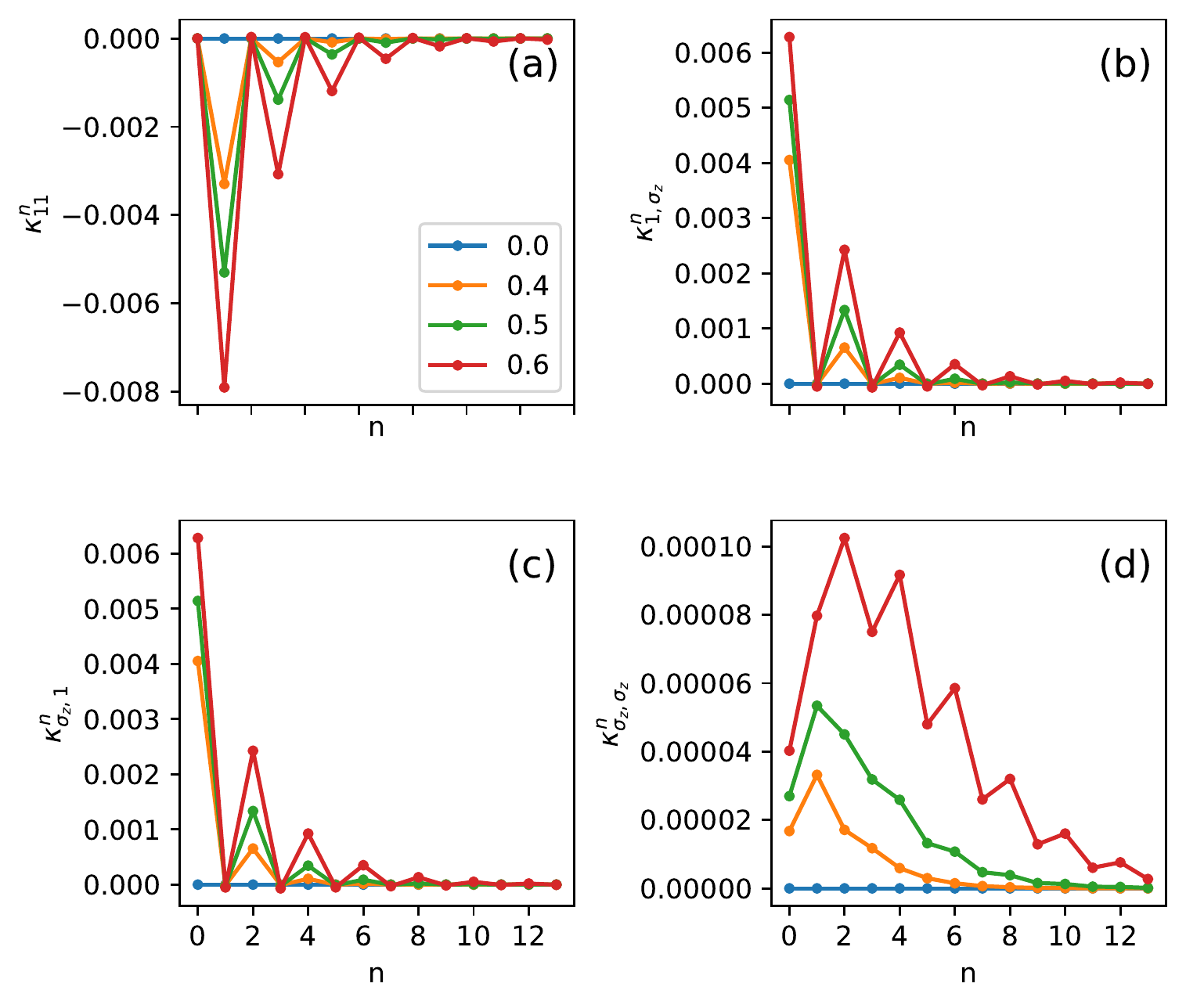}
    \caption{The memory Kernel for the (stable) TMS dynamics, Eq.~\eqref{XY_TMS} with $\lambda_s = 0.1$ and different values of $\lambda_e$. 
    Each curve corresponds to a different entry of Eq.~\eqref{theta_mem_kernel}; namely, $\kappa_{11}^n$, $\kappa_{1,\sigma_z}^n$, $\kappa_{\sigma_z,1}^n$ and $\kappa_{\sigma_z,\sigma_z}^n$. 
    }
    \label{fig:MK_TMS_1}
\end{figure}

Next we turn to the TMS case. 
In this case it is found that there are, in total, 
\begin{equation}
    \mathcal{K}_n(\theta) = \kappa_{11}^n \theta + \kappa_{1z}^n \theta \sigma_z + \kappa_{z1}^n \sigma_z \theta + \kappa_{zz}^n \sigma_z \theta \sigma_z. 
\end{equation}
These quantities are plotted in Fig.~\ref{fig:MK_TMS_1}  for the stable dynamics ($\nu_e < \nu_e^\text{crit}$), with $\lambda_s = 0.1$.
All four coefficients are found to decay in time in an oscillatory fashion.

\begin{figure}
    \centering
    \includegraphics[width=0.45\textwidth]{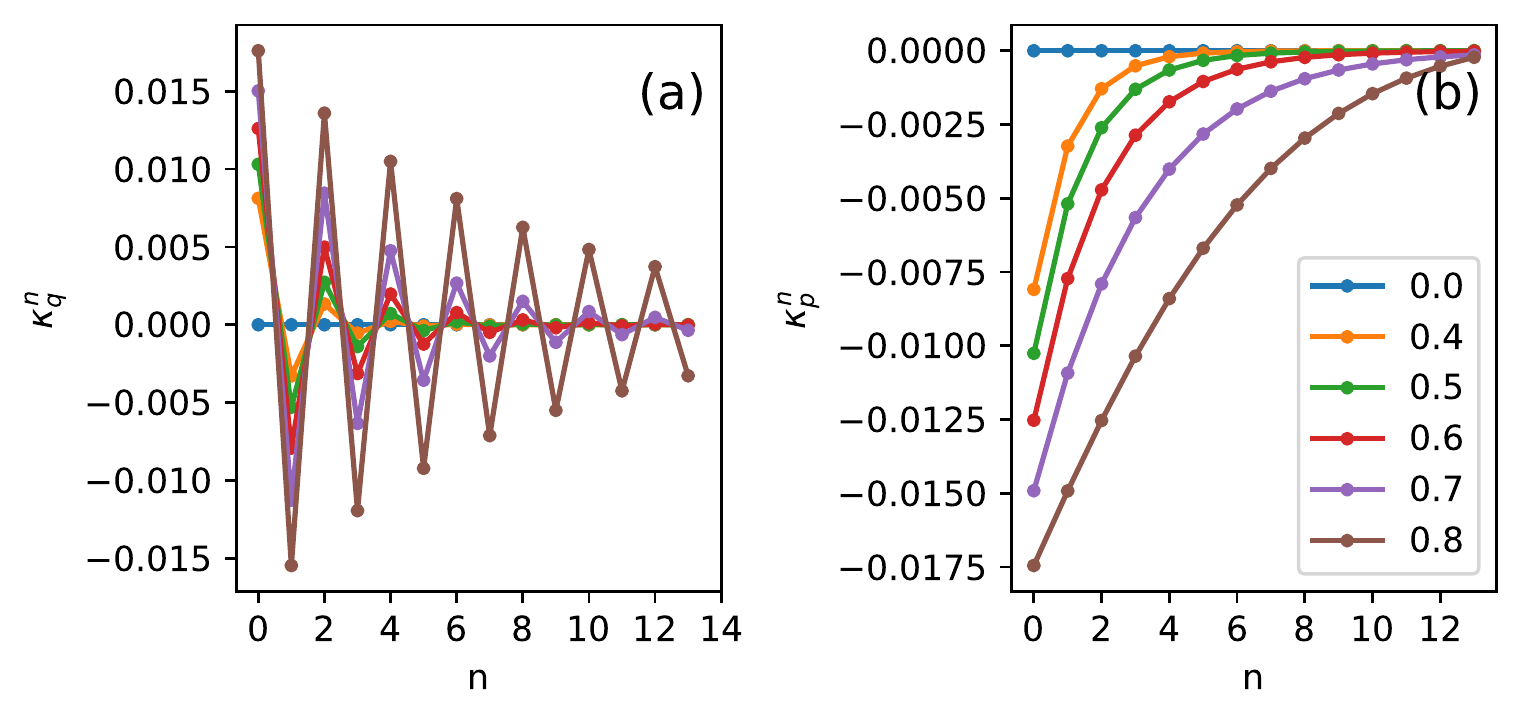}
    \caption{The MK for $\langle Q^2 \rangle$ and $\langle P^2 \rangle$, Eq.~\eqref{MK_TMS_two_contributions}, for the TMS dynamics. Other parameters are the same as Fig.~\ref{fig:MK_TMS_1}.
    }
    \label{fig:MK_TMS_QP}
\end{figure}

The physics of each coefficient, however, is not necessarily transparent. 
In order to gain better intuition, let us focus on the diagonal entries of $\theta^n$. In this case one finds that 
\begin{IEEEeqnarray}{rCl}
    \bigg( \mathcal{K}_n(\theta) \bigg)_{11} &=& \big(\kappa_{11}^n + \kappa_{1z}^n + \kappa_{z1}^n + \kappa_{zz}^n\big) \theta_{11}^n := \kappa_q^n \theta_{11}^n,\nonumber \\
    && \label{MK_TMS_two_contributions}\\[0.2cm]
    \bigg( \mathcal{K}_n(\theta) \bigg)_{22} &=& \big(\kappa_{11}^n - \kappa_{1z}^n - \kappa_{z1}^n + \kappa_{zz}^n\big) \theta_{22}^n := \kappa_p^n \theta_{22}^n. \nonumber
\end{IEEEeqnarray}
The coefficients $\kappa_q^n$ and $\kappa_p^n$ therefore describe the individual memory kernels of $\langle Q^2 \rangle$ and $\langle P^2 \rangle$, which are different in the TMS dynamics. 

\begin{figure*}[!t]
    \centering
    \includegraphics[width=\textwidth]{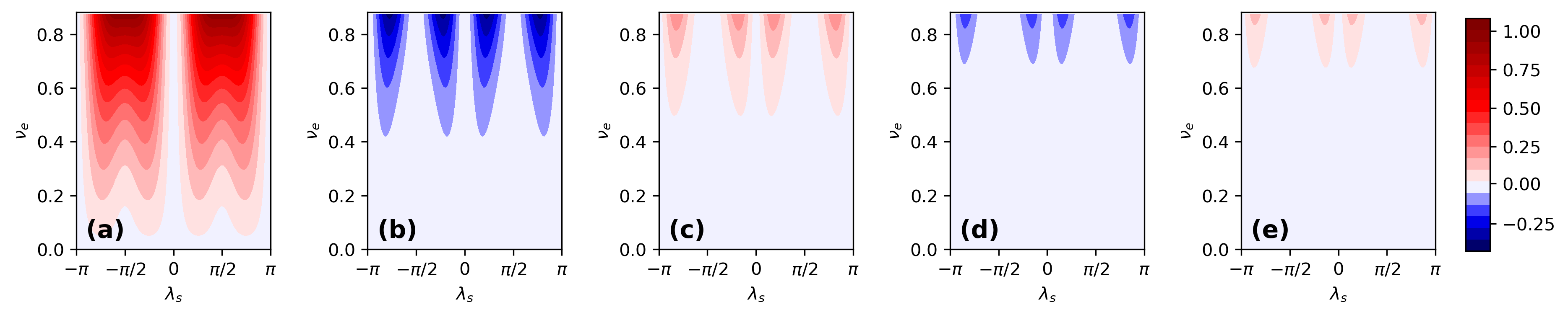}
    \caption{Diagrams for the memory kernel coefficient $\kappa_q$ [Eq.~\eqref{MK_TMS_two_contributions}] of the TMS dynamics, in the $(\lambda_s, \nu_e)$ place, for $n = 0, \ldots, 4$.
    \label{fig:MK_TMS_diagram_Q}
    }
\end{figure*}

\begin{figure*}[!t]
    \centering
    \includegraphics[width=\textwidth]{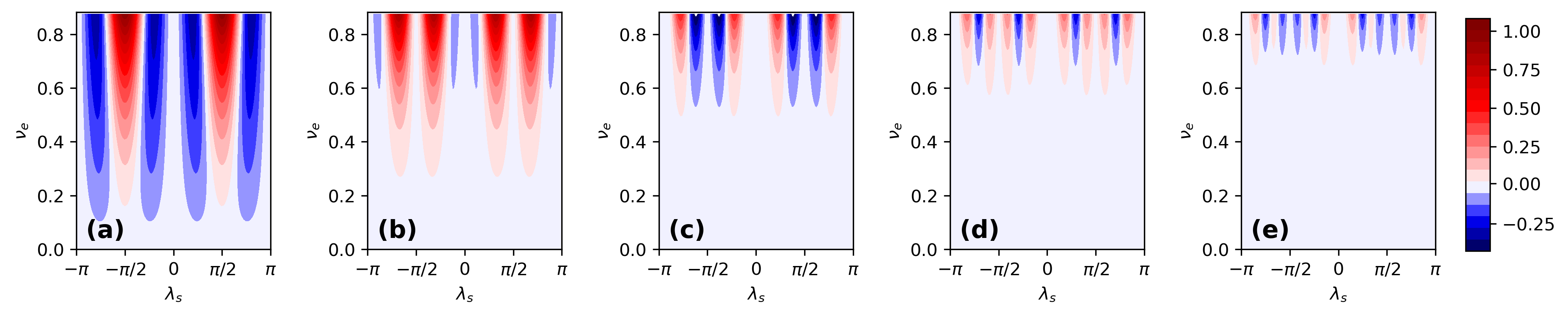}
    \caption{Similar to Fig.~\ref{fig:MK_TMS_diagram_Q}, but for $\kappa_p$.
    \label{fig:MK_TMS_diagram_P}
    }
\end{figure*}

These two contributions are shown in Fig.~\ref{fig:MK_TMS_QP}, for the same parameters as in Fig.~\ref{fig:MK_TMS_1}.
We also present diagrams in the $(\lambda_s, \nu_e)$ plane in Figs.~\ref{fig:MK_TMS_diagram_Q} and \ref{fig:MK_TMS_diagram_P}.
The plots in Fig.~\ref{fig:MK_TMS_QP} reveal an extremely interesting asymmetry between the two quadratures. 
We see that the memory associated with $\langle Q^2\rangle$ is oscillatory, whereas that associated with  $\langle P^2 \rangle$ is always negative and decays monotonically. 
This asymmetry is a consequence of our choice of two-mode squeezing in the TMS interaction~\eqref{V_TMS}. 
Figs.~\ref{fig:MK_TMS_diagram_Q} and \ref{fig:MK_TMS_diagram_P}, however, show that the situation is more intricate. 
Indeed, for fixed $(\lambda_s, \nu_e)$, $\kappa_q$ is found to oscillate with $n$. But for $\kappa_p$ this is not necessarily the case.


\begin{figure}
    \centering
    \includegraphics[width=0.45\textwidth]{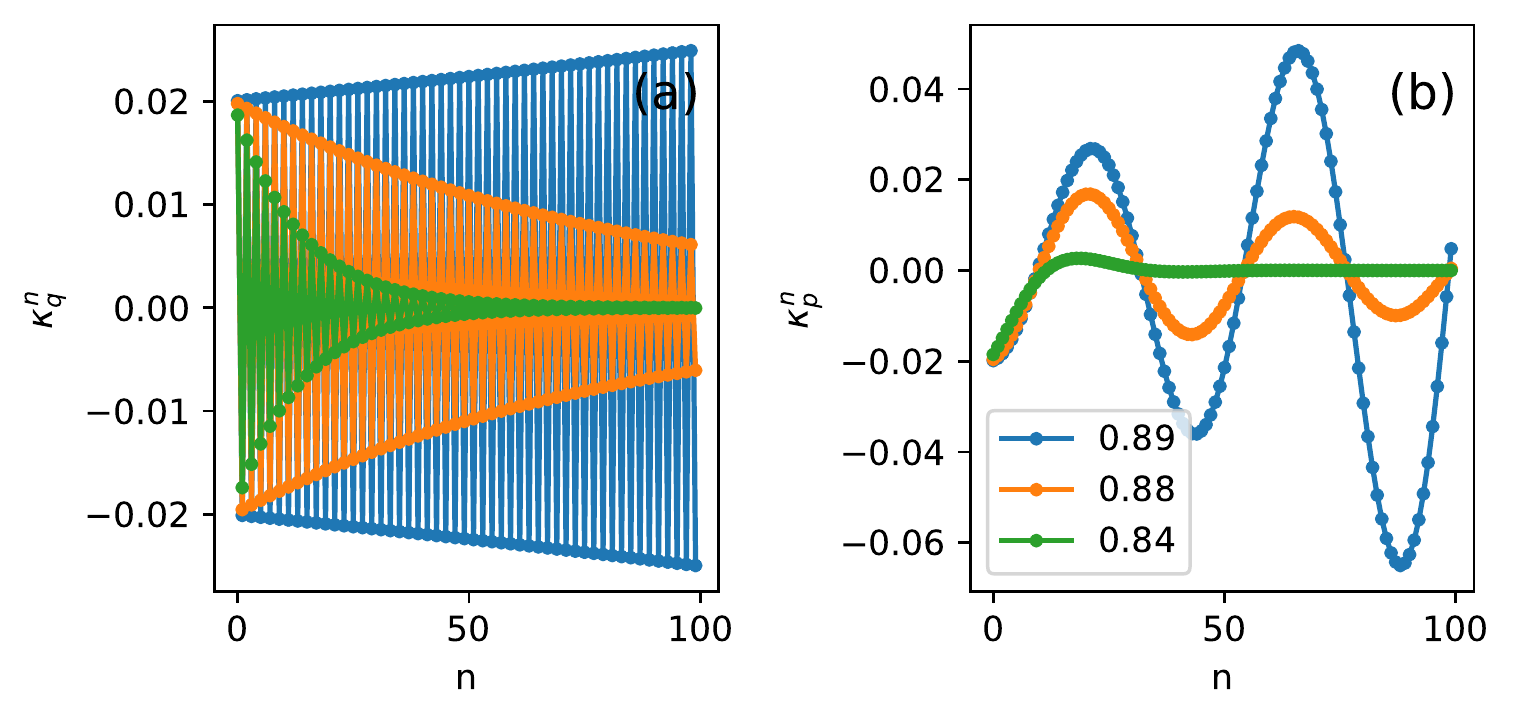}
    \caption{Similar to Fig.~\ref{fig:MK_TMS_QP}, but for values of $\nu_e$ close to, and larger than, $\nu_e^\text{crit} =0.8813$.}
    \label{fig:MK_TMS_QP_divergent}
\end{figure}

Finally, in Fig.~\ref{fig:MK_TMS_QP_divergent} we compare the previous result with the case of $\nu_e$ in the vicinity, and larger than, $\nu_e^\text{crit} =0.8813$; i.e., in the situation where the dynamics diverges. 
As can be seen, in this case both $\kappa_q$ and $\kappa_p$ diverge as well (notice the different scale of the horizontal axis). 
This is therefore contrary to our usual notion of memory: It means that the system retains a stronger memory from events in the distant past, than those in the recent one. Or, put it differently, the relative importance of past events accumulate. 

%
%
\section{\label{sec:divisibility}Gaussian CP divisibility}
%
%

Even though the MK explicitly shows the dependence on  previous states, this alone does not necessarily imply a non-Markovian dynamic~\cite{mazzola2010phenomenological}. It is therefore important to contrast the MK with an actual test of non-Markovianity.
Here we focus on CP-divisibility of intermediate maps. 
This was formulated for Gaussian dynamics, at the level of the covariance matrix, in Refs.~\cite{torre2015non,liuzzo2017non}.
Any Gaussian CPTP map must have the form 
\[
\theta \to \mathcal{X} \theta \mathcal{X}\trans + \mathcal{Y},
\] 
where $\mathcal{X}$ and $\mathcal{Y}$ are matrices satisfying~\cite{lindblad2000cloning,serafini2017quantum} 
\begin{equation}\label{CP_CPTP_condition}
    \mathcal{M}[\mathcal{X}, \mathcal{Y}] :=  2 \mathcal{Y} + i\Omega - i\mathcal{X} \Omega \mathcal{X}\trans   \geq 0, 
\end{equation}
with $\Omega = i\sigma_y$ the symplectic form.
Here $\mathcal{M}\geq0$ means the matrix must be positive semidefinite. 

In our case, the evolution of the system's CM, from time 0 to $n$, must therefore also be of this form:
\begin{equation}\label{CP_basic_map}
    \theta^n = \mathcal{X}_n \theta^0 \mathcal{X}_n\trans + \mathcal{Y}_n.
\end{equation}
The matrices $\mathcal{X}_n$ and $\mathcal{Y}_n$ can be read from the $(1,1)$ block of the general solution~\eqref{evolution_complete_map} and are independent of the initial state $\theta^0$; viz.,
\begin{IEEEeqnarray}{rCl}
    \mathcal{X}_n &=& (X^n)_{11}, \\[0.2cm]
    \mathcal{Y}_n &=& (X^n)_{12} \epsilon ({X^n}\trans)_{12} + \sum\limits_{r=0}^{n-1} \bigg[X^{n-r-1} Y (X\trans)^{n-r-1}\bigg]_{11},
\end{IEEEeqnarray}
where the subscripts $i,j$ refer here to specific blocks. 
This easiness in reading of the corresponding map matrices is another significant advantage of the Markovian embedding representation~\eqref{gamma_embedding}. 

To probe whether the dynamics is divisible, we consider the map taking the system from $n$ to $m>n$.
Assuming that $\mathcal{X}_n$ and $\mathcal{Y}_n$ are invertible, which is true in our case, this will have the form~\cite{torre2015non} 
\begin{equation}\label{CP_theta_m_n}
    \theta^m = \mathcal{X}_{mn} \theta^n \mathcal{X}_{mn}\trans + \mathcal{Y}_{mn},
\end{equation}
where 
\begin{equation}\label{CP_Xmn_Ymn}
    \mathcal{X}_{mn} = \mathcal{X}_m \mathcal{X}_n^{-1}, 
    \qquad 
    \mathcal{Y}_{mn} = \mathcal{Y}_m - \mathcal{X}_{mn} \mathcal{Y}_n \mathcal{X}_{mn}\trans.
\end{equation}
See  Fig.~\ref{fig:diagram}(d).
The dynamics is then considered divisible when the intermediate maps~\eqref{CP_theta_m_n} are a proper CPTP Gaussian map. 
That is, when $\mathcal{M}[\mathcal{X}_{mn}, \mathcal{Y}_{mn}] \geq 0$ [Eq.~\eqref{CP_CPTP_condition}].

The above criteria can be used not only as a dichotomic measure of divisibility, but also as a figure of merit~\cite{torre2015non}.
This is accomplished by defining 
\begin{equation}\label{CP_Nmn}
    \mathcal{N}_{mn} = \sum\limits_{k} \frac{|m_k|-m_k}{2},
    \quad 
    \{m_k\} = \text{eigs} \Big(\mathcal{M}[\mathcal{X}_{mn}, \mathcal{Y}_{mn}]\Big).
\end{equation}
This quantity is always non-negative and the map is divisible iff $\mathcal{N}_{mn}\equiv 0$ for all $m,n$. Otherwise, the magnitude of $\mathcal{N}_{mn}$ quantifies the extent to which divisibility is broken for that choice of $m,n$.

%
%
\subsection{BS dynamics}
%
%

\begin{figure}
    \centering
    \includegraphics[width=0.45\textwidth]{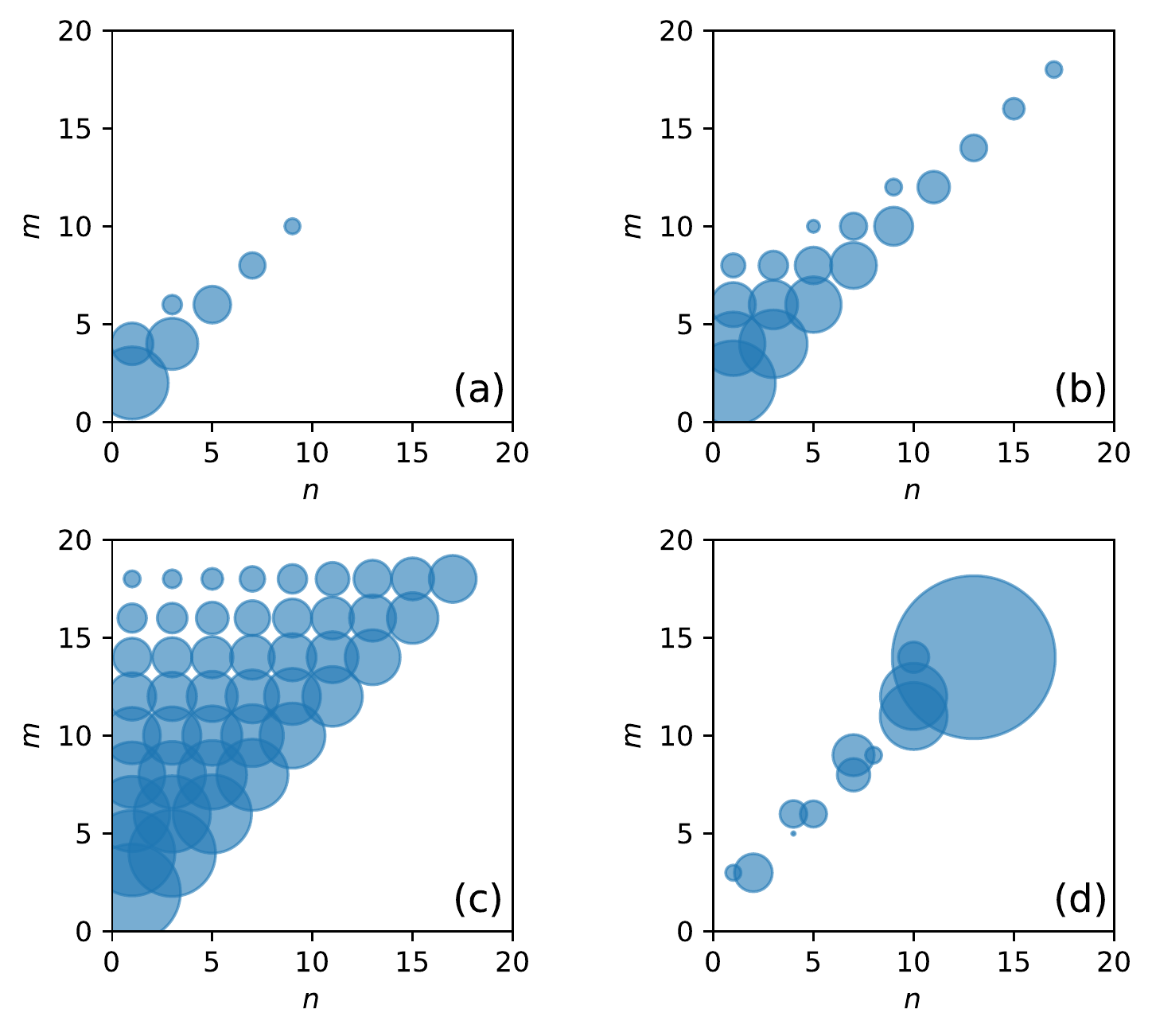}
    \caption{
    Example of the divisibility criteria for the BS dynamics. 
    The plots show $\mathcal{N}_{mn}$ in the $(n,m)$ plane, with the size of each point reflecting the magnitude of $\mathcal{N}_{mn}$.
    All curves are for $\lambda_s = 1.1$ and 
    (a) $\lambda_e = 0.75$, (b) 0.9, (c) 1.1 and (d) -0.7.
    }
    \label{fig:CP_example}
\end{figure}

\begin{figure*}[!t]
    \centering
    \includegraphics[width=\textwidth]{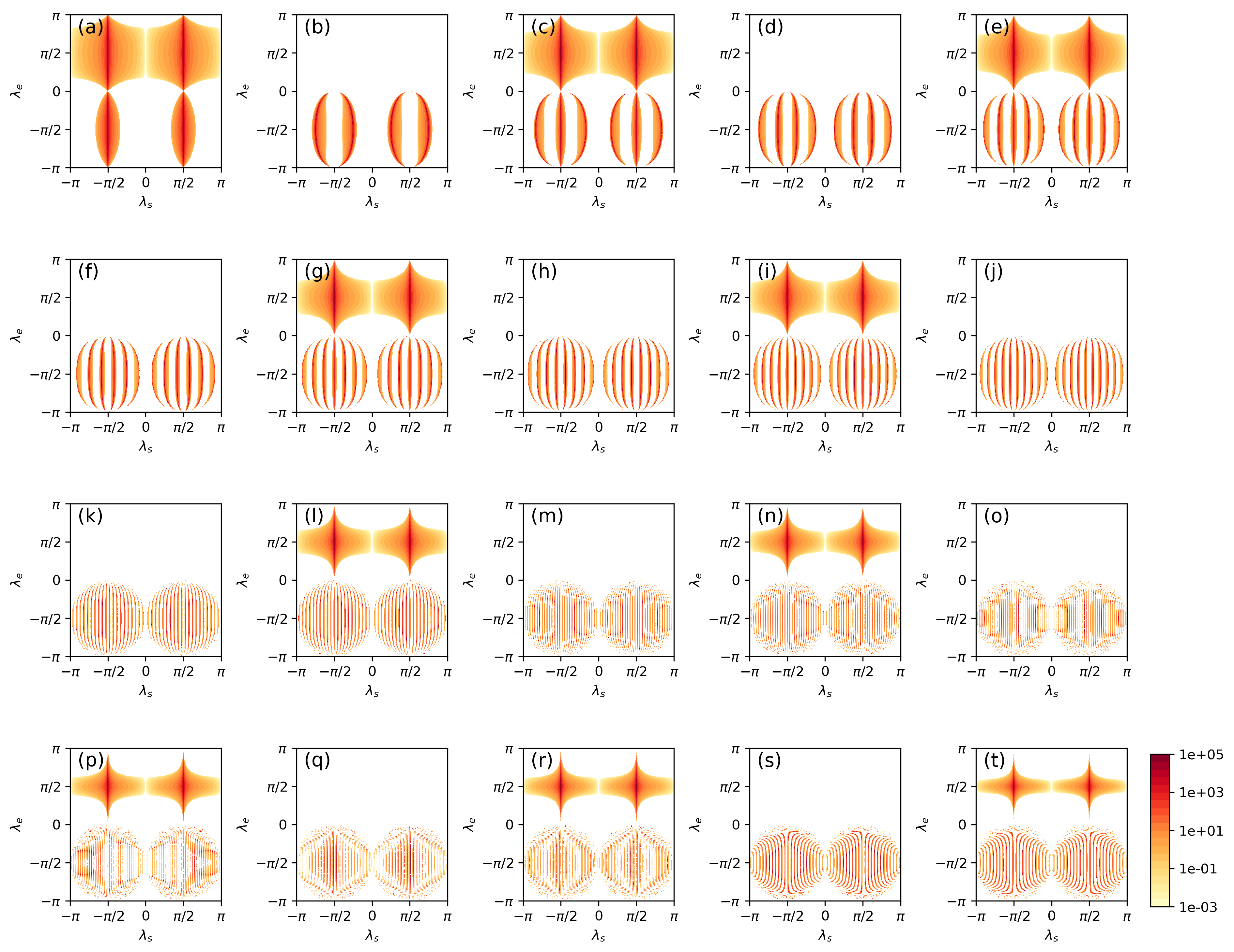}
    \caption{CP-divisibility measure $\mathcal{N}_{n+1,n}$ [Eq.~\eqref{CP_Nmn}] in the $(\lambda_s, \lambda_e)$ plane, for the BS dynamics. Each plot corresponds to a different values of $n$: in the first 2 lines, $n$ ranges from 1 to 10 in steps of 1. In the 3rd and 4th lines, $n = 20, 21$, $30, 31$, $40, 41$, $50,51$ and $100, 101$. 
    }
    \label{fig:CP_BS_diagram}
\end{figure*}

We begin our investigation of $\mathcal{N}_{mn}$ by focusing on the BS dynamics [Eq.~\eqref{XY_BS}]. 
An example of the behaviour of~\eqref{CP_Nmn} is shown in Fig.~\ref{fig:CP_example}, where we plot $\mathcal{N}_{mn}$ in the $(n,m)$ plane, with fixed $\lambda_s = 1.1$ and different values of $\lambda_e$. 
The magnitude of $\mathcal{N}_{mn}$ is represented by the size of each point. 
These diagrams are interpreted as follows. 
We start with Fig.~\ref{fig:CP_example}(a).
In this case we see that, for $n = 1$,  $\mathcal{N}_{mn}$ is non-zero only for $m = 2$ and $m = 4$, being smaller in the latter. 
For $n = 3$ the map is always divisible. And for $n = 3$, it is not divisible only for $m = 4$ and 6. 
These irregularities are a consequence of the oscillatory character of the parameters appearing, e.g., in Eq.~\eqref{XY_BS}. 
Still concerning Fig.~\ref{fig:CP_example}(a), we see notwithstanding that as $n$ gets large, the map tends to be Markovian for all $m$. 
As we increase $\lambda_e$, however, as in Figs.~\ref{fig:CP_example}(b) and (c), we see that overall the regions where $\mathcal{N}_{mn} > 0$ tend to increase. They increase both as a function of $n$, as well as a function of $m$ for fixed $n$.

When $\lambda_e <0$, however, strange things happen [Fig.~\ref{fig:CP_example}(d)]. In this case we find that there can be highly irregular values of $(n,m)$ which yield non-zero $\mathcal{N}_{mn}$ which, in fact, can reach significantly large values. 
For instance, the largest value plotted in Fig.~\ref{fig:CP_example}(d) is for $n=13$, $m=14$ and has the value $\mathcal{N} \sim 69.7$. 
For $n = 16$, $m = 17$, however, one finds $\mathcal{N} \sim 10309$ (not shown).
This is to be contrasted with Fig.~\ref{fig:CP_example}(a), whose largest value is $\mathcal{N} = 3.42$. 
We present these results simply to emphasize that $\mathcal{N}_{mn}$ can oscillate violently. 
The reason is due to the term $\mathcal{X}_n^{-1}$ in Eq.~\eqref{CP_Xmn_Ymn}, which can blow up for certain values of $\lambda_s, \lambda_e$ and $n$.

\begin{figure}
    \centering
    \includegraphics[width=0.45\textwidth]{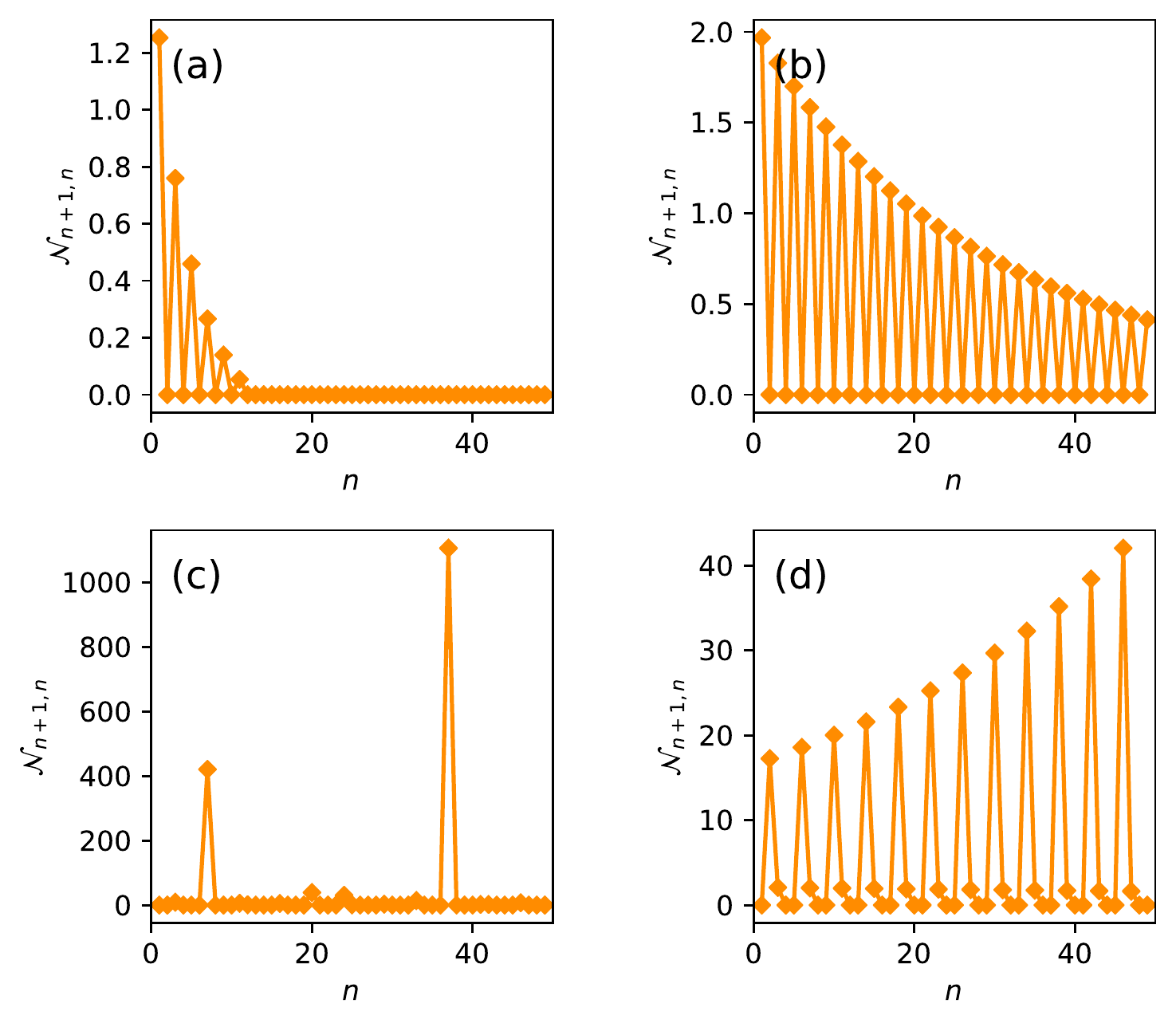}
    \caption{
    CP divisibility measure $\mathcal{N}_{n+1,n}$ as a function of $n$, for the BS dynamics with $\lambda_s = 0.8$ and  $\lambda_e = 0.9, 1.3,-0.5, -0.8$. Complements Fig.~\eqref{fig:CP_BS_diagram}.
    }
    \label{fig:CP_example_func_n}
\end{figure}

Next we turn to the divisibility of a single collision; that is, with $m = n+1$.
Plots of $\mathcal{N}_{n+1,n}$ in the $(\lambda_s, \lambda_e)$ plane are shown in  Fig.~\ref{fig:CP_BS_diagram}.
The overall behaviour is found to alternate with even and odd $n$.
For $n$ even, the map is always divisible for $\lambda_e>0$ and potentially non-divisible within certain regions of $\lambda_e <0$. 
Conversely, for $n$ odd, one finds that divisibility breaks down in significant portions of the $(\lambda_s, \lambda_e)$ plane.
An additional illustration of the complex dependence of $\mathcal{N}_{n+1,n}$ on $\lambda_s, \lambda_e, n$ is provided in Fig.~\ref{fig:CP_example_func_n}, where we plot $\mathcal{N}_{n+1,n}$ as a function of $n$ for selected values of $\lambda_s$ and $\lambda_e$.
From this figure, both the even/odd behavior, as well as the dramatic variations in the $(\lambda_s, \lambda_e)$ plane can be more clearly appreciated. 

The behavior of $\mathcal{N}_{n+1,n}$ in Fig.~\ref{fig:CP_BS_diagram} is exacerbated close to the special points $\lambda_{s(e)} = \pi/2$.
For instance, in the vicinity of $\lambda_s = \pi/2$, the dynamics is  non-divisible even for infinitesimally small $\lambda_e$. 
This occurs because $\lambda_s = \pi/2$ corresponds to the full SWAP, where the CM of the system is completely transferred to the ancilla. 
As a consequence, when then next ancilla arrives to interact with the system, it will always contain a significant amount of information about it. 
We therefore expect that in the limit $n\to\infty$ the diagrams in Fig.~\ref{fig:CP_BS_diagram} should converge to narrow lines going through these special points (although, unfortunately, we cannot actually verify this since the simulation cost become prohibitive for extremely large $n$).

\begin{figure*}[!t]
    \centering
    \includegraphics[width=\textwidth]{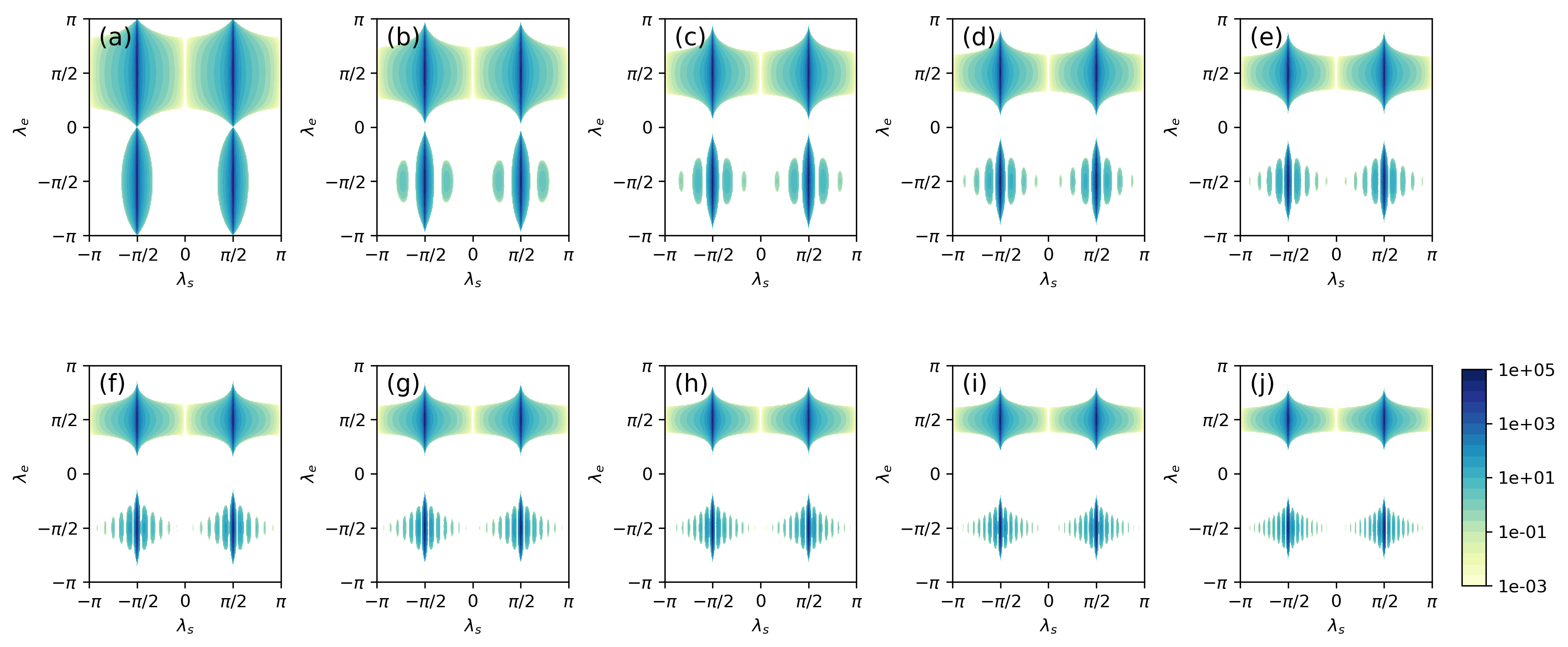}
    \caption{CP-divisibility measure, $\mathcal{N}_{m,1}$ [Eq.~\eqref{CP_Nmn}] in the $(\lambda_s, \lambda_e)$ plane, for the BS dynamics. Each plot corresponds to a different values of $m$, from $m = 2$ to 30 in steps of 2. 
    }
    \label{fig:CP_BS_diagram_11pm}
\end{figure*}

We may also study similar diagrams for collisions that are more broadly spaced in time. 
In Fig.~\ref{fig:CP_BS_diagram_11pm} we present results for $\mathcal{N}_{1,1+m}$ for different values of $m$ (we focus on even values, $m = 2, 4, \ldots$). 
This therefore describes the long-term memory of the map, concerning the first collision. 
Two features stand out from this figure. 
First, as one would expect, the overall region in the $(\lambda_s, \lambda_e)$ plane where the map is CP-divisible tends to shrink with increasing $m$. 
However, the regions around $\lambda_s = \pm \pi/2$ tend to be remarkably persistent, remaining highly non-divisible even for large $m$.

\begin{figure}
    \centering
    \includegraphics[width=0.4\textwidth]{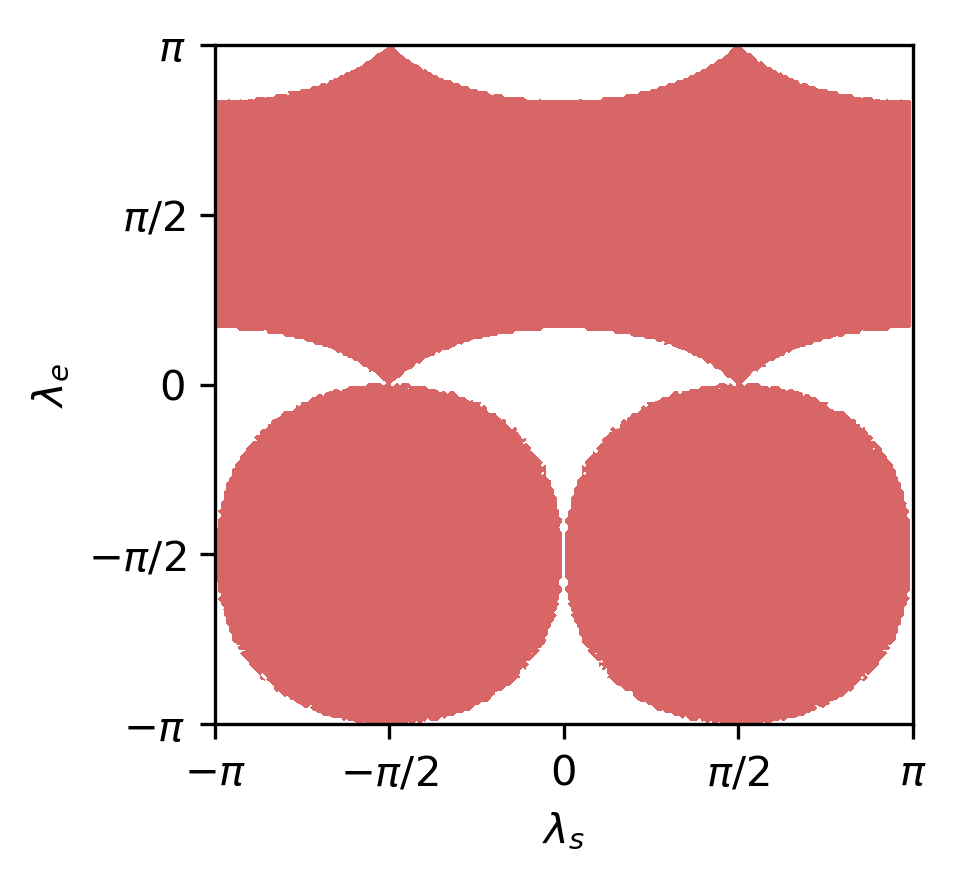}
    \caption{Regions in the $(\lambda_s, \lambda_e)$ plane where the BS dynamics is \emph{not} CP-divisible for at least one choice of $(n,m)$. }
    \label{fig:CP_BS_consolidated}
\end{figure}

The results in Figs.~\ref{fig:CP_BS_diagram} and \ref{fig:CP_BS_diagram_11pm} refer to divisibility for specific times $(n,m)$. We can also combine all data and ask, for which regions in 
the $(\lambda_s, \lambda_e)$ plane, the BS dynamics is divisible for all $(n,m)$. 
This is shown in Fig.~\ref{fig:CP_BS_consolidated}. 
As expected, for most choices of parameters, the map will not be CP-divisible for some $(n,m)$. 
Notwithstanding, there are regions where the map is always divisible. 
These regions tend to be concentrated close to $\lambda_e = 0$ (or $\lambda_e = \pi$, which is equivalent). 
And they exist even for large values of $\lambda_s$. 

A direct comparison with the memory kernel, Sec.~\ref{sec:Memory_Kernel}, is not generally possible since both refer to different physical aspects of the problem. 
But if we focus on $\mathcal{N}_{n+1,n}$, then some comparison is possible. 
Recall that the MK describes how the dynamics from $n\to n+1$ is affected by previous times.
Thus, regions where the memory kernel is large tend to be accompanied by regions where $\mathcal{N}_{n+1,n} > 0$. 
This is indeed the case, as can be seen by comparing Fig.~\ref{fig:CP_BS_diagram} with \ref{fig:MK_BS_diagram}.

%
%
\subsection{TMS dynamics}
%
%

\begin{figure*}
    \centering
    \includegraphics[width=\textwidth]{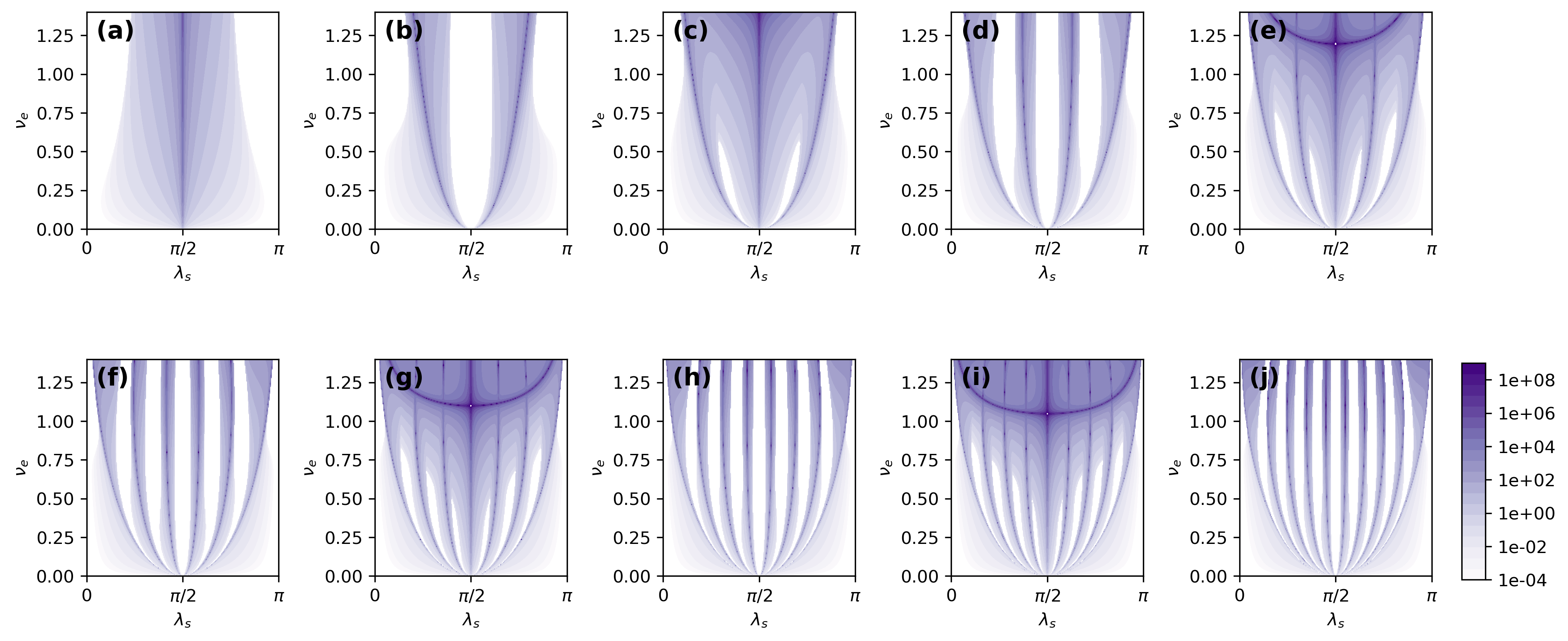}
    \caption{CP-divisibility measure, $\mathcal{N}_{n+1,n}$ [Eq.~\eqref{CP_Nmn}] in the $(\lambda_s, \nu_e)$ plane, for the TMS dynamics. Each plot corresponds to a different values of $n$, from 1 to 10 in steps of 1.
      }
    \label{fig:CP_TMS_diagram}
\end{figure*}

The situation for the TMS dynamics is dramatically different. 
Diagrams for $\mathcal{N}_{n+1,n}$ in the $(\lambda_s, \nu_e)$ plane are shown in Fig.~\ref{fig:CP_TMS_diagram} for different values of $n$. 
In contrast to the BS maps, now most of parameter space is non-divisible.
Moreover, the region where it is non-divisible increases for longer times.
And finally,  what is perhaps the least intuitive, the regions where the map is non-divisible are denser for \emph{small}, instead of large, $\nu_e$ (although the values of $\mathcal{N}_{n+1,n}$ are smaller correspondingly smaller).
This is a consequence of the fact that the TMS dynamics spontaneously creates excitations in the system, which implies that for large $\nu_e$ a substantial amount of noise is introduced, making the map more likely to be divisible. If $\nu_e = 0$ the map is, of course, divisible by construction. 
However, the results in Fig.~\ref{fig:CP_TMS_diagram} show that for arbitrarily small, but non-zero $\nu_e$, the map is already non-divisible, albeit with a small $\mathcal{N}_{n+1,n}$. 
As with the BS dynamics, one could also combine all these diagrams to ask whether there are regions in the $(\lambda_s, \nu_e)$ where the map is always divisible, for all $(n,m)$. 

The answer to this question is, in this case, negative: for the TMS dynamics the dynamics is never divisible, except for the trivial line $\nu_e = 0$.
This represents a major difference in comparison with teh BS dynamics and, once again, is ultimately a property of the entangling nature of the two-mode squeezing interaction~\eqref{V_TMS}.

%
%
\section{\label{sec:discussion}Discussion}
%
%

\subsection{Summary of main results}

The goal of this paper was to provide a robust framework for studying non-Markovianity from multiple angles. We did this using two main ingredients. 
First, collisional models, which allow us to introduce non-Markovianity in a fully controllable way. 
And second, continuous-variable Gaussian operations, which replace the (generally complicated) dynamics of the density matrix into a much simpler map for the covariance matrix. 
We showed that the non-Markovian dynamics can be fully encapsulated into a Markovian embedding, from which all relevant properties and quantifiers can be neatly derived. 
In order to gain  physical insight into what is, generally, a very complicated problem, our exposition was example-oriented. 
We focused on two types of interactions, with very distinct physical properties. 
Our framework, however, is  general. 
The main results can be summarized as follows: 
\begin{itemize}
    \item The global evolution at the level of the density matrix [Eq.~\eqref{basic_map}] is converted into an equation for the global covariance matrix [Eq.~\eqref{basic_map_sigma}]. Unitaries are replaced by symplectic matrices. 
    \item To fully describe the dynamics of  $S$, it suffices to keep track of the joint state (including correlations) of $S$ and only one of the ancillas; namely $E_{n+1}$ at time $n$. This is the matrix $\gamma^n$ in [Eq.~\eqref{gamma_def}].
    \item The dynamics of $\gamma^n$ is now Markovian and obeys the standard Gaussian CP map~\eqref{gamma_embedding} (Markovian embedding). The matrices $X$, $Y$ are related to the entries of the symplectic matrices $S_n$ and $S_{n+1,n}$ according to Eqs.~\eqref{S2_general}, \eqref{S12_general} and~\eqref{XY_general}. 
    \item The mutual information~\eqref{MI}, between the system and ancilla, immediately before they interacted, provides an intuitive measure of information backflow and can be readily computed from the symplectic eigenvalues of $\gamma^n$ (see attached python code).
    \item The time-non-local dynamics defining the memory kernel, Eq.~\eqref{mem_kernel}, can be rewritten at the level of the system covariance matrix as in~\eqref{theta_mem_kernel}. The memory kernel depends \emph{only} on the matrix $X$ and can be computed using Eq.~\eqref{MK_matrix}. One can also write a Kraus decomposition of the MK, Eq.~\eqref{theta_mem_kernel}. The coefficients $\kappa_{ij}^n$ are found from Eq.~\eqref{kappa_recipe}. 
    \item The intermediate map, taking the system from time $n$ to time $m$ is given by Eqs.~\eqref{CP_theta_m_n} and~\eqref{CP_Xmn_Ymn}. 
    When this map is CP, we say the dynamics is CP-divisible. A monotone of CP-divisibility is given by Eq.~\eqref{CP_Nmn} and depends only on the matrices $X$ and $Y$. 
\end{itemize}

\subsection{Main conclusions for the BS and TMS dynamics}

We have focused on two types of maps. 
The system-ancilla interaction was always fixed to be of beam-splitter-type (partial SWAP). 
But the ancilla-ancilla interaction could be either beam-splitter or a two-mode squeezing. 
The behaviour of the two are dramatically different. 

For the former, we have found that the combination of the two beam-splitter interactions lead to strong resonance effects that cause most quantities to oscillate in time and also depend sensibly on the relative signs of the interaction strengths (c.f. Figs.~\ref{fig:MK_BS_diagram} or~\ref{fig:CP_BS_diagram}). 
For the BS dynamics, there is also a non-negligible portion of parameter space in which the dynamics is always Markovian (Fig.~\ref{fig:CP_BS_consolidated}).

Conversely, in the TMS dynamics excitations are constantly being generated in the system. As a consequence, the dynamics is only stable for certain values of the interaction strength (Fig.~\ref{fig:ada}(d)). 
If the interaction is too strong, the occupations in the system diverge (never reach a steady-state). Interestingly, this is also reflected in the memory kernel, which acquires infinitely long memory (Fig.~\ref{fig:MK_TMS_QP_divergent}). 
The TMS dynamics is also always non-Markovian (never CP-divisible; Fig.~\ref{fig:CP_TMS_diagram}), unless the ancilla-ancilla interaction is strictly zero. 
This reflects the entangling nature of the two-mode squeezing. 
The magnitude of the non-Markovianity, of course, is small for weak interactions. This is clearly seen, for instance, in the memory kernel, Fig.~\ref{fig:MK_TMS_diagram_Q}.

\subsection{Possible extensions}

Our framework can be readily extended to a broad range of scenarios. We being by mentioning problems which are straightforward extensions of our results. 
Throughout the paper, we have focused on ancillas initially prepared in the vacuum state. Studying different initial preparations would be interesting since the memory kernel does not depend on this, but CP-divisibility does. It would be particularly interesting to study the introduction of single-mode squeezing in the ancillas.  

Another natural extension would be to consider different types of interactions, as in Refs~\cite{jin2018non, ccakmak2017non}. In particular, one thing that we have not explored are interactions that lead to ``non-diagonal'' memory kernels. 
As discussed below Eq.~\eqref{theta_mem_kernel}, a MK involving the identity or $\sigma_z$ is always diagonal, meaning that each entry of $\theta^n$ is only affected by the same entry at past times. A memory kernel involving $\sigma_\pm$, however, would imply, for instance, that $\langle Q^2 \rangle_n$ could be affected by past values of $\langle P^2 \rangle_n$. This could, in principle, generate a plethora of interesting effects. 
Another possibility would be the inclusion of stochastic SWAPs, as in Refs.~\cite{ciccarello2013collision,ciccarello2013quantum}.

Concerning less trivial extensions, throughout this paper we have assumed that the Markov memory length is 1. That is, each ancilla $E_n$ only propagates information to its nearest neighbor. The extension to arbitrary memory length, as studied in Refs.~\cite{jin2018non,ccakmak2017non}, would be quite interesting. And it is also amenable to our framework, provided one extends the Markovian embedding to have longer memory. 

Finally, we mention that the basic ideas set up in this paper could also serve as a starting point for exploring the Gaussian formulation of process tensors~\cite{pollock2018operational,pollock2018non,taranto2019structure}, which provide an alternative, and much broader, way of characterizing non-Markovianity. In fact, this could perhaps also be used as a way to bridge process tensors and the memory kernel.

\emph{Acknowledgements - }
The authors acknowledge fruitful discussions with J. P. Santos, C. B. Maria, S. Campbell and B. S. de Mendonca. 
G.T.L. acknowledges the hospitality of Apt44, where part of this work was developed. 
G.T.L. acknowledges the S\~ao Paulo Research Foundation (grants 2017/07973-5, 2017/50304-7 and 2018/12813-0). 
R.R.C. acknowledges the Brazilian funding agency CNPq (grant 157168/2018-2).

\appendix
\section{\label{app:A} Stability Theory}

We are interested in studying the fixed point stability of the Markovian embedding equation~\eqref{gamma_embedding}, i.e. solutions that satisfy $\gamma^{n+1}=\gamma^n$. To this end, we use the vectorized form~\eqref{gamma_vec} and label the vectorized fixed point solution as $\vec{\gamma^*}$:
\begin{equation}
   \vec{\gamma^*} = X \otimes X \vec{\gamma^*} + \vec{Y}. 
\end{equation}
As long as $\det(\mathbb{I}-X\otimes X)\neq 0$ a fixed point solution can be readily found as
\begin{equation}
    \vec{\gamma^*}=(\mathbb{I}-X\otimes X)^{-1} \vec{Y}.
\end{equation}

The stability of $\vec{\gamma^*}$ will be associated to the eigenvalues of the $X\otimes X$ matrix. 
Or, what is equivalent, the eigenvalues of $X$.
If their modulus are below $1$,  the fixed point will be a globally asymptotic state (GAS) and all trajectories will converge to $\gamma^*$ for large enough $n$. Otherwise, it may diverge. 

The eigenvalues of the matrix $X$ for the BS channel, Eq.~\eqref{XY_BS}, read
\begin{equation}
    \frac{1}{2}\Big( -wx + x \pm \sqrt{(w+1)^2x^2+4wy^2}\Big).
\end{equation}
Using the $(\lambda_s,\lambda_e)$ parametrization, one finds that the only values not satisfying the GAS conditions are $\lambda_e=\pm \pi/2 $ or $\lambda_s=0,\pi$, which represent, respectively, the case where no particle flow to the ancillas and when the system does not interact at all. Excluding those points, the fixed point is a GAS given by:
\begin{equation}
    \gamma^*_{BS}=\begin{pmatrix}
	\epsilon	& 	0 \\[0.2cm]
	0	&	\epsilon.
\end{pmatrix}
\end{equation}
That is, the map tends to homogenize the system to the same initial state of the ancillas. This, of course, is what is expected of a beam-splitter/partial SWAP dynamics.
It is notwithstanding interesting that it remains true even in the case of ancilla-ancilla interactions and non-Markovian dynamics.

Similarly, the eigenvalues of $X$ for the TMS case, Eq.~\eqref{XY_TMS}, read
\begin{equation}
    \begin{aligned}
    &\frac{1}{2}\Big( (1+\tilde{w})x \pm \sqrt{(\tilde{w}-1)^2x^2-4\tilde{w}y^2}\Big), \\ &\frac{1}{2}\Big( (1-\tilde{w})x \pm \sqrt{(\tilde{w}+1)^2x^2+4\tilde{w}y^2}\Big).
    \end{aligned}
\end{equation}
These eigenvalues only fulfill the GAS requirements in the interval where $\nu_e \in \left[0, \sinh{^{-1}(1)} \right]$.
This therefore defines the critical value $\nu_e^\text{crit} = \sinh^{-1}(1)$, after which the dynamics diverges.
Inside this interval, the fixed point is a GAS given by 
\begin{equation}
   \gamma^*_{TMS}= 
   \begin{pmatrix}
	\left(\frac{2\sinh^2(\nu_e)}{1-\sinh^2(\nu_e)} + 1 \right) \epsilon & 0 \\[0.2cm] 
	0 & \left(\frac{2\sinh^2(\nu_e)}{1-\sinh^2(\nu_e)} + 1 \right) \epsilon 
\end{pmatrix}.
\end{equation} 
Thus, we see that system and ancilla once again tend to homogenize. However, the ancilla initial state $\epsilon$ is now amplified by a factor which is always larger than unity and diverges when $\nu_e = \nu_e^\text{crit}$. 
We also call attention to the fact that $\gamma^*_{TMS}$ is a product state, so that no correlations survive in the long-time limit.
    
    
\section{\label{app:MK_BS}Memory Kernel for the BS dynamics}

In this appendix we discuss how to obtain a more compact expression for the memory kernel~\eqref{theta_mem_kernel}, in the case of the BS dynamics. 
This case is simpler because the only non-zero coefficient is $\kappa_{11}^n$, which is proportional to the identity map. That is to say, in this case the MK is actually just a $c$-number, instead of a superoperator.  

To accomplish this, we exploit in more detail the tensor structure of the matrices used in Sec.~\ref{sec:Memory_Kernel} (now all specialized to $N_S = N_E = 1$). 
We being by noting that the matrix $X$ of the BS dynamics, Eq.~\eqref{XY_BS}, can also be written as 
\begin{equation}\label{app_MK_BS_X}
    X = \chi \otimes \mathbb{I}, 
    \qquad 
    \chi = \begin{pmatrix} x & y \\[0.2cm] yw & -x w\end{pmatrix},
\end{equation}
where $\chi$ is now a simple $2\times 2$ matrix and, in this appendix, $\mathbb{I}$ will always refer to the identity of dimension 2. 
Similarly, the projection operator $P_S$ in Eq.~\eqref{MK_PS} can be written as 
\begin{equation}\label{app_MK_BS_P_S}
    P_S = p_s \otimes \mathbb{I}, 
    \qquad p_s = \begin{pmatrix} 1 & 0 \\ 0 & 0 \end{pmatrix}.
\end{equation}
Thus, the matrix $P$ in Eq.~\eqref{MK_P} becomes
\begin{equation}
    P = p_s \otimes \mathbb{I} \otimes p_s \otimes \mathbb{I}. 
\end{equation}
This type of tensor structure, favouring slots 1 and 3, is simply a consequence of the vectorization procedure, Eq.~\eqref{vec_property}. 

The matrix $p_s$  can be further decomposed as 
\begin{equation}\label{app_MK_BS_ps}
    p_s = |0 \rangle \langle 0 |, 
    \qquad 
    |0\rangle = \begin{pmatrix} 1\\0 \end{pmatrix}.
\end{equation}
Dirac's notation is introduced here just for clarity; the state $|0\rangle$ is completely unrelated to the actual Hilbert space basis of the system. 
The advantage of this decomposition is that it allows us to write the isometry $\pi$, in Eq.~\eqref{MK_pi}, as
\begin{equation}\label{app_MK_BS_pi}
    \pi = \langle 0 | \otimes \mathbb{I} \otimes \langle 0 | \otimes \mathbb{I}. 
\end{equation}
This now clearly shows that $\pi$ contracts slots 1 and 3, while acting trivially on 2 and 4. 

At this point, it is convenient to simplify the notation and introduce indices $1,2,3,4$, to refer to which slow of the tensor product the operators act. 
Thus, for instance, we will henceforth write \begin{equation}\label{app_MK_BS_XotimesX}
X\otimes X = \chi \otimes \mathbb{I} \otimes \chi \otimes \mathbb{I} := \chi_1 \chi_3,
\end{equation}
meaning $\chi_1$ acts on slot 1 and $\chi_3$ on slot 3. Similarly, $P = p_s^1 p_s^3$ and, therefore, $Q = 1- p_s^1 p_s^3 := Q_{13}$ is a matrix acting only on slots 1 and 3 (we emphasize that $Q_{13}$ cannot be written as a simple product of an operator acting on 1 and another acting on 3). 
Notice how the special structure appearing in Eq.~\eqref{app_MK_BS_XotimesX} is unique of the BS dynamics. For other types of dynamics, $X\otimes X$ would in general act non-trivially on all four slots. 
Due to this simplification,  the quantity appearing inside $\pi(\ldots) \pi\trans$ in Eq.~\eqref{MK_matrix} will  be an operator acting only on slots 1 and 3.

Next we turn to Eq.~\eqref{kappa_recipe}, describing the coefficients $\kappa_{ij}^n$. 
The contraction $\pi(\ldots) \pi\trans$  eliminates slots 1 and 3, so that $(M_j\trans \otimes M_i\trans)$ is effectively multiplying matrices from slots 2 and 4. 
Thus, one may equivalently write 
\[
(M_j\trans \otimes M_i\trans) \pi(\ldots) \pi\trans = \pi \Big[ (\mathbb{I} \otimes M_j\trans \otimes \mathbb{I} \otimes M_i\trans) \ldots \Big] \pi\trans, 
\]
where $(\ldots)$ refers to all terms inside $\pi(\ldots)\pi\trans$ in Eq.~\eqref{MK_matrix}.
But from the arguments above, these quantities act only on slots 1 and 3. 
Combining this with the fact that $\tr(A\otimes B) = \tr(A) \tr(B)$ explains why, in the BS case, the only non-trivial coefficient will be $\kappa_{11}^n$, corresponding to $M_i = M_j = \mathbb{I}$.
This coefficient may then be written as 
\[
\kappa_{11}^n = \tr_{13}\bigg\{ \pi_{13} \Big[\chi_1 \chi_3 (Q_{13} \chi_1 \chi_3 Q_{13})^n \chi_1 \chi_3 \Big] \pi_{13}\trans\bigg\},
\]
where the remaining trace is now only over slots 1 and 3. 
Finally, we use Eq.~\eqref{app_MK_BS_pi} to express $\pi$ in terms of $\langle 0 |$. 
This allows us to write 
\begin{equation}\label{app_MK_BF_kappa_final}
    \kappa_{11}^n = \langle 00| \, \bar{\chi} \, \big(\bar{Q}\, \bar{\chi}\, \bar{Q}\big)^n \, \bar{\chi} \, | 00\rangle,
\end{equation}
where $|00\rangle = |0\rangle\otimes |0\rangle$, $\bar{\chi} = \chi\otimes \chi$ and $\bar{Q} = \mathbb{I}_4 - p_s \otimes p_s$ are all objects of dimension 4. 
Eq.~\eqref{app_MK_BF_kappa_final} therefore provides a compact representation of the memory Kernel for the BS dynamics. It is expressed solely in terms of $|0\rangle$, $\chi$ and $p_s$, [Eqs.~\eqref{app_MK_BS_X} and \eqref{app_MK_BS_ps}]. 
And it requires exponentiating only operators of dimension 4, in comparison with~\eqref{MK_matrix} which would have dimension 16.

\bibliography{Bib}

\begin{thebibliography}{70}%
\makeatletter
\providecommand \@ifxundefined [1]{%
 \@ifx{#1\undefined}
}%
\providecommand \@ifnum [1]{%
 \ifnum #1\expandafter \@firstoftwo
 \else \expandafter \@secondoftwo
 \fi
}%
\providecommand \@ifx [1]{%
 \ifx #1\expandafter \@firstoftwo
 \else \expandafter \@secondoftwo
 \fi
}%
\providecommand \natexlab [1]{#1}%
\providecommand \enquote  [1]{``#1''}%
\providecommand \bibnamefont  [1]{#1}%
\providecommand \bibfnamefont [1]{#1}%
\providecommand \citenamefont [1]{#1}%
\providecommand \href@noop [0]{\@secondoftwo}%
\providecommand \href [0]{\begingroup \@sanitize@url \@href}%
\providecommand \@href[1]{\@@startlink{#1}\@@href}%
\providecommand \@@href[1]{\endgroup#1\@@endlink}%
\providecommand \@sanitize@url [0]{\catcode `\\12\catcode `\$12\catcode
  `\&12\catcode `\#12\catcode `\^12\catcode `\_12\catcode `\%12\relax}%
\providecommand \@@startlink[1]{}%
\providecommand \@@endlink[0]{}%
\providecommand \url  [0]{\begingroup\@sanitize@url \@url }%
\providecommand \@url [1]{\endgroup\@href {#1}{\urlprefix }}%
\providecommand \urlprefix  [0]{URL }%
\providecommand \Eprint [0]{\href }%
\providecommand \doibase [0]{http://dx.doi.org/}%
\providecommand \selectlanguage [0]{\@gobble}%
\providecommand \bibinfo  [0]{\@secondoftwo}%
\providecommand \bibfield  [0]{\@secondoftwo}%
\providecommand \translation [1]{[#1]}%
\providecommand \BibitemOpen [0]{}%
\providecommand \bibitemStop [0]{}%
\providecommand \bibitemNoStop [0]{.\EOS\space}%
\providecommand \EOS [0]{\spacefactor3000\relax}%
\providecommand \BibitemShut  [1]{\csname bibitem#1\endcsname}%
\let\auto@bib@innerbib\@empty
\bibitem [{\citenamefont {Doob}(1990)}]{doob1990stochastic}%
  \BibitemOpen
  \bibfield  {author} {\bibinfo {author} {\bibfnamefont {J.}~\bibnamefont
  {Doob}},\ }\href {https://books.google.com.br/books?id=NrsrAAAAYAAJ} {\emph
  {\bibinfo {title} {Stochastic processes}}},\ Wiley publications in
  statistics\ (\bibinfo  {publisher} {Wiley},\ \bibinfo {year}
  {1990})\BibitemShut {NoStop}%
\bibitem [{\citenamefont {Binder}\ \emph {et~al.}(2018)\citenamefont {Binder},
  \citenamefont {Thompson},\ and\ \citenamefont {Gu}}]{binder2018practical}%
  \BibitemOpen
  \bibfield  {author} {\bibinfo {author} {\bibfnamefont {F.~C.}\ \bibnamefont
  {Binder}}, \bibinfo {author} {\bibfnamefont {J.}~\bibnamefont {Thompson}}, \
  and\ \bibinfo {author} {\bibfnamefont {M.}~\bibnamefont {Gu}},\ }\href@noop
  {} {\bibfield  {journal} {\bibinfo  {journal} {Physical review letters}\
  }\textbf {\bibinfo {volume} {120}},\ \bibinfo {pages} {240502} (\bibinfo
  {year} {2018})}\BibitemShut {NoStop}%
\bibitem [{\citenamefont {Rivas}\ \emph {et~al.}(2014)\citenamefont {Rivas},
  \citenamefont {Huelga},\ and\ \citenamefont {Plenio}}]{rivas2014quantum}%
  \BibitemOpen
  \bibfield  {author} {\bibinfo {author} {\bibfnamefont {{\'A}.}~\bibnamefont
  {Rivas}}, \bibinfo {author} {\bibfnamefont {S.~F.}\ \bibnamefont {Huelga}}, \
  and\ \bibinfo {author} {\bibfnamefont {M.~B.}\ \bibnamefont {Plenio}},\
  }\href@noop {} {\bibfield  {journal} {\bibinfo  {journal} {Reports on
  Progress in Physics}\ }\textbf {\bibinfo {volume} {77}},\ \bibinfo {pages}
  {094001} (\bibinfo {year} {2014})}\BibitemShut {NoStop}%
\bibitem [{\citenamefont {Breuer}\ \emph {et~al.}(2016)\citenamefont {Breuer},
  \citenamefont {Laine}, \citenamefont {Piilo},\ and\ \citenamefont
  {Vacchini}}]{breuer2016colloquium}%
  \BibitemOpen
  \bibfield  {author} {\bibinfo {author} {\bibfnamefont {H.-P.}\ \bibnamefont
  {Breuer}}, \bibinfo {author} {\bibfnamefont {E.-M.}\ \bibnamefont {Laine}},
  \bibinfo {author} {\bibfnamefont {J.}~\bibnamefont {Piilo}}, \ and\ \bibinfo
  {author} {\bibfnamefont {B.}~\bibnamefont {Vacchini}},\ }\href@noop {}
  {\bibfield  {journal} {\bibinfo  {journal} {Reviews of Modern Physics}\
  }\textbf {\bibinfo {volume} {88}},\ \bibinfo {pages} {021002} (\bibinfo
  {year} {2016})}\BibitemShut {NoStop}%
\bibitem [{\citenamefont {Breuer}\ \emph {et~al.}(2009)\citenamefont {Breuer},
  \citenamefont {Laine},\ and\ \citenamefont {Piilo}}]{breuer2009measure}%
  \BibitemOpen
  \bibfield  {author} {\bibinfo {author} {\bibfnamefont {H.-P.}\ \bibnamefont
  {Breuer}}, \bibinfo {author} {\bibfnamefont {E.-M.}\ \bibnamefont {Laine}}, \
  and\ \bibinfo {author} {\bibfnamefont {J.}~\bibnamefont {Piilo}},\
  }\href@noop {} {\bibfield  {journal} {\bibinfo  {journal} {Physical review
  letters}\ }\textbf {\bibinfo {volume} {103}},\ \bibinfo {pages} {210401}
  (\bibinfo {year} {2009})}\BibitemShut {NoStop}%
\bibitem [{\citenamefont {Chru{\'s}ci{\'n}ski}\ \emph
  {et~al.}(2018)\citenamefont {Chru{\'s}ci{\'n}ski}, \citenamefont {Rivas},\
  and\ \citenamefont {St{\o}rmer}}]{chruscinski2018divisibility}%
  \BibitemOpen
  \bibfield  {author} {\bibinfo {author} {\bibfnamefont {D.}~\bibnamefont
  {Chru{\'s}ci{\'n}ski}}, \bibinfo {author} {\bibfnamefont
  {{\'A}.}~\bibnamefont {Rivas}}, \ and\ \bibinfo {author} {\bibfnamefont
  {E.}~\bibnamefont {St{\o}rmer}},\ }\href@noop {} {\bibfield  {journal}
  {\bibinfo  {journal} {Physical review letters}\ }\textbf {\bibinfo {volume}
  {121}},\ \bibinfo {pages} {080407} (\bibinfo {year} {2018})}\BibitemShut
  {NoStop}%
\bibitem [{\citenamefont {Vasile}\ \emph {et~al.}(2011)\citenamefont {Vasile},
  \citenamefont {Maniscalco}, \citenamefont {Paris}, \citenamefont {Breuer},\
  and\ \citenamefont {Piilo}}]{vasile2011quantifying}%
  \BibitemOpen
  \bibfield  {author} {\bibinfo {author} {\bibfnamefont {R.}~\bibnamefont
  {Vasile}}, \bibinfo {author} {\bibfnamefont {S.}~\bibnamefont {Maniscalco}},
  \bibinfo {author} {\bibfnamefont {M.~G.}\ \bibnamefont {Paris}}, \bibinfo
  {author} {\bibfnamefont {H.-P.}\ \bibnamefont {Breuer}}, \ and\ \bibinfo
  {author} {\bibfnamefont {J.}~\bibnamefont {Piilo}},\ }\href@noop {}
  {\bibfield  {journal} {\bibinfo  {journal} {Physical Review A}\ }\textbf
  {\bibinfo {volume} {84}},\ \bibinfo {pages} {052118} (\bibinfo {year}
  {2011})}\BibitemShut {NoStop}%
\bibitem [{\citenamefont {Laine}\ \emph {et~al.}(2010)\citenamefont {Laine},
  \citenamefont {Piilo},\ and\ \citenamefont {Breuer}}]{laine2010measure}%
  \BibitemOpen
  \bibfield  {author} {\bibinfo {author} {\bibfnamefont {E.-M.}\ \bibnamefont
  {Laine}}, \bibinfo {author} {\bibfnamefont {J.}~\bibnamefont {Piilo}}, \ and\
  \bibinfo {author} {\bibfnamefont {H.-P.}\ \bibnamefont {Breuer}},\
  }\href@noop {} {\bibfield  {journal} {\bibinfo  {journal} {Physical Review
  A}\ }\textbf {\bibinfo {volume} {81}},\ \bibinfo {pages} {062115} (\bibinfo
  {year} {2010})}\BibitemShut {NoStop}%
\bibitem [{\citenamefont {Rivas}\ \emph {et~al.}(2010)\citenamefont {Rivas},
  \citenamefont {Huelga},\ and\ \citenamefont
  {Plenio}}]{rivas2010entanglement}%
  \BibitemOpen
  \bibfield  {author} {\bibinfo {author} {\bibfnamefont {{\'A}.}~\bibnamefont
  {Rivas}}, \bibinfo {author} {\bibfnamefont {S.~F.}\ \bibnamefont {Huelga}}, \
  and\ \bibinfo {author} {\bibfnamefont {M.~B.}\ \bibnamefont {Plenio}},\
  }\href@noop {} {\bibfield  {journal} {\bibinfo  {journal} {Physical review
  letters}\ }\textbf {\bibinfo {volume} {105}},\ \bibinfo {pages} {050403}
  (\bibinfo {year} {2010})}\BibitemShut {NoStop}%
\bibitem [{\citenamefont {Hou}\ \emph {et~al.}(2011)\citenamefont {Hou},
  \citenamefont {Yi}, \citenamefont {Yu},\ and\ \citenamefont
  {Oh}}]{hou2011alternative}%
  \BibitemOpen
  \bibfield  {author} {\bibinfo {author} {\bibfnamefont {S.}~\bibnamefont
  {Hou}}, \bibinfo {author} {\bibfnamefont {X.}~\bibnamefont {Yi}}, \bibinfo
  {author} {\bibfnamefont {S.}~\bibnamefont {Yu}}, \ and\ \bibinfo {author}
  {\bibfnamefont {C.}~\bibnamefont {Oh}},\ }\href@noop {} {\bibfield  {journal}
  {\bibinfo  {journal} {Physical Review A}\ }\textbf {\bibinfo {volume} {83}},\
  \bibinfo {pages} {062115} (\bibinfo {year} {2011})}\BibitemShut {NoStop}%
\bibitem [{\citenamefont {Luo}\ \emph {et~al.}(2012)\citenamefont {Luo},
  \citenamefont {Fu},\ and\ \citenamefont {Song}}]{luo2012quantifying}%
  \BibitemOpen
  \bibfield  {author} {\bibinfo {author} {\bibfnamefont {S.}~\bibnamefont
  {Luo}}, \bibinfo {author} {\bibfnamefont {S.}~\bibnamefont {Fu}}, \ and\
  \bibinfo {author} {\bibfnamefont {H.}~\bibnamefont {Song}},\ }\href@noop {}
  {\bibfield  {journal} {\bibinfo  {journal} {Physical Review A}\ }\textbf
  {\bibinfo {volume} {86}},\ \bibinfo {pages} {044101} (\bibinfo {year}
  {2012})}\BibitemShut {NoStop}%
\bibitem [{\citenamefont {Chru{\'s}ci{\'n}ski}\ and\ \citenamefont
  {Kossakowski}(2012)}]{chruscinski2012markovianity}%
  \BibitemOpen
  \bibfield  {author} {\bibinfo {author} {\bibfnamefont {D.}~\bibnamefont
  {Chru{\'s}ci{\'n}ski}}\ and\ \bibinfo {author} {\bibfnamefont
  {A.}~\bibnamefont {Kossakowski}},\ }\href@noop {} {\bibfield  {journal}
  {\bibinfo  {journal} {Journal of Physics B: Atomic, Molecular and Optical
  Physics}\ }\textbf {\bibinfo {volume} {45}},\ \bibinfo {pages} {154002}
  (\bibinfo {year} {2012})}\BibitemShut {NoStop}%
\bibitem [{\citenamefont {Chru{\'s}ci{\'n}ski}\ and\ \citenamefont
  {Kossakowski}(2014)}]{chruscinski2014witnessing}%
  \BibitemOpen
  \bibfield  {author} {\bibinfo {author} {\bibfnamefont {D.}~\bibnamefont
  {Chru{\'s}ci{\'n}ski}}\ and\ \bibinfo {author} {\bibfnamefont
  {A.}~\bibnamefont {Kossakowski}},\ }\href@noop {} {\bibfield  {journal}
  {\bibinfo  {journal} {The European Physical Journal D}\ }\textbf {\bibinfo
  {volume} {68}},\ \bibinfo {pages} {7} (\bibinfo {year} {2014})}\BibitemShut
  {NoStop}%
\bibitem [{\citenamefont {Costa}\ \emph {et~al.}(2014)\citenamefont {Costa},
  \citenamefont {Angelo},\ and\ \citenamefont {Beims}}]{costa2014monogamy}%
  \BibitemOpen
  \bibfield  {author} {\bibinfo {author} {\bibfnamefont {A.}~\bibnamefont
  {Costa}}, \bibinfo {author} {\bibfnamefont {R.}~\bibnamefont {Angelo}}, \
  and\ \bibinfo {author} {\bibfnamefont {M.}~\bibnamefont {Beims}},\
  }\href@noop {} {\bibfield  {journal} {\bibinfo  {journal} {Physical Review
  A}\ }\textbf {\bibinfo {volume} {90}},\ \bibinfo {pages} {012322} (\bibinfo
  {year} {2014})}\BibitemShut {NoStop}%
\bibitem [{\citenamefont {Strasberg}\ and\ \citenamefont
  {Esposito}(2018)}]{strasberg2018response}%
  \BibitemOpen
  \bibfield  {author} {\bibinfo {author} {\bibfnamefont {P.}~\bibnamefont
  {Strasberg}}\ and\ \bibinfo {author} {\bibfnamefont {M.}~\bibnamefont
  {Esposito}},\ }\href@noop {} {\bibfield  {journal} {\bibinfo  {journal}
  {Physical review letters}\ }\textbf {\bibinfo {volume} {121}},\ \bibinfo
  {pages} {040601} (\bibinfo {year} {2018})}\BibitemShut {NoStop}%
\bibitem [{\citenamefont {Souza}\ \emph {et~al.}(2015)\citenamefont {Souza},
  \citenamefont {Dhar}, \citenamefont {Bera}, \citenamefont {Liuzzo-Scorpo},\
  and\ \citenamefont {Adesso}}]{souza2015gaussian}%
  \BibitemOpen
  \bibfield  {author} {\bibinfo {author} {\bibfnamefont {L.~A.}\ \bibnamefont
  {Souza}}, \bibinfo {author} {\bibfnamefont {H.~S.}\ \bibnamefont {Dhar}},
  \bibinfo {author} {\bibfnamefont {M.~N.}\ \bibnamefont {Bera}}, \bibinfo
  {author} {\bibfnamefont {P.}~\bibnamefont {Liuzzo-Scorpo}}, \ and\ \bibinfo
  {author} {\bibfnamefont {G.}~\bibnamefont {Adesso}},\ }\href@noop {}
  {\bibfield  {journal} {\bibinfo  {journal} {Physical Review A}\ }\textbf
  {\bibinfo {volume} {92}},\ \bibinfo {pages} {052122} (\bibinfo {year}
  {2015})}\BibitemShut {NoStop}%
\bibitem [{\citenamefont {Fanchini}\ \emph {et~al.}(2014)\citenamefont
  {Fanchini}, \citenamefont {Karpat}, \citenamefont {{\c{C}}akmak},
  \citenamefont {Castelano}, \citenamefont {Aguilar}, \citenamefont
  {Far{\'\i}as}, \citenamefont {Walborn}, \citenamefont {Ribeiro},\ and\
  \citenamefont {De~Oliveira}}]{fanchini2014non}%
  \BibitemOpen
  \bibfield  {author} {\bibinfo {author} {\bibfnamefont {F.~F.}\ \bibnamefont
  {Fanchini}}, \bibinfo {author} {\bibfnamefont {G.}~\bibnamefont {Karpat}},
  \bibinfo {author} {\bibfnamefont {B.}~\bibnamefont {{\c{C}}akmak}}, \bibinfo
  {author} {\bibfnamefont {L.}~\bibnamefont {Castelano}}, \bibinfo {author}
  {\bibfnamefont {G.}~\bibnamefont {Aguilar}}, \bibinfo {author} {\bibfnamefont
  {O.~J.}\ \bibnamefont {Far{\'\i}as}}, \bibinfo {author} {\bibfnamefont
  {S.}~\bibnamefont {Walborn}}, \bibinfo {author} {\bibfnamefont {P.~S.}\
  \bibnamefont {Ribeiro}}, \ and\ \bibinfo {author} {\bibfnamefont
  {M.}~\bibnamefont {De~Oliveira}},\ }\href@noop {} {\bibfield  {journal}
  {\bibinfo  {journal} {Physical Review Letters}\ }\textbf {\bibinfo {volume}
  {112}},\ \bibinfo {pages} {210402} (\bibinfo {year} {2014})}\BibitemShut
  {NoStop}%
\bibitem [{\citenamefont {Lu}\ \emph {et~al.}(2010)\citenamefont {Lu},
  \citenamefont {Wang},\ and\ \citenamefont {Sun}}]{lu2010quantum}%
  \BibitemOpen
  \bibfield  {author} {\bibinfo {author} {\bibfnamefont {X.-M.}\ \bibnamefont
  {Lu}}, \bibinfo {author} {\bibfnamefont {X.}~\bibnamefont {Wang}}, \ and\
  \bibinfo {author} {\bibfnamefont {C.}~\bibnamefont {Sun}},\ }\href@noop {}
  {\bibfield  {journal} {\bibinfo  {journal} {Physical Review A}\ }\textbf
  {\bibinfo {volume} {82}},\ \bibinfo {pages} {042103} (\bibinfo {year}
  {2010})}\BibitemShut {NoStop}%
\bibitem [{\citenamefont {Barnett}\ and\ \citenamefont
  {Stenholm}(2001)}]{barnett2001hazards}%
  \BibitemOpen
  \bibfield  {author} {\bibinfo {author} {\bibfnamefont {S.~M.}\ \bibnamefont
  {Barnett}}\ and\ \bibinfo {author} {\bibfnamefont {S.}~\bibnamefont
  {Stenholm}},\ }\href@noop {} {\bibfield  {journal} {\bibinfo  {journal}
  {Physical Review A}\ }\textbf {\bibinfo {volume} {64}},\ \bibinfo {pages}
  {033808} (\bibinfo {year} {2001})}\BibitemShut {NoStop}%
\bibitem [{\citenamefont {Shabani}\ and\ \citenamefont
  {Lidar}(2005)}]{shabani2005completely}%
  \BibitemOpen
  \bibfield  {author} {\bibinfo {author} {\bibfnamefont {A.}~\bibnamefont
  {Shabani}}\ and\ \bibinfo {author} {\bibfnamefont {D.~A.}\ \bibnamefont
  {Lidar}},\ }\href@noop {} {\bibfield  {journal} {\bibinfo  {journal}
  {Physical Review A}\ }\textbf {\bibinfo {volume} {71}},\ \bibinfo {pages}
  {020101} (\bibinfo {year} {2005})}\BibitemShut {NoStop}%
\bibitem [{\citenamefont {Hall}\ \emph {et~al.}(2014)\citenamefont {Hall},
  \citenamefont {Cresser}, \citenamefont {Li},\ and\ \citenamefont
  {Andersson}}]{hall2014canonical}%
  \BibitemOpen
  \bibfield  {author} {\bibinfo {author} {\bibfnamefont {M.~J.}\ \bibnamefont
  {Hall}}, \bibinfo {author} {\bibfnamefont {J.~D.}\ \bibnamefont {Cresser}},
  \bibinfo {author} {\bibfnamefont {L.}~\bibnamefont {Li}}, \ and\ \bibinfo
  {author} {\bibfnamefont {E.}~\bibnamefont {Andersson}},\ }\href@noop {}
  {\bibfield  {journal} {\bibinfo  {journal} {Physical Review A}\ }\textbf
  {\bibinfo {volume} {89}},\ \bibinfo {pages} {042120} (\bibinfo {year}
  {2014})}\BibitemShut {NoStop}%
\bibitem [{\citenamefont {Mazzola}\ \emph
  {et~al.}(2010{\natexlab{a}})\citenamefont {Mazzola}, \citenamefont {Laine},
  \citenamefont {Breuer}, \citenamefont {Maniscalco},\ and\ \citenamefont
  {Piilo}}]{Mazzola2010}%
  \BibitemOpen
  \bibfield  {author} {\bibinfo {author} {\bibfnamefont {L.}~\bibnamefont
  {Mazzola}}, \bibinfo {author} {\bibfnamefont {E.~M.}\ \bibnamefont {Laine}},
  \bibinfo {author} {\bibfnamefont {H.~P.}\ \bibnamefont {Breuer}}, \bibinfo
  {author} {\bibfnamefont {S.}~\bibnamefont {Maniscalco}}, \ and\ \bibinfo
  {author} {\bibfnamefont {J.}~\bibnamefont {Piilo}},\ }\href {\doibase
  10.1103/PhysRevA.81.062120} {\bibfield  {journal} {\bibinfo  {journal}
  {Physical Review A - Atomic, Molecular, and Optical Physics}\ }\textbf
  {\bibinfo {volume} {81}},\ \bibinfo {pages} {062120} (\bibinfo {year}
  {2010}{\natexlab{a}})},\ \Eprint {http://arxiv.org/abs/1003.3817}
  {arXiv:1003.3817} \BibitemShut {NoStop}%
\bibitem [{\citenamefont {Liu}\ \emph {et~al.}(2019)\citenamefont {Liu},
  \citenamefont {Zhou},\ and\ \citenamefont {Zhou}}]{Liu2019a}%
  \BibitemOpen
  \bibfield  {author} {\bibinfo {author} {\bibfnamefont {F.}~\bibnamefont
  {Liu}}, \bibinfo {author} {\bibfnamefont {X.}~\bibnamefont {Zhou}}, \ and\
  \bibinfo {author} {\bibfnamefont {Z.~W.}\ \bibnamefont {Zhou}},\ }\href
  {\doibase 10.1103/PhysRevA.99.052119} {\bibfield  {journal} {\bibinfo
  {journal} {Physical Review A}\ }\textbf {\bibinfo {volume} {99}},\ \bibinfo
  {pages} {052119} (\bibinfo {year} {2019})}\BibitemShut {NoStop}%
\bibitem [{\citenamefont {Pollock}\ \emph
  {et~al.}(2018{\natexlab{a}})\citenamefont {Pollock}, \citenamefont
  {Rodr{\'\i}guez-Rosario}, \citenamefont {Frauenheim}, \citenamefont
  {Paternostro},\ and\ \citenamefont {Modi}}]{pollock2018operational}%
  \BibitemOpen
  \bibfield  {author} {\bibinfo {author} {\bibfnamefont {F.~A.}\ \bibnamefont
  {Pollock}}, \bibinfo {author} {\bibfnamefont {C.}~\bibnamefont
  {Rodr{\'\i}guez-Rosario}}, \bibinfo {author} {\bibfnamefont {T.}~\bibnamefont
  {Frauenheim}}, \bibinfo {author} {\bibfnamefont {M.}~\bibnamefont
  {Paternostro}}, \ and\ \bibinfo {author} {\bibfnamefont {K.}~\bibnamefont
  {Modi}},\ }\href@noop {} {\bibfield  {journal} {\bibinfo  {journal} {Physical
  review letters}\ }\textbf {\bibinfo {volume} {120}},\ \bibinfo {pages}
  {040405} (\bibinfo {year} {2018}{\natexlab{a}})}\BibitemShut {NoStop}%
\bibitem [{\citenamefont {Pollock}\ \emph
  {et~al.}(2018{\natexlab{b}})\citenamefont {Pollock}, \citenamefont
  {Rodr{\'\i}guez-Rosario}, \citenamefont {Frauenheim}, \citenamefont
  {Paternostro},\ and\ \citenamefont {Modi}}]{pollock2018non}%
  \BibitemOpen
  \bibfield  {author} {\bibinfo {author} {\bibfnamefont {F.~A.}\ \bibnamefont
  {Pollock}}, \bibinfo {author} {\bibfnamefont {C.}~\bibnamefont
  {Rodr{\'\i}guez-Rosario}}, \bibinfo {author} {\bibfnamefont {T.}~\bibnamefont
  {Frauenheim}}, \bibinfo {author} {\bibfnamefont {M.}~\bibnamefont
  {Paternostro}}, \ and\ \bibinfo {author} {\bibfnamefont {K.}~\bibnamefont
  {Modi}},\ }\href@noop {} {\bibfield  {journal} {\bibinfo  {journal} {Physical
  Review A}\ }\textbf {\bibinfo {volume} {97}},\ \bibinfo {pages} {012127}
  (\bibinfo {year} {2018}{\natexlab{b}})}\BibitemShut {NoStop}%
\bibitem [{\citenamefont {Taranto}\ \emph {et~al.}(2019)\citenamefont
  {Taranto}, \citenamefont {Milz}, \citenamefont {Pollock},\ and\ \citenamefont
  {Modi}}]{taranto2019structure}%
  \BibitemOpen
  \bibfield  {author} {\bibinfo {author} {\bibfnamefont {P.}~\bibnamefont
  {Taranto}}, \bibinfo {author} {\bibfnamefont {S.}~\bibnamefont {Milz}},
  \bibinfo {author} {\bibfnamefont {F.~A.}\ \bibnamefont {Pollock}}, \ and\
  \bibinfo {author} {\bibfnamefont {K.}~\bibnamefont {Modi}},\ }\href@noop {}
  {\bibfield  {journal} {\bibinfo  {journal} {Physical Review A}\ }\textbf
  {\bibinfo {volume} {99}},\ \bibinfo {pages} {042108} (\bibinfo {year}
  {2019})}\BibitemShut {NoStop}%
\bibitem [{\citenamefont {Rau}(1963)}]{Rau1963}%
  \BibitemOpen
  \bibfield  {author} {\bibinfo {author} {\bibfnamefont {J.}~\bibnamefont
  {Rau}},\ }\href {\doibase 10.1103/PhysRev.129.1880} {\bibfield  {journal}
  {\bibinfo  {journal} {Physical Review}\ }\textbf {\bibinfo {volume} {129}},\
  \bibinfo {pages} {1880} (\bibinfo {year} {1963})}\BibitemShut {NoStop}%
\bibitem [{\citenamefont {Scarani}\ \emph {et~al.}(2002)\citenamefont
  {Scarani}, \citenamefont {Ziman}, \citenamefont {{\v{S}}telmachovi{\v{c}}},
  \citenamefont {Gisin}, \citenamefont {Bu{\v{z}}ek},\ and\ \citenamefont
  {Bu{\v{z}}ek}}]{Scarani2002}%
  \BibitemOpen
  \bibfield  {author} {\bibinfo {author} {\bibfnamefont {V.}~\bibnamefont
  {Scarani}}, \bibinfo {author} {\bibfnamefont {M.}~\bibnamefont {Ziman}},
  \bibinfo {author} {\bibfnamefont {P.}~\bibnamefont
  {{\v{S}}telmachovi{\v{c}}}}, \bibinfo {author} {\bibfnamefont
  {N.}~\bibnamefont {Gisin}}, \bibinfo {author} {\bibfnamefont
  {V.}~\bibnamefont {Bu{\v{z}}ek}}, \ and\ \bibinfo {author} {\bibfnamefont
  {V.}~\bibnamefont {Bu{\v{z}}ek}},\ }\href {\doibase
  10.1103/PhysRevLett.88.097905} {\bibfield  {journal} {\bibinfo  {journal}
  {Physical Review Letters}\ }\textbf {\bibinfo {volume} {88}},\ \bibinfo
  {pages} {097905} (\bibinfo {year} {2002})},\ \Eprint
  {http://arxiv.org/abs/0110088} {arXiv:0110088 [quant-ph]} \BibitemShut
  {NoStop}%
\bibitem [{\citenamefont {Ziman}\ \emph {et~al.}(2002)\citenamefont {Ziman},
  \citenamefont {{\v{S}}telmachovi{\v{c}}}, \citenamefont {Buz{\v{z}}ek},
  \citenamefont {Hillery}, \citenamefont {Scarani},\ and\ \citenamefont
  {Gisin}}]{Ziman2002}%
  \BibitemOpen
  \bibfield  {author} {\bibinfo {author} {\bibfnamefont {M.}~\bibnamefont
  {Ziman}}, \bibinfo {author} {\bibfnamefont {P.}~\bibnamefont
  {{\v{S}}telmachovi{\v{c}}}}, \bibinfo {author} {\bibfnamefont
  {V.}~\bibnamefont {Buz{\v{z}}ek}}, \bibinfo {author} {\bibfnamefont
  {M.}~\bibnamefont {Hillery}}, \bibinfo {author} {\bibfnamefont
  {V.}~\bibnamefont {Scarani}}, \ and\ \bibinfo {author} {\bibfnamefont
  {N.}~\bibnamefont {Gisin}},\ }\href {\doibase 10.1103/PhysRevA.65.042105}
  {\bibfield  {journal} {\bibinfo  {journal} {Physical Review A. Atomic,
  Molecular, and Optical Physics}\ }\textbf {\bibinfo {volume} {65}},\ \bibinfo
  {pages} {042105} (\bibinfo {year} {2002})}\BibitemShut {NoStop}%
\bibitem [{\citenamefont {Englert}\ and\ \citenamefont
  {Morigi}(2002)}]{Englert2002}%
  \BibitemOpen
  \bibfield  {author} {\bibinfo {author} {\bibfnamefont {B.-G.}\ \bibnamefont
  {Englert}}\ and\ \bibinfo {author} {\bibfnamefont {G.}~\bibnamefont
  {Morigi}},\ }in\ \href@noop {} {\emph {\bibinfo {booktitle} {Coherent
  Evolution in Noisy Environments - Lecture Notes in Physics}}},\ \bibinfo
  {editor} {edited by\ \bibinfo {editor} {\bibfnamefont {A.}~\bibnamefont
  {Buchleitner}}\ and\ \bibinfo {editor} {\bibfnamefont {K.}~\bibnamefont
  {Hornberger}}}\ (\bibinfo  {publisher} {Springer},\ \bibinfo {address}
  {Berlin, Heidelberg},\ \bibinfo {year} {2002})\ p.\ \bibinfo {pages} {611},\
  \Eprint {http://arxiv.org/abs/0206116} {arXiv:0206116 [quant-ph]}
  \BibitemShut {NoStop}%
\bibitem [{\citenamefont {Attal}\ and\ \citenamefont
  {Pautrat}(2006)}]{Attal2006a}%
  \BibitemOpen
  \bibfield  {author} {\bibinfo {author} {\bibfnamefont {S.}~\bibnamefont
  {Attal}}\ and\ \bibinfo {author} {\bibfnamefont {Y.}~\bibnamefont
  {Pautrat}},\ }\href {\doibase 10.1007/s00023-005-0242-8} {\bibfield
  {journal} {\bibinfo  {journal} {Annales Henri Poincar{\'{e}}}\ }\textbf
  {\bibinfo {volume} {7}},\ \bibinfo {pages} {59} (\bibinfo {year} {2006})},\
  \Eprint {http://arxiv.org/abs/0311002} {arXiv:0311002 [math-ph]} \BibitemShut
  {NoStop}%
\bibitem [{\citenamefont {Pellegrini}\ and\ \citenamefont
  {Petruccione}(2009)}]{Pellegrini2009}%
  \BibitemOpen
  \bibfield  {author} {\bibinfo {author} {\bibfnamefont {C.}~\bibnamefont
  {Pellegrini}}\ and\ \bibinfo {author} {\bibfnamefont {F.}~\bibnamefont
  {Petruccione}},\ }\href {\doibase 10.1088/1751-8113/42/42/425304} {\bibfield
  {journal} {\bibinfo  {journal} {Journal of Physics A: Mathematical and
  Theoretical}\ }\textbf {\bibinfo {volume} {42}},\ \bibinfo {pages} {425304}
  (\bibinfo {year} {2009})},\ \Eprint {http://arxiv.org/abs/0903.3859}
  {arXiv:0903.3859} \BibitemShut {NoStop}%
\bibitem [{\citenamefont {Karevski}\ and\ \citenamefont
  {Platini}(2009)}]{karevski2009quantum}%
  \BibitemOpen
  \bibfield  {author} {\bibinfo {author} {\bibfnamefont {D.}~\bibnamefont
  {Karevski}}\ and\ \bibinfo {author} {\bibfnamefont {T.}~\bibnamefont
  {Platini}},\ }\href@noop {} {\bibfield  {journal} {\bibinfo  {journal}
  {Physical review letters}\ }\textbf {\bibinfo {volume} {102}},\ \bibinfo
  {pages} {207207} (\bibinfo {year} {2009})}\BibitemShut {NoStop}%
\bibitem [{\citenamefont {Landi}\ \emph {et~al.}(2014)\citenamefont {Landi},
  \citenamefont {Novais}, \citenamefont {de~Oliveira},\ and\ \citenamefont
  {Karevski}}]{landi2014flux}%
  \BibitemOpen
  \bibfield  {author} {\bibinfo {author} {\bibfnamefont {G.~T.}\ \bibnamefont
  {Landi}}, \bibinfo {author} {\bibfnamefont {E.}~\bibnamefont {Novais}},
  \bibinfo {author} {\bibfnamefont {M.~J.}\ \bibnamefont {de~Oliveira}}, \ and\
  \bibinfo {author} {\bibfnamefont {D.}~\bibnamefont {Karevski}},\ }\href@noop
  {} {\bibfield  {journal} {\bibinfo  {journal} {Physical Review E}\ }\textbf
  {\bibinfo {volume} {90}},\ \bibinfo {pages} {042142} (\bibinfo {year}
  {2014})}\BibitemShut {NoStop}%
\bibitem [{\citenamefont {Giovannetti}\ and\ \citenamefont
  {Palma}(2012)}]{Giovannetti2012}%
  \BibitemOpen
  \bibfield  {author} {\bibinfo {author} {\bibfnamefont {V.}~\bibnamefont
  {Giovannetti}}\ and\ \bibinfo {author} {\bibfnamefont {G.~M.}\ \bibnamefont
  {Palma}},\ }\href {\doibase 10.1103/PhysRevLett.108.040401} {\bibfield
  {journal} {\bibinfo  {journal} {Physical Review Letters}\ }\textbf {\bibinfo
  {volume} {108}},\ \bibinfo {pages} {040401} (\bibinfo {year}
  {2012})}\BibitemShut {NoStop}%
\bibitem [{\citenamefont {Strasberg}\ \emph {et~al.}(2017)\citenamefont
  {Strasberg}, \citenamefont {Schaller}, \citenamefont {Brandes},\ and\
  \citenamefont {Esposito}}]{strasberg2017quantum}%
  \BibitemOpen
  \bibfield  {author} {\bibinfo {author} {\bibfnamefont {P.}~\bibnamefont
  {Strasberg}}, \bibinfo {author} {\bibfnamefont {G.}~\bibnamefont {Schaller}},
  \bibinfo {author} {\bibfnamefont {T.}~\bibnamefont {Brandes}}, \ and\
  \bibinfo {author} {\bibfnamefont {M.}~\bibnamefont {Esposito}},\ }\href@noop
  {} {\bibfield  {journal} {\bibinfo  {journal} {Physical Review X}\ }\textbf
  {\bibinfo {volume} {7}},\ \bibinfo {pages} {021003} (\bibinfo {year}
  {2017})}\BibitemShut {NoStop}%
\bibitem [{\citenamefont {Barra}(2015)}]{barra2015thermodynamic}%
  \BibitemOpen
  \bibfield  {author} {\bibinfo {author} {\bibfnamefont {F.}~\bibnamefont
  {Barra}},\ }\href@noop {} {\bibfield  {journal} {\bibinfo  {journal}
  {Scientific reports}\ }\textbf {\bibinfo {volume} {5}},\ \bibinfo {pages}
  {14873} (\bibinfo {year} {2015})}\BibitemShut {NoStop}%
\bibitem [{\citenamefont {De~Chiara}\ \emph {et~al.}(2018)\citenamefont
  {De~Chiara}, \citenamefont {Landi}, \citenamefont {Hewgill}, \citenamefont
  {Reid}, \citenamefont {Ferraro}, \citenamefont {Roncaglia},\ and\
  \citenamefont {Antezza}}]{de2018reconciliation}%
  \BibitemOpen
  \bibfield  {author} {\bibinfo {author} {\bibfnamefont {G.}~\bibnamefont
  {De~Chiara}}, \bibinfo {author} {\bibfnamefont {G.}~\bibnamefont {Landi}},
  \bibinfo {author} {\bibfnamefont {A.}~\bibnamefont {Hewgill}}, \bibinfo
  {author} {\bibfnamefont {B.}~\bibnamefont {Reid}}, \bibinfo {author}
  {\bibfnamefont {A.}~\bibnamefont {Ferraro}}, \bibinfo {author} {\bibfnamefont
  {A.~J.}\ \bibnamefont {Roncaglia}}, \ and\ \bibinfo {author} {\bibfnamefont
  {M.}~\bibnamefont {Antezza}},\ }\href@noop {} {\bibfield  {journal} {\bibinfo
   {journal} {New Journal of Physics}\ }\textbf {\bibinfo {volume} {20}},\
  \bibinfo {pages} {113024} (\bibinfo {year} {2018})}\BibitemShut {NoStop}%
\bibitem [{\citenamefont {Ryb{\'a}r}\ \emph {et~al.}(2012)\citenamefont
  {Ryb{\'a}r}, \citenamefont {Filippov}, \citenamefont {Ziman},\ and\
  \citenamefont {Bu{\v{z}}ek}}]{rybar2012simulation}%
  \BibitemOpen
  \bibfield  {author} {\bibinfo {author} {\bibfnamefont {T.}~\bibnamefont
  {Ryb{\'a}r}}, \bibinfo {author} {\bibfnamefont {S.~N.}\ \bibnamefont
  {Filippov}}, \bibinfo {author} {\bibfnamefont {M.}~\bibnamefont {Ziman}}, \
  and\ \bibinfo {author} {\bibfnamefont {V.}~\bibnamefont {Bu{\v{z}}ek}},\
  }\href@noop {} {\bibfield  {journal} {\bibinfo  {journal} {Journal of Physics
  B: Atomic, Molecular and Optical Physics}\ }\textbf {\bibinfo {volume}
  {45}},\ \bibinfo {pages} {154006} (\bibinfo {year} {2012})}\BibitemShut
  {NoStop}%
\bibitem [{\citenamefont {Bernardes}\ \emph {et~al.}(2014)\citenamefont
  {Bernardes}, \citenamefont {Carvalho}, \citenamefont {Monken},\ and\
  \citenamefont {Santos}}]{bernardes2014environmental}%
  \BibitemOpen
  \bibfield  {author} {\bibinfo {author} {\bibfnamefont {N.}~\bibnamefont
  {Bernardes}}, \bibinfo {author} {\bibfnamefont {A.}~\bibnamefont {Carvalho}},
  \bibinfo {author} {\bibfnamefont {C.}~\bibnamefont {Monken}}, \ and\ \bibinfo
  {author} {\bibfnamefont {M.~F.}\ \bibnamefont {Santos}},\ }\href@noop {}
  {\bibfield  {journal} {\bibinfo  {journal} {Physical Review A}\ }\textbf
  {\bibinfo {volume} {90}},\ \bibinfo {pages} {032111} (\bibinfo {year}
  {2014})}\BibitemShut {NoStop}%
\bibitem [{\citenamefont {Bernardes}\ \emph {et~al.}(2017)\citenamefont
  {Bernardes}, \citenamefont {Carvalho}, \citenamefont {Monken},\ and\
  \citenamefont {Santos}}]{bernardes2017coarse}%
  \BibitemOpen
  \bibfield  {author} {\bibinfo {author} {\bibfnamefont {N.~K.}\ \bibnamefont
  {Bernardes}}, \bibinfo {author} {\bibfnamefont {A.~R.}\ \bibnamefont
  {Carvalho}}, \bibinfo {author} {\bibfnamefont {C.}~\bibnamefont {Monken}}, \
  and\ \bibinfo {author} {\bibfnamefont {M.~F.}\ \bibnamefont {Santos}},\
  }\href@noop {} {\bibfield  {journal} {\bibinfo  {journal} {Physical Review
  A}\ }\textbf {\bibinfo {volume} {95}},\ \bibinfo {pages} {032117} (\bibinfo
  {year} {2017})}\BibitemShut {NoStop}%
\bibitem [{\citenamefont {Mascarenhas}\ and\ \citenamefont
  {De~Vega}(2017)}]{mascarenhas2017quantum}%
  \BibitemOpen
  \bibfield  {author} {\bibinfo {author} {\bibfnamefont {E.}~\bibnamefont
  {Mascarenhas}}\ and\ \bibinfo {author} {\bibfnamefont {I.}~\bibnamefont
  {De~Vega}},\ }\href@noop {} {\bibfield  {journal} {\bibinfo  {journal}
  {Physical Review A}\ }\textbf {\bibinfo {volume} {96}},\ \bibinfo {pages}
  {062117} (\bibinfo {year} {2017})}\BibitemShut {NoStop}%
\bibitem [{\citenamefont {Man}\ \emph {et~al.}(2018)\citenamefont {Man},
  \citenamefont {Xia},\ and\ \citenamefont {Franco}}]{man2018temperature}%
  \BibitemOpen
  \bibfield  {author} {\bibinfo {author} {\bibfnamefont {Z.-X.}\ \bibnamefont
  {Man}}, \bibinfo {author} {\bibfnamefont {Y.-J.}\ \bibnamefont {Xia}}, \ and\
  \bibinfo {author} {\bibfnamefont {R.~L.}\ \bibnamefont {Franco}},\
  }\href@noop {} {\bibfield  {journal} {\bibinfo  {journal} {Physical Review
  A}\ }\textbf {\bibinfo {volume} {97}},\ \bibinfo {pages} {062104} (\bibinfo
  {year} {2018})}\BibitemShut {NoStop}%
\bibitem [{\citenamefont {Ciccarello}\ \emph {et~al.}(2013)\citenamefont
  {Ciccarello}, \citenamefont {Palma},\ and\ \citenamefont
  {Giovannetti}}]{ciccarello2013collision}%
  \BibitemOpen
  \bibfield  {author} {\bibinfo {author} {\bibfnamefont {F.}~\bibnamefont
  {Ciccarello}}, \bibinfo {author} {\bibfnamefont {G.}~\bibnamefont {Palma}}, \
  and\ \bibinfo {author} {\bibfnamefont {V.}~\bibnamefont {Giovannetti}},\
  }\href@noop {} {\bibfield  {journal} {\bibinfo  {journal} {Physical Review
  A}\ }\textbf {\bibinfo {volume} {87}},\ \bibinfo {pages} {040103} (\bibinfo
  {year} {2013})}\BibitemShut {NoStop}%
\bibitem [{\citenamefont {Ciccarello}\ and\ \citenamefont
  {Giovannetti}(2013)}]{ciccarello2013quantum}%
  \BibitemOpen
  \bibfield  {author} {\bibinfo {author} {\bibfnamefont {F.}~\bibnamefont
  {Ciccarello}}\ and\ \bibinfo {author} {\bibfnamefont {V.}~\bibnamefont
  {Giovannetti}},\ }\href@noop {} {\bibfield  {journal} {\bibinfo  {journal}
  {Physica Scripta}\ }\textbf {\bibinfo {volume} {2013}},\ \bibinfo {pages}
  {014010} (\bibinfo {year} {2013})}\BibitemShut {NoStop}%
\bibitem [{\citenamefont {McCloskey}\ and\ \citenamefont
  {Paternostro}(2014)}]{mccloskey2014non}%
  \BibitemOpen
  \bibfield  {author} {\bibinfo {author} {\bibfnamefont {R.}~\bibnamefont
  {McCloskey}}\ and\ \bibinfo {author} {\bibfnamefont {M.}~\bibnamefont
  {Paternostro}},\ }\href@noop {} {\bibfield  {journal} {\bibinfo  {journal}
  {Physical Review A}\ }\textbf {\bibinfo {volume} {89}},\ \bibinfo {pages}
  {052120} (\bibinfo {year} {2014})}\BibitemShut {NoStop}%
\bibitem [{\citenamefont {{\c{C}}akmak}\ \emph {et~al.}(2017)\citenamefont
  {{\c{C}}akmak}, \citenamefont {Pezzutto}, \citenamefont {Paternostro},\ and\
  \citenamefont {M{\"u}stecapl{\i}o{\u{g}}lu}}]{ccakmak2017non}%
  \BibitemOpen
  \bibfield  {author} {\bibinfo {author} {\bibfnamefont {B.}~\bibnamefont
  {{\c{C}}akmak}}, \bibinfo {author} {\bibfnamefont {M.}~\bibnamefont
  {Pezzutto}}, \bibinfo {author} {\bibfnamefont {M.}~\bibnamefont
  {Paternostro}}, \ and\ \bibinfo {author} {\bibfnamefont {{\"O}.}~\bibnamefont
  {M{\"u}stecapl{\i}o{\u{g}}lu}},\ }\href@noop {} {\bibfield  {journal}
  {\bibinfo  {journal} {Physical Review A}\ }\textbf {\bibinfo {volume} {96}},\
  \bibinfo {pages} {022109} (\bibinfo {year} {2017})}\BibitemShut {NoStop}%
\bibitem [{\citenamefont {Kretschmer}\ \emph {et~al.}(2016)\citenamefont
  {Kretschmer}, \citenamefont {Luoma},\ and\ \citenamefont
  {Strunz}}]{kretschmer2016collision}%
  \BibitemOpen
  \bibfield  {author} {\bibinfo {author} {\bibfnamefont {S.}~\bibnamefont
  {Kretschmer}}, \bibinfo {author} {\bibfnamefont {K.}~\bibnamefont {Luoma}}, \
  and\ \bibinfo {author} {\bibfnamefont {W.~T.}\ \bibnamefont {Strunz}},\
  }\href@noop {} {\bibfield  {journal} {\bibinfo  {journal} {Physical Review
  A}\ }\textbf {\bibinfo {volume} {94}},\ \bibinfo {pages} {012106} (\bibinfo
  {year} {2016})}\BibitemShut {NoStop}%
\bibitem [{\citenamefont {Campbell}\ \emph {et~al.}(2018)\citenamefont
  {Campbell}, \citenamefont {Ciccarello}, \citenamefont {Palma},\ and\
  \citenamefont {Vacchini}}]{campbell2018system}%
  \BibitemOpen
  \bibfield  {author} {\bibinfo {author} {\bibfnamefont {S.}~\bibnamefont
  {Campbell}}, \bibinfo {author} {\bibfnamefont {F.}~\bibnamefont
  {Ciccarello}}, \bibinfo {author} {\bibfnamefont {G.~M.}\ \bibnamefont
  {Palma}}, \ and\ \bibinfo {author} {\bibfnamefont {B.}~\bibnamefont
  {Vacchini}},\ }\href@noop {} {\bibfield  {journal} {\bibinfo  {journal}
  {Physical Review A}\ }\textbf {\bibinfo {volume} {98}},\ \bibinfo {pages}
  {012142} (\bibinfo {year} {2018})}\BibitemShut {NoStop}%
\bibitem [{\citenamefont {Lorenzo}\ \emph {et~al.}(2017)\citenamefont
  {Lorenzo}, \citenamefont {Ciccarello},\ and\ \citenamefont
  {Palma}}]{lorenzo2017composite}%
  \BibitemOpen
  \bibfield  {author} {\bibinfo {author} {\bibfnamefont {S.}~\bibnamefont
  {Lorenzo}}, \bibinfo {author} {\bibfnamefont {F.}~\bibnamefont {Ciccarello}},
  \ and\ \bibinfo {author} {\bibfnamefont {G.~M.}\ \bibnamefont {Palma}},\
  }\href@noop {} {\bibfield  {journal} {\bibinfo  {journal} {Physical Review
  A}\ }\textbf {\bibinfo {volume} {96}},\ \bibinfo {pages} {032107} (\bibinfo
  {year} {2017})}\BibitemShut {NoStop}%
\bibitem [{\citenamefont {Jin}\ and\ \citenamefont {Yu}(2018)}]{jin2018non}%
  \BibitemOpen
  \bibfield  {author} {\bibinfo {author} {\bibfnamefont {J.}~\bibnamefont
  {Jin}}\ and\ \bibinfo {author} {\bibfnamefont {C.-s.}\ \bibnamefont {Yu}},\
  }\href@noop {} {\bibfield  {journal} {\bibinfo  {journal} {New Journal of
  Physics}\ }\textbf {\bibinfo {volume} {20}},\ \bibinfo {pages} {053026}
  (\bibinfo {year} {2018})}\BibitemShut {NoStop}%
\bibitem [{\citenamefont {Serafini}(2017)}]{serafini2017quantum}%
  \BibitemOpen
  \bibfield  {author} {\bibinfo {author} {\bibfnamefont {A.}~\bibnamefont
  {Serafini}},\ }\href@noop {} {\emph {\bibinfo {title} {Quantum continuous
  variables: a primer of theoretical methods}}}\ (\bibinfo  {publisher} {CRC
  Press},\ \bibinfo {year} {2017})\BibitemShut {NoStop}%
\bibitem [{\citenamefont {Ferraro}\ \emph {et~al.}(2005)\citenamefont
  {Ferraro}, \citenamefont {Olivares},\ and\ \citenamefont
  {Paris}}]{ferraro2005gaussian}%
  \BibitemOpen
  \bibfield  {author} {\bibinfo {author} {\bibfnamefont {A.}~\bibnamefont
  {Ferraro}}, \bibinfo {author} {\bibfnamefont {S.}~\bibnamefont {Olivares}}, \
  and\ \bibinfo {author} {\bibfnamefont {M.~G.}\ \bibnamefont {Paris}},\
  }\href@noop {} {\bibfield  {journal} {\bibinfo  {journal} {arXiv preprint
  quant-ph/0503237}\ } (\bibinfo {year} {2005})}\BibitemShut {NoStop}%
\bibitem [{\citenamefont {Adesso}\ and\ \citenamefont
  {Illuminati}(2007)}]{adesso2007entanglement}%
  \BibitemOpen
  \bibfield  {author} {\bibinfo {author} {\bibfnamefont {G.}~\bibnamefont
  {Adesso}}\ and\ \bibinfo {author} {\bibfnamefont {F.}~\bibnamefont
  {Illuminati}},\ }\href@noop {} {\bibfield  {journal} {\bibinfo  {journal}
  {Journal of Physics A: Mathematical and Theoretical}\ }\textbf {\bibinfo
  {volume} {40}},\ \bibinfo {pages} {7821} (\bibinfo {year}
  {2007})}\BibitemShut {NoStop}%
\bibitem [{\citenamefont {Adesso}\ \emph {et~al.}(2014)\citenamefont {Adesso},
  \citenamefont {Ragy},\ and\ \citenamefont {Lee}}]{adesso2014continuous}%
  \BibitemOpen
  \bibfield  {author} {\bibinfo {author} {\bibfnamefont {G.}~\bibnamefont
  {Adesso}}, \bibinfo {author} {\bibfnamefont {S.}~\bibnamefont {Ragy}}, \ and\
  \bibinfo {author} {\bibfnamefont {A.~R.}\ \bibnamefont {Lee}},\ }\href@noop
  {} {\bibfield  {journal} {\bibinfo  {journal} {Open Systems \& Information
  Dynamics}\ }\textbf {\bibinfo {volume} {21}},\ \bibinfo {pages} {1440001}
  (\bibinfo {year} {2014})}\BibitemShut {NoStop}%
\bibitem [{\citenamefont {Holevo}(2007)}]{holevo2007one}%
  \BibitemOpen
  \bibfield  {author} {\bibinfo {author} {\bibfnamefont {A.~S.}\ \bibnamefont
  {Holevo}},\ }\href@noop {} {\bibfield  {journal} {\bibinfo  {journal}
  {Problems of Information Transmission}\ }\textbf {\bibinfo {volume} {43}},\
  \bibinfo {pages} {1} (\bibinfo {year} {2007})}\BibitemShut {NoStop}%
\bibitem [{\citenamefont {Caruso}\ \emph {et~al.}(2006)\citenamefont {Caruso},
  \citenamefont {Giovannetti},\ and\ \citenamefont {Holevo}}]{caruso2006one}%
  \BibitemOpen
  \bibfield  {author} {\bibinfo {author} {\bibfnamefont {F.}~\bibnamefont
  {Caruso}}, \bibinfo {author} {\bibfnamefont {V.}~\bibnamefont {Giovannetti}},
  \ and\ \bibinfo {author} {\bibfnamefont {A.~S.}\ \bibnamefont {Holevo}},\
  }\href@noop {} {\bibfield  {journal} {\bibinfo  {journal} {New Journal of
  Physics}\ }\textbf {\bibinfo {volume} {8}},\ \bibinfo {pages} {310} (\bibinfo
  {year} {2006})}\BibitemShut {NoStop}%
\bibitem [{\citenamefont {Simon}\ \emph {et~al.}(1987)\citenamefont {Simon},
  \citenamefont {Sudarshan},\ and\ \citenamefont
  {Mukunda}}]{simon1987gaussian}%
  \BibitemOpen
  \bibfield  {author} {\bibinfo {author} {\bibfnamefont {R.}~\bibnamefont
  {Simon}}, \bibinfo {author} {\bibfnamefont {E.}~\bibnamefont {Sudarshan}}, \
  and\ \bibinfo {author} {\bibfnamefont {N.}~\bibnamefont {Mukunda}},\
  }\href@noop {} {\bibfield  {journal} {\bibinfo  {journal} {Physical Review
  A}\ }\textbf {\bibinfo {volume} {36}},\ \bibinfo {pages} {3868} (\bibinfo
  {year} {1987})}\BibitemShut {NoStop}%
\bibitem [{\citenamefont {Simon}\ \emph {et~al.}(1988)\citenamefont {Simon},
  \citenamefont {Sudarshan},\ and\ \citenamefont
  {Mukunda}}]{simon1988gaussian}%
  \BibitemOpen
  \bibfield  {author} {\bibinfo {author} {\bibfnamefont {R.}~\bibnamefont
  {Simon}}, \bibinfo {author} {\bibfnamefont {E.}~\bibnamefont {Sudarshan}}, \
  and\ \bibinfo {author} {\bibfnamefont {N.}~\bibnamefont {Mukunda}},\
  }\href@noop {} {\bibfield  {journal} {\bibinfo  {journal} {Physical Review
  A}\ }\textbf {\bibinfo {volume} {37}},\ \bibinfo {pages} {3028} (\bibinfo
  {year} {1988})}\BibitemShut {NoStop}%
\bibitem [{\citenamefont {Simon}\ \emph {et~al.}(1994)\citenamefont {Simon},
  \citenamefont {Mukunda},\ and\ \citenamefont {Dutta}}]{simon1994quantum}%
  \BibitemOpen
  \bibfield  {author} {\bibinfo {author} {\bibfnamefont {R.}~\bibnamefont
  {Simon}}, \bibinfo {author} {\bibfnamefont {N.}~\bibnamefont {Mukunda}}, \
  and\ \bibinfo {author} {\bibfnamefont {B.}~\bibnamefont {Dutta}},\
  }\href@noop {} {\bibfield  {journal} {\bibinfo  {journal} {Physical Review
  A}\ }\textbf {\bibinfo {volume} {49}},\ \bibinfo {pages} {1567} (\bibinfo
  {year} {1994})}\BibitemShut {NoStop}%
\bibitem [{Note1()}]{Note1}%
  \BibitemOpen
  \bibinfo {note} {\protect \href
  {https://github.com/gtlandi/gaussianonmark}{https://github.com/gtlandi/gaussianonmark}}\BibitemShut
  {NoStop}%
\bibitem [{\citenamefont {Kraus}(1983)}]{Kraus1983}%
  \BibitemOpen
  \bibfield  {author} {\bibinfo {author} {\bibfnamefont {K.}~\bibnamefont
  {Kraus}},\ }\href@noop {} {\emph {\bibinfo {title} {{States, Effects, and
  Operations: Fundamental Notions of Quantum Theory}}}},\ edited by\ \bibinfo
  {editor} {\bibfnamefont {A.}~\bibnamefont {B{\"{o}}hm}}, \bibinfo {editor}
  {\bibfnamefont {J.~D.}\ \bibnamefont {Dollard}}, \ and\ \bibinfo {editor}
  {\bibfnamefont {W.~H.}\ \bibnamefont {Wooters}}\ (\bibinfo  {publisher}
  {Springer},\ \bibinfo {address} {Heidelberg},\ \bibinfo {year} {1983})\ p.\
  \bibinfo {pages} {154}\BibitemShut {NoStop}%
\bibitem [{\citenamefont {Nielsen}\ and\ \citenamefont
  {Chuang}(2000)}]{Nielsen}%
  \BibitemOpen
  \bibfield  {author} {\bibinfo {author} {\bibfnamefont {M.~A.}\ \bibnamefont
  {Nielsen}}\ and\ \bibinfo {author} {\bibfnamefont {I.~L.}\ \bibnamefont
  {Chuang}},\ }\href@noop {} {\emph {\bibinfo {title} {{Quantum Computation and
  Quantum Information}}}}\ (\bibinfo  {publisher} {Cambridge University
  Press},\ \bibinfo {year} {2000})\BibitemShut {NoStop}%
\bibitem [{\citenamefont {Nakajima}(1958)}]{Nakajima1958}%
  \BibitemOpen
  \bibfield  {author} {\bibinfo {author} {\bibfnamefont {S.}~\bibnamefont
  {Nakajima}},\ }\href {\doibase 10.1143/ptp.20.948} {\bibfield  {journal}
  {\bibinfo  {journal} {Progress of Theoretical Physics}\ }\textbf {\bibinfo
  {volume} {20}},\ \bibinfo {pages} {948} (\bibinfo {year} {1958})}\BibitemShut
  {NoStop}%
\bibitem [{\citenamefont {Zwanzig}(1960)}]{Zwanzig1960}%
  \BibitemOpen
  \bibfield  {author} {\bibinfo {author} {\bibfnamefont {R.}~\bibnamefont
  {Zwanzig}},\ }\href {\doibase 10.1063/1.1731409} {\bibfield  {journal}
  {\bibinfo  {journal} {The Journal of Chemical Physics}\ }\textbf {\bibinfo
  {volume} {33}},\ \bibinfo {pages} {1338} (\bibinfo {year}
  {1960})}\BibitemShut {NoStop}%
\bibitem [{\citenamefont {Turkington}(2013)}]{Vectorization}%
  \BibitemOpen
  \bibfield  {author} {\bibinfo {author} {\bibfnamefont {D.~A.}\ \bibnamefont
  {Turkington}},\ }\href@noop {} {\emph {\bibinfo {title} {{Generalized
  Vectorization, Cross-Products, and Matrix Calculus}}}}\ (\bibinfo
  {publisher} {Cambridge University Press},\ \bibinfo {address} {Cambridge},\
  \bibinfo {year} {2013})\ p.\ \bibinfo {pages} {275}\BibitemShut {NoStop}%
\bibitem [{\citenamefont {Mazzola}\ \emph
  {et~al.}(2010{\natexlab{b}})\citenamefont {Mazzola}, \citenamefont {Laine},
  \citenamefont {Breuer}, \citenamefont {Maniscalco},\ and\ \citenamefont
  {Piilo}}]{mazzola2010phenomenological}%
  \BibitemOpen
  \bibfield  {author} {\bibinfo {author} {\bibfnamefont {L.}~\bibnamefont
  {Mazzola}}, \bibinfo {author} {\bibfnamefont {E.-M.}\ \bibnamefont {Laine}},
  \bibinfo {author} {\bibfnamefont {H.-P.}\ \bibnamefont {Breuer}}, \bibinfo
  {author} {\bibfnamefont {S.}~\bibnamefont {Maniscalco}}, \ and\ \bibinfo
  {author} {\bibfnamefont {J.}~\bibnamefont {Piilo}},\ }\href@noop {}
  {\bibfield  {journal} {\bibinfo  {journal} {Physical Review A}\ }\textbf
  {\bibinfo {volume} {81}},\ \bibinfo {pages} {062120} (\bibinfo {year}
  {2010}{\natexlab{b}})}\BibitemShut {NoStop}%
\bibitem [{\citenamefont {Torre}\ \emph {et~al.}(2015)\citenamefont {Torre},
  \citenamefont {Roga},\ and\ \citenamefont {Illuminati}}]{torre2015non}%
  \BibitemOpen
  \bibfield  {author} {\bibinfo {author} {\bibfnamefont {G.}~\bibnamefont
  {Torre}}, \bibinfo {author} {\bibfnamefont {W.}~\bibnamefont {Roga}}, \ and\
  \bibinfo {author} {\bibfnamefont {F.}~\bibnamefont {Illuminati}},\
  }\href@noop {} {\bibfield  {journal} {\bibinfo  {journal} {Physical Review
  Letters}\ }\textbf {\bibinfo {volume} {115}},\ \bibinfo {pages} {070401}
  (\bibinfo {year} {2015})}\BibitemShut {NoStop}%
\bibitem [{\citenamefont {Liuzzo-Scorpo}\ \emph {et~al.}(2017)\citenamefont
  {Liuzzo-Scorpo}, \citenamefont {Roga}, \citenamefont {Souza}, \citenamefont
  {Bernardes},\ and\ \citenamefont {Adesso}}]{liuzzo2017non}%
  \BibitemOpen
  \bibfield  {author} {\bibinfo {author} {\bibfnamefont {P.}~\bibnamefont
  {Liuzzo-Scorpo}}, \bibinfo {author} {\bibfnamefont {W.}~\bibnamefont {Roga}},
  \bibinfo {author} {\bibfnamefont {L.~A.}\ \bibnamefont {Souza}}, \bibinfo
  {author} {\bibfnamefont {N.~K.}\ \bibnamefont {Bernardes}}, \ and\ \bibinfo
  {author} {\bibfnamefont {G.}~\bibnamefont {Adesso}},\ }\href@noop {}
  {\bibfield  {journal} {\bibinfo  {journal} {Physical Review Letters}\
  }\textbf {\bibinfo {volume} {118}},\ \bibinfo {pages} {050401} (\bibinfo
  {year} {2017})}\BibitemShut {NoStop}%
\bibitem [{\citenamefont {Lindblad}(2000)}]{lindblad2000cloning}%
  \BibitemOpen
  \bibfield  {author} {\bibinfo {author} {\bibfnamefont {G.}~\bibnamefont
  {Lindblad}},\ }\href@noop {} {\bibfield  {journal} {\bibinfo  {journal}
  {Journal of Physics A: Mathematical and General}\ }\textbf {\bibinfo {volume}
  {33}},\ \bibinfo {pages} {5059} (\bibinfo {year} {2000})}\BibitemShut
  {NoStop}%
\end{thebibliography}%
\end{document}